\documentclass{raa_twocolumn} 

\usepackage{graphicx,times}             
\usepackage{natbib}
\usepackage{amssymb,amsmath}
\usepackage{longtable}
\usepackage{subcaption}
\usepackage{graphicx}
\usepackage{multirow}
\usepackage{cancel}
\usepackage{pdflscape}
\usepackage{grffile}
\usepackage{float}
\usepackage{threeparttable}
\usepackage{xcolor} 
\bibpunct{(}{)}{;}{a}{}{,}
\usepackage[a4paper=true,pagebackref=true]{hyperref}
\hypersetup{colorlinks = true, linkcolor = blue, anchorcolor = blue, citecolor = blue, filecolor = blue, pagecolor = blue, urlcolor = blue}

\begin{document}

  \title{FAST observations of an extremely active episode of FRB\,20201124A: \\ \uppercase\expandafter{\romannumeral4}. Spin Period Search}

   \volnopage{Vol.0 (20xx) No.0, 000--000}      
   \setcounter{page}{1}          
  \author{Jia-Rui Niu \inst{1,2}, Wei-Wei Zhu \inst{1,3}, Bing Zhang \inst{4}, Mao Yuan \inst{1,2}, De-Jiang Zhou \inst{1,2}, Yong-Kun Zhang \inst{1,2}, Jin-Chen Jiang \inst{1,6}, J.~L. Han \inst{1,2}, Di Li \inst{1},  Ke-Jia Lee \inst{1,5,6}, Pei Wang \inst{1}, Yi Feng \inst{7}, Dong-Zi Li \inst{8}, Rui Luo \inst{9}, Fa-Yin Wang \inst{10,11}, Zi-Gao Dai \inst{10,12}, Chen-Chen Miao \inst{1,2},  Chen-Hui Niu \inst{1} , Heng Xu \inst{5,6},  Chun-Feng Zhang \inst{5,6}, Wei-Yang Wang \inst{5,6}, Bo-Jun Wang \inst{5,6}, Jiang-Wei Xu \inst{5,6}}

   \institute{National Astronomical Observatories, Chinese Academy of Sciences,
             Beijing 100101, China; {\it zhuww@nao.cas.cn}\\
        \and
             University of Chinese Academy of Sciences, Beijing, China.\\
        \and                
             Institute for Frontier in Astronomy and Astrophysics, Beijing Normal University,  Beijing 102206, China. \\
        \and
             Department of Physics and Astronomy, University of Nevada, Las Vegas, NV 89154, USA; {\it bing.zhang@unlv.edu}\\    
        \and
             Kavli Institute for Astronomy and Astrophysics, Peking University, Beijing 100871, China \\
        \and     
            Department of Astronomy, Peking University, Beijing 100871, China \\
        \and
            Research Center for Intelligent Computing Platforms, Zhejiang Laboratory, Hangzhou 311100, China\\
        \and             
            Cahill Center for Astronomy and Astrophysics, MC 249-17 California Institute of Technology, Pasadena CA 91125, USA\\ 
        \and
            CSIRO Space and Astronomy, PO Box 76, Epping, NSW 1710, Australia\\
        \and 
            School of Astronomy and Space Science, Nanjing University, Nanjing 210093, China\\
        \and 
            Key Laboratory of Modern Astronomy and Astrophysics (Nanjing University), Ministry of Education, China\\
        \and
            Department of Astronomy, University of Science and Technology of China, Hefei 230026, China\\
\vs\no
   {\small Received 20xx month day; accepted 20xx month day}}
\abstract{
We report the properties of more than 800 bursts detected from the repeating fast radio burst (FRB) source FRB\,20201124A with the Five-hundred-meter Aperture Spherical radio telescope (FAST) during an extremely active episode on UTC September 25th-28th, 2021 in a series of four papers.
In this fourth paper of the series, we present a systematic search of the spin period and linear acceleration of the source object from both 996 individual pulse peaks and the dedispersed time series.
No credible spin period was found from this data set.
We rule out the presence of significant periodicity in the range between 1~ms to 100~s with a pulse duty cycle $\textless$ 0.49$\pm0.08$ (when the profile is defined by a von-Mises function, not a boxcar function) and linear acceleration up to $300$~m~s$^{-2}$ in each of the four one-hour observing sessions, and up to $0.6$~m~s$^{-2}$ in all 4 days.
These searches contest theoretical scenarios involving a 1~ms to 100~s isolated magnetar/pulsar with surface magnetic field $<10^{15}$~G and a small duty cycle (such as in a polar-cap emission mode) or a pulsar with a companion star or black hole up to 100\,M$_{\rm \odot}$ and $P_b>10$~hours.
We also perform a periodicity search of the fine structures and identify 53 unrelated millisecond-timescale "periods" in multi-components with the highest significance of 3.9\,$\sigma$.
The "periods" recovered from the fine structures are neither consistent nor harmonically related. Thus they are not likely to come from a spin period. We caution against claiming spin periodicity with significance below $\sim$ 4\,$\sigma$ with multi-components from one-off FRBs. We discuss the implications of our results and the possible connections between FRB multi-components and pulsar micro-structures.  
}

   \authorrunning{Jia-Rui Niu, et al.}            
   \titlerunning{FAST observation of FRB\,20201124A: \\ \uppercase\expandafter{\romannumeral4}. Periodicity}  

\maketitle
\section{Introduction}           
\label{sect:intro}
Fast radio bursts~(FRBs) are bright radio transients with milliseconds duration discovered since 2007 \citep{lorimer2007bright}. 
Most of the known FRBs are one-off events, but a small group of them repeat and are called repeating FRBs \citep{spitler2016repeating}. 
Their origin is not yet fully understood. 
The detection of FRB\,200428 associated with the galactic magnetar SGR\,J1935+2154 suggests that magnetars could produce at least some FRBs \citep{bochenek2020fast,Chime2020Nature,lin2020no,lick2021NA,tavani2021x}.

Active repeating FRBs are observed to emit apparently sporadic pulses separated by seconds. Finding a spin period ($P$) from a repeating FRB source could be the smoking gun evidence for a neutron star origin. 
However, no apparent periods have been detected in the millisecond to second range for either FRB\,20121102A or FRB\,20201124A, two of the most well-studied repeaters, despite thousands of pulses have been detected \citep{li2021bimodal,caleb2020simultaneous,cruces2021repeating,gourdji2019sample, xu2021fast}. The same applies to FRB\,20180916B \citep{2020Natur.582..351C,chawla2020detection,marthi2020detection}, which has a 16-day longer period. 
\cite{caleb2020simultaneous} used the \texttt{PRESTO}\footnote{\url{https://github.com/scottransom/presto}}\citep{ransom2011presto} to perform a Fast Fourier Transform (FFT) periodic search of FRB\,20121102A using MeerKAT on the 6th and 8th of October 2019. After folding and visual checking, no credible period was found. They also performed a Fast Folding Algorithm (FFA) \citep{staelin1969fast, morello2020optimal} to search for periods ranging from 500\,ms to 10\,min, but did not detect any significant periodic pulsations above the S/N threshold of 8.
\cite{cruces2021repeating} performed an extensive multi-wavelength observation in radio, optical, X-ray and $\gamma$-ray of FRB\,20121102A. They searched with DM trials for the acceleration search and the FFA search \citep{staelin1969fast, morello2020optimal}. No source was found among the candidates down to an S/N of 6. \cite{li2021bimodal} observed FRB\,20121102A using the FAST telescope and performed a period search of $P$ in 1\,ms -- 1000\,s and $\dot{P}$ in $10^{-12}$ -- $10^{-2}$ $s\,s^{-1}$ with the Phase-Folding and the Lomb-Scargle Periodogram (LSP) methods. They did not detect any periodicity or quasi-periodicity and proposed that this challenges the theory related to single rotating compact object.
\cite{jahns2022frb} studied the period and waiting time characteristics of FRB\,20121102A with the Arecibo Telescope. They used the Pearson $\chi^2$ test and the LSP to search for periods between 10\,ms and 100\,s. They found no statistically significant periods from two observations, one with 218 bursts detected in 1~hr and one with 227 in 1.38~hr. \cite{https://doi.org/10.48550/arxiv.2206.03759} observed FRB\,20200120E using the Effelsberg telescope. The highest event rate reached 53 bursts in 40 minutes. They used \texttt{PRESTO}’s \texttt{accelsearch}, brute force search and fold search with $\chi^2$ 
statistic. They found no strict periodicity in the burst arrival times during the storm.

While a clear second-level spin-related periodicity has not yet been observed from any FRB,
some repeating FRBs show an active-dormant cycle on the order of tens to hundreds of days, e.g. FRB\,20180916B -- 16 days \citep{2020Natur.582..351C} and FRB\,20121102A -- 160 days \citep{rajwade2020possible,cruces2021repeating,li2021bimodal}. 
The theoretical interpretations for these long cycles for FRBs include binary systems \citep{ioka2020binary,zhang2020binary,sridhar2021periodic}, magnetar precession \citep{yang2020periodic,levin2020precessing,zanazzi2020periodic,sob2020periodic,tong2020periodicity}, asteroid interactions \cite{dai2020periodic,voisin2021periodic,du2021geometry}, and slow rotating magnetars \citep{beniamini2020periodicity}.

In addition to these long-term cycles, some sub-second periodic fine pulse structures, also called multi-components, have been observed from single FRB events. It was first reported in FRB\,20121102A \citep{hessels2019frb,gourdji2019sample} and later seen in FRB\,180814B \citep{amiri2019second} and many other bursts. 
\cite{andersen2021sub} discovered three FRB signals with multiple components that show a tentative periodicity. The most significant signal comes from FRB\,20191221A, with a period of 216.8(1)\,ms and a significance of $6.5\,\sigma$. 
The authors speculated that the intriguing periodic multi-components could be related to the spin of the source object if the same-period multi-component bursts can be reliably observed to repeat themselves in the future. Alternatively, \citep{chen2021frbs} proposed that the micro-lensing effect could cause such a phenomenon. The superposition of pulses from individual micro-images produces a light curve that appears as multi-peak FRBs.
It should be noted that similar fine structures were also observed from some normal and millisecond pulsars \citep{cordes1990quasiperiodic,de2016detection,liukuo2022}.

In summary, repeating FRBs have shown temporal behaviors on three different time scales: 
1. Pulses were emitted with seconds separations when the source is most active and the separation time distribution appears sporadic; 
2. Active and dormant cycles of tens to hundreds of days have been observed from some FRBs such as FRB\,20121102A \citep{rajwade2020possible} and FRB\,20180916B \citep{2020Natur.582..351C};
3. Millisecond-timescale micro-component fine structures have been observed in some bursts within a duration of hundreds of milliseconds.

FRB\,20201124A was discovered by CHIME \citep{Chime2021ATel14497} and located by giant meter-wave radio telescope (uGMRT) \citep{Wharton2021ATel14538} and European VLBI Networ (EVN) \citep{Nimmo2021arXiv211101600N}. Subsequently, \cite{lanman2021sudden} published the morphology, fluences, and arrival times from the follow-up observations by CHIME.
They did not find any evidence of cyclical behavior on timescales between 1 and 178.5 days.
\citet{hilmarsson2021polarization} published polarization observations by using the Effelsberg 100-m radio telescope at 1.36\,GHz. 
\citet{marthi2021burst} observed FRB\,20201124A in the incoherent array mode using the upgraded giant uGMRT. 
They found persistent radio emissions associated with the host galaxy and detected 48\,bursts at 550-750\,MHz. 
They also searched for spin periodicity by using different period search methods, including \texttt{PRESTO} \citep{ransom2011presto} \texttt{accelsearch} and jerk search \citep{andersen2018fourier}, Fast Folding Algorithm\,(FFA) analysis \citep{staelin1969fast, morello2020optimal}, but no significant periodic candidates were found.
FAST observed bursts of FRB\,20201124A from April 1st, 2021 to June 11th, 2021. \cite{xu2021fast} reported the data and discovered significant and irregular variations of the Faraday Rotation Measurement (RM) and the evidence of polarization Faraday conversion. They searched for periodicity using LSP between 30\,ms and 10\,days and found no obvious periodic signals.

\begin{figure*}[htb!] \centering \includegraphics[width=0.85\linewidth]{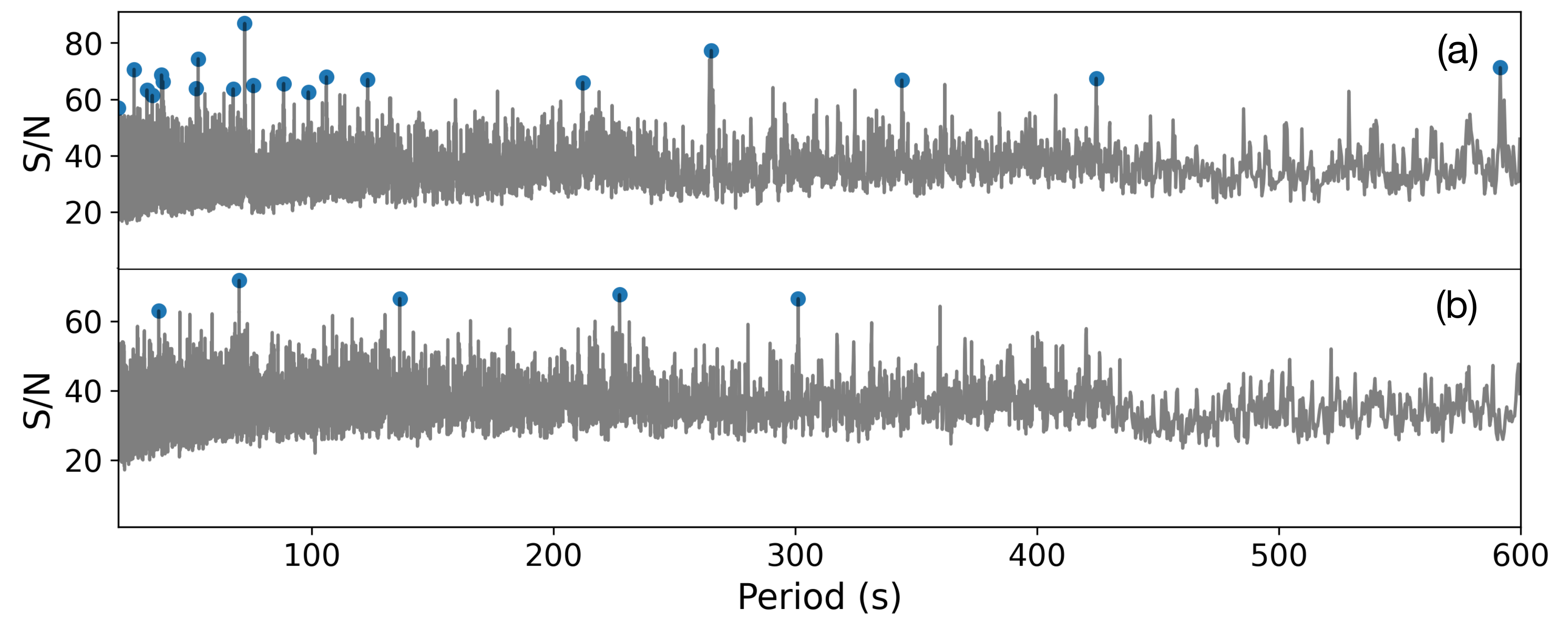} \caption{FFA periodicity search in September 27th (a) and 28th (b) from 20\,s to 600\,s. Comparing the results of these two days, there are no significant matching periods.} \label{tab:ffapsearch} 
\end{figure*}

In the September 2021 episode, FRB\,20201124A exhibited an outburst and reported by CHIME\footnote{\url{https://www.chime-frb.ca/repeaters/FRB 20201124A}}  and Effelsberg telescope \citep{Main2021ATel14933}.
We observed the source during its outburst using FAST. We detected 25 pulses on September 25th, 59 on September 26th, 274 pulses on September 27th and 638 pulses on September 28th. In this paper, we refer to the two rounds of FAST observations of April as the active episode\,1 and September as the active episode\,2. 

The number of pulses reported in this paper differs from those reported in our companion papers (Zhou et al. 2022, Zhang et al. 2022 and Jiang et al. 2022) because we use different criteria for isolating the pulses which suit our purpose of searching for a period.
This extremely active episode enables us to search for the millisecond to second-level spin period from FRB\,20201124A with an unprecedentedly large and high-hourly-rate event sample. We also perform a period acceleration search both in the raw data and from the pulse arrival times to detect or rule out putative binary systems.

In this paper, we employ the following criteria to determine whether the signals have a credible spin period: 1) It has a high statistical significance despite the number of trials searched; 2) It appears in multiple epochs of observations either as identical, close or harmonically related signals.
Abiding by these two criteria, we do not detect any credible periodic signal from the extremely active episode of this active repeating FRB. 

The paper is organized as follows: Section \ref{sect:Observation} describes the observations. In Section \ref{sect:Result}, we present the results of the periodicity search using different methods, including searches using dedispersed time series, using TOAs, using the TOAs of special pulses, and using the multi-components fine structures.
Finally, we briefly discuss and summarize our results in Section \ref{sect:summary}.

\begin{table}[]
\renewcommand\arraystretch{1.5}
\centering
\caption{Observation table.} \label{tab:obs2}
\begin{tabular}{lllll}
\hline\\
\multicolumn{1}{c}{\begin{tabular}[c]{@{}c@{}}Date\\ \\ (Y-M-D)\end{tabular}} & \multicolumn{2}{c}{\begin{tabular}[c]{@{}c@{}}Start Time\\ \\ UTC     ~~~~~~~~~~~~~~~      MJD\end{tabular}} & \multicolumn{1}{c}{\begin{tabular}[c]{@{}c@{}}Duration\\ \\ (s)\end{tabular}} & \multicolumn{1}{c}{Pulses} \\  \\
\hline \\
2021-09-25                                                                 & 22:37:01                                   & 59482.942361                                  & 3600                                                                       & 25                         \\
2021-09-26                                                                 & 20:41:00                                   & 59483.861805                                  & 3600                                                                       & 59                        \\
2021-09-27                                                                 & 19:31:00                                   & 59484.813194                                  & 3600                                                                       & 274                       \\
2021-09-28                                                                 & 18:46:00                                   & 59485.781944                                  & 3600                                                                       & 638                      \\
2021-09-29                                                                 & 18:48:00                                   & 59486.783333                                  & 3600                                                                       & 0                          \\
2021-09-30                                                                 & 19:58:00                                   & 59487.831944                                  & 3600                                                                       & 0                          \\
2021-10-01                                                                 & 21:08:00                                   & 59488.880555                                  & 3600                                                                       & 0                          \\
2021-10-02                                                                 & 20:29:00                                   & 59489.853472                                  & 2357                                                                       & 0                          \\
2021-10-02                                                                 & 21:18:55                                   & 59489.888136                                  & 7814                                                                       & 0                          \\
2021-10-07                                                                 & 22:20:00                                   & 59494.930555                                  & 3600                                                                       & 0                          \\
2021-10-08                                                                 & 21:50:00                                   & 59495.909722                                  & 3600                                                                       & 0                          \\
2021-10-09                                                                 & 21:11:00                                   & 59496.924305                                 & 3600                                                                       & 0                          \\
2021-10-10                                                                 & 21:42:00                                   & 59497.904166                                 & 3600                                                                       & 0                          \\
2021-10-11                                                                 & 21:52:00                                   & 59498.911111                                 & 3600                                                                       & 0                          \\
2021-10-12                                                                 & 17:44:00                                   & 59499.738888                                & 3600                                                                       & 0                          \\
2021-10-13                                                                 & 18:23:00                                   & 59500.765972                                & 3600                                                                       & 0                          \\
2021-10-14                                                                 & 21:29:00                                   & 59501.895138                              & 3600                                                                       & 0                          \\
2021-10-17                                                                 & 21:07:00                                   & 59504.879861                                & 3600                                                                       & 0                          
   \\   \\ \hline

\end{tabular}
\end{table}

\section{Observations}
\label{sect:Observation}
Observation data used in this paper are taken using the FAST telescope with its 19 beam receivers in 1.05-1.45\,GHz \citep{nan2011five,li2018fast, jiang2019commissioning}. Data is recorded with the re-configurable open architecture computing hardware version 2 (ROACH2) System \citep{hickish2016decade}. The recorded data is in PSRFITS format with four polarizations, 49.152\,$\mu$s sampling interval and 4096 channels in search mode.

We utilize the \texttt{psrfits\_utils}\footnote{\url{https://github.com/demorest/psrfits\_utils}} software package to merge the four polarization raw data into one single channel 8-bit intensity data. The program also adjusts the levels of the data according to its running statistics. We then dedisperse and search the data using the \texttt{PRESTO}\footnote{\url{https://github.com/scottransom/presto}}\citep{ransom2011presto} software suite using a single fixed DM value of 413.5 pc\,cm$^{-3}$ and (see Zhou et al. 2022 for the {detailed analysis}of the DM distribution of the pulses)  an S/N threshold of 6.5. We further search into the fine structures of the detected pulses by {using other methods}. We used the \texttt{DSPSR}\footnote{\url{http://dspsr.sourceforge.net/}} \citep{van2011dspsr} program and the \texttt{PSRCHIVE}\footnote{\url{http://psrchive.sourceforge.net/}} \citep{van2012psrchive} software package for polarization analysis. The details of the single pulse analysis will be presented in Section \ref{sec:PsearchTOA}. Observation information is listed in the Table \ref{tab:obs2}.

In our period search based on TOAs, we select special pulses by their polarization properties for part of our period search. 
The FAST observations was taken with four polarizations and calibrated by using a winking 10\,k noise signal injected into the feed with an on-off period of 0.2 seconds in the first minute of our observation. 
We extracted the polarizations and correct for Faraday rotation of the pulses using the software \texttt{DSPSR}\footnote{\url{http://dspsr.sourceforge.net/}} \citep{van2011dspsr} and \texttt{PSRCHIVE}\footnote{\url{http://psrchive.sourceforge.net/}} \citep{van2012psrchive}. The RM of FRB\,20201124A has changed dramatically over time, as observed in the preceding study \citep{xu2021fast}. 
The first active episode's RM change ranges from -887\,rad m$^{-2}$ to -363\,rad m$^{-2}$\citep{xu2021fast}, while the second time's RM change ranges from -650\,rad m$^{-2}$ to {-545\,rad m$^{-2}$} (Jiang et al. 2022). The majority of the pulses are strongly linearly polarized, with a narrow range of PA variation. 
For periodicity search using special pulses, we extracted the polarization signals from the detected pulses by using their best-fit RM value.

\section{Result}
\label{sect:Result}

\subsection{Periodicity search with raw data}
We utilize \texttt{PRESTO}\,\citep{ransom2011presto} to search the dedispersed time series from data taken on the 27th and 28th with high burst rates and employ Fourier transform and \texttt{accelsearch} on the time series. The specific steps and parameters are as follows. 
Firstly, we mask the interference on the original data using \texttt{rfifind} with a time block size of 1~s. Secondly, we dedisperse and retrieve the time series with DM 413.5~pc\,cm$^{-3}$ using \texttt{prepdata}. Thirdly, we run \texttt{accelsearch} on the time series. We search for periodicity between 0.1~ms and 20~s using a sigma cutoff of 5, \texttt{zmax} parameter of 1200, \texttt{wmax} to 4000 to account for the potential Doppler acceleration with the maximum number of harmonics set to 32. Both \texttt{zmax} and \texttt{wmax} are set to the maximum allowed value of \texttt{accelsearch}. 
Based on such setting, we can search Doppler accelerations up to $a/c={\rm zmax}$~$P/T_{\rm obs}^2 = 1\times$10$\rm ^{-4}$~s$^{-1}$ and acceleration derivatives up to $\dot{a}/c = {\rm wmax}$~$P/T_{\rm obs}^3 = 8.6\times10^{-8}$~$\rm s^{-2}$ for a sinusoidal signal of period 1~s and an $a/c=9\times$10$\rm ^{-8}$~s$^{-1}$, $\dot{a}= 8.6\times10^{-11}\rm s^{-2}$ for a period of 1~ms and an observing time of $T_{\rm obs}=$3600~s \citep{andersen2018fourier}.
For putative pulsars that rotate slower than 1~ms, our \texttt{PRESTO} acceleration search should detect their periodicity (assuming \texttt{PRESTO} $\sigma > 5$) if the pulsar is in a binary with a companion lighter than 1~$M_{\rm \odot}$ and an orbit longer than 10~hr. 
However, we find no significant period candidate after removing candidates caused by radio frequency interference (RFI).
Candidates caused by RFI are identified with the best DM of 0 pc\,cm$^{-3}$ or signal coming from a narrow band.
We list the remaining marginal candidates in Table \ref{tab:candlistpresto}. 
Meanwhile, we also use the FFA algorithm \texttt{riptide}\footnote{\url{https://github.com/v-morello/riptide}}\citep{morello2020optimal} to search the time series for periods in the range 20\,s to 600\,s. 
In the FFA search, the signal to noise ratio (SNR) is equivalent to the best estimator of Z-statistic. When dealing with ideal data containing only Gaussian noise and periodic signals, the critical threshold $\eta$ can be expressed as $\eta_{z}(\alpha)=\bar{\Phi}^{-1}(\alpha)$ \citep{morello2020optimal}, where $\bar{\Phi}^{-1}$ is the inverse survival function of the standard normal distribution. $\eta_{z}$ represents the number of Gaussian sigmas associated with the probability $\alpha$.
However, real data often contains more than simple Gaussian noise and the signal-to-noise ratio of periodograms often increase with the period and may be significantly higher than the theoretical threshold. 
In this paper, we use the \texttt{find\_peaks}\footnote{\url{https://docs.scipy.org/doc/scipy/reference/generated/scipy.signal.find_peaks.html}} method to pick out signals that stand out from the surrounding baseline. The \texttt{find\_peaks} program has one parameter -- prominence, which is defined as the vertical distance between the peak and its lowest profile line. We use observation taken on September 29, 2021 in which we detected no pulses. We inject a mock periodic signal into this empty data set, and obtained an artificial periodic signal 10 $\sigma$ higher than the baseline. We apply the FFA search on the mock data and get a SNR of 41 for the artificial signal, so we set the \texttt{find\_peaks} prominence parameter to 41 in our search for the September 25th-28th data.
We find few prominent peaks that appear in individual days but none simultaneously appears in the data from September 27th and September 28th, as shown in Figure \ref{tab:ffapsearch}. We list these candidates in Table \ref{tab:candlist}.

\begin{figure*}[!h] \centering \includegraphics[width=0.9\linewidth]{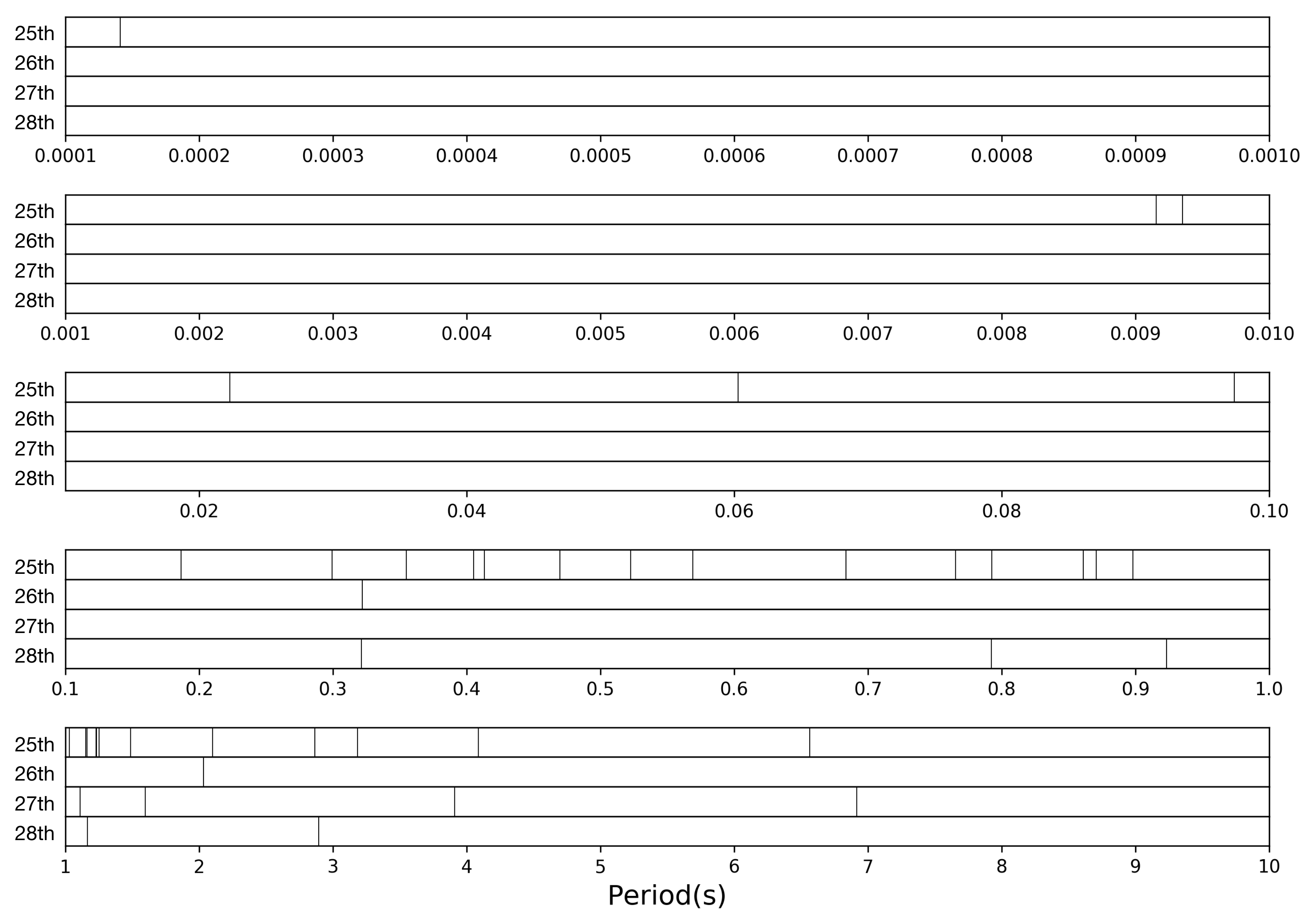} \caption{Lomb-Scargle periodogram search result from all TOAs from September 25th-28th. Different sub-plots correspond to different period ranges. Each subplot shows the stack of search results from 25th-28th from top to bottom. The vertical lines in the plot represent a significant peak in LSP identified by FAP$=0.1$. We stack the detections from different days together to check for recurrence of the same signal on different days. 
} \label{tab:lssearch0928}  \end{figure*}

\subsection{Lomb-Scargle Periodicity search with TOAs \label{sec:PsearchTOA}}

The DM of FRB\,20201124A is known to fluctuate, ranging from 409 pc\,cm$^{-3}$ to 420 pc\,cm$^{-3}$ in FAST observation of active episode 1 \citep{xu2021fast}, and few pulses varied considerably. 
When obtaining TOAs from the data, 
we fix the DM to 413.5~pc\,cm$^{-3}$, a DM value that is found to optimally dedisperse many of the pulses (Zhou et al. 2022), to avoid shifting our TOAs randomly due to their varying pulse morphology.
We treat the multi-components (pulse fine structure) as multiple individual pulses to obtain a more finely resolved list of TOAs. After the initial search, we obtain the time-frequency waterfall plot of the pulses. We use the \texttt{find\_peaks} algorithm to find significant peaks of the pulse profile.
The strength of different components varies greatly in complex multi-components. We adjust the threshold of \texttt{find\_peaks} to obtain the peaks of all components. 
Then, we perform a multi-Gaussian fitting whose initial parameter is the position and strength of peak values from \texttt{find\_peaks} to determine the best fit arrival time.

\begin{figure}[!hbt]
    \centering
    \begin{subfigure}[b]{0.24\textwidth}
        \centering
        \includegraphics[height=2.5in]{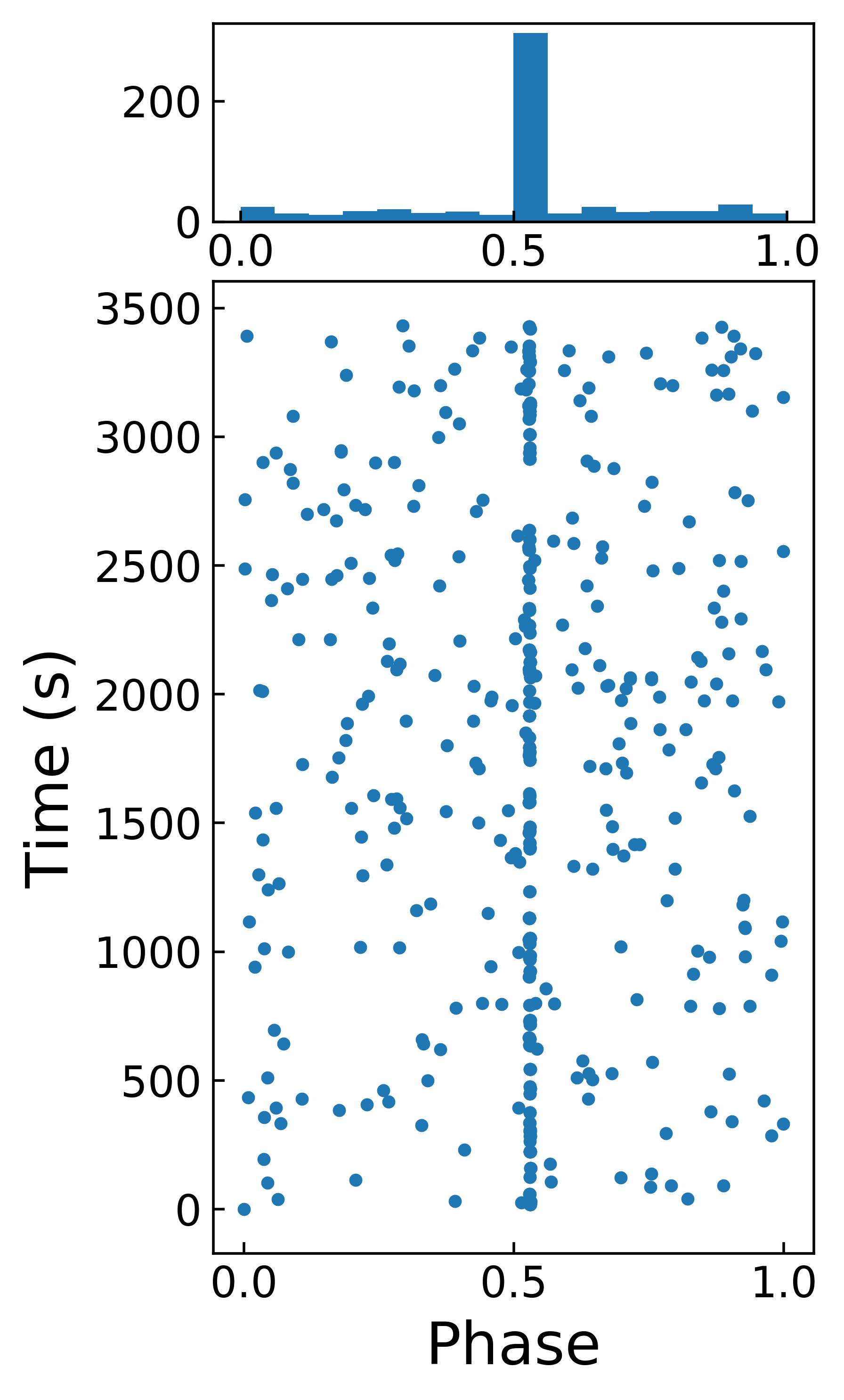}
        \caption{}
    \end{subfigure}%
    \hfill
    \begin{subfigure}[b]{0.24\textwidth}
        \centering
        \includegraphics[height=2.5in]{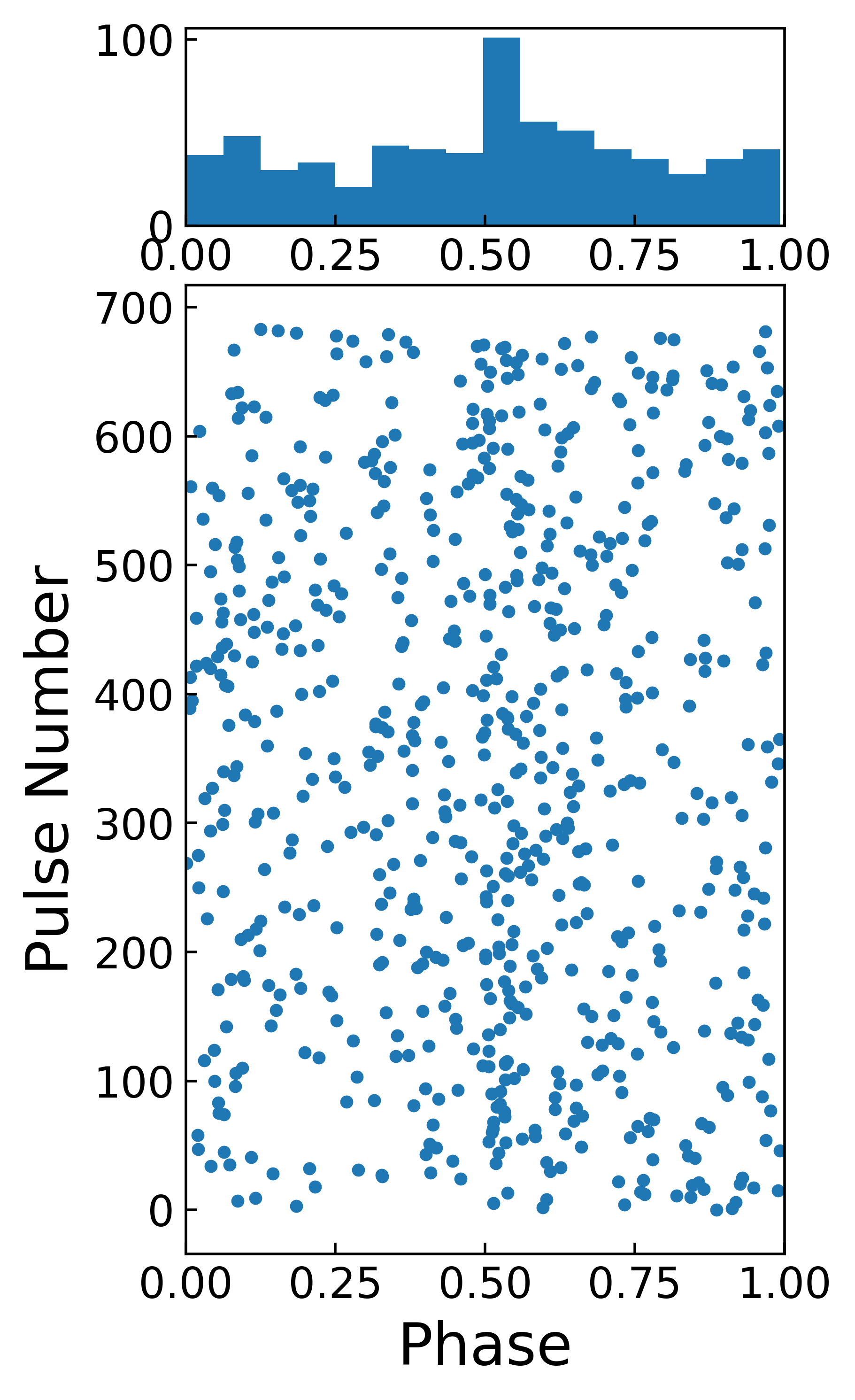}
        \caption{}
    \end{subfigure}
    \caption{(a): An example of a periodic signal detected from the September 28th observation at 2543.13151\,Hz that is likely a harmonic of the sampling frequency. This one is 8 times the time resolution instead of the real period. (b): An example of a periodic signal detected from four-day search at 293.883889\,Hz ($P = 3.4027$\,ms). See Table \ref{tab:candlistppdot} for specific information. The four-day time span is large, so the number of pulses is used on the y-axis.}
\label{tab:exampleTOAsearch} 
\end{figure}

\begin{figure}[!hbt]
    \centering
    \begin{subfigure}[b]{0.5\textwidth}
    \includegraphics[height=2.2in]{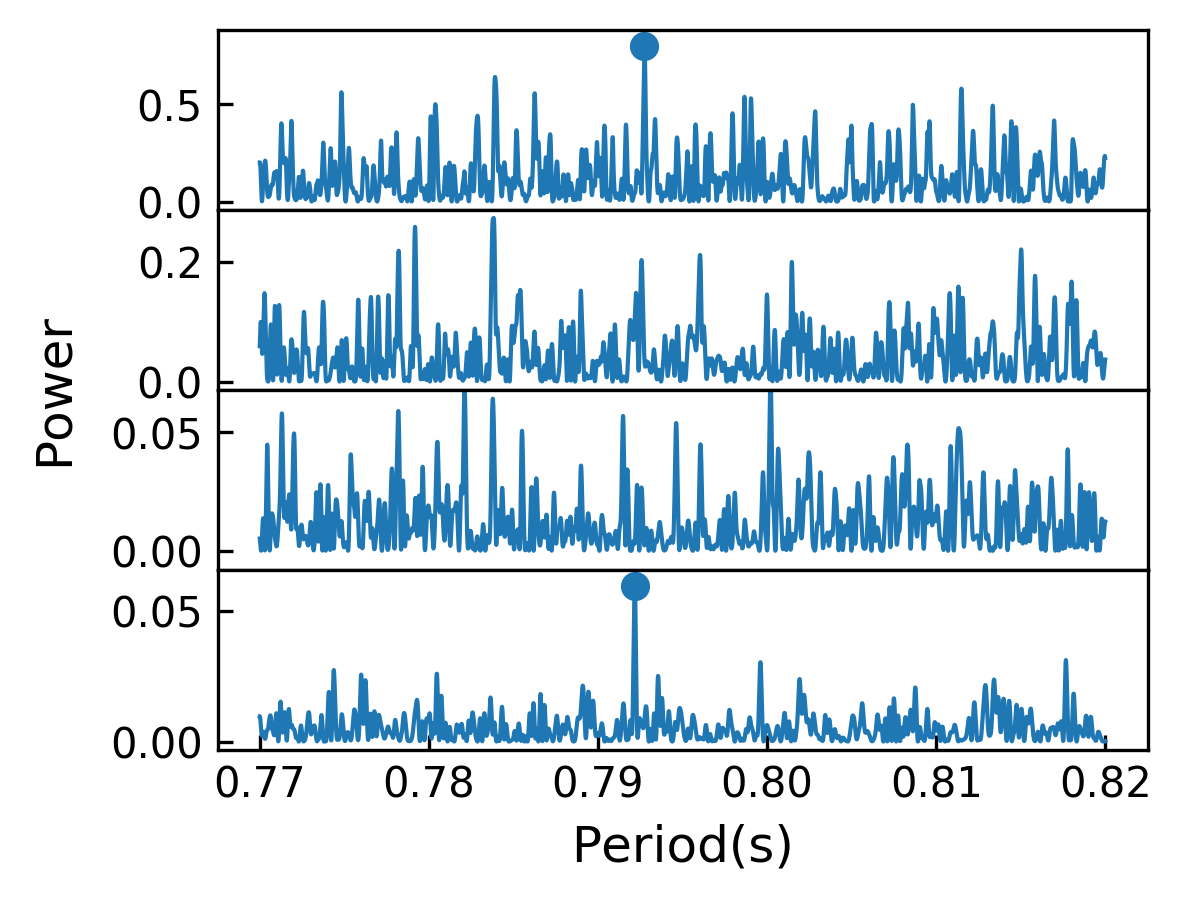}
    \caption{In the LSP, we find same periods showing significant peaks on the 25th and 28th in $P=0.7924(3)$\,s with ${\rm FAP}=0.1$.}
    \end{subfigure}
    \hfill
    \begin{subfigure}[b]{0.5\textwidth}
    \includegraphics[height=2.2in]{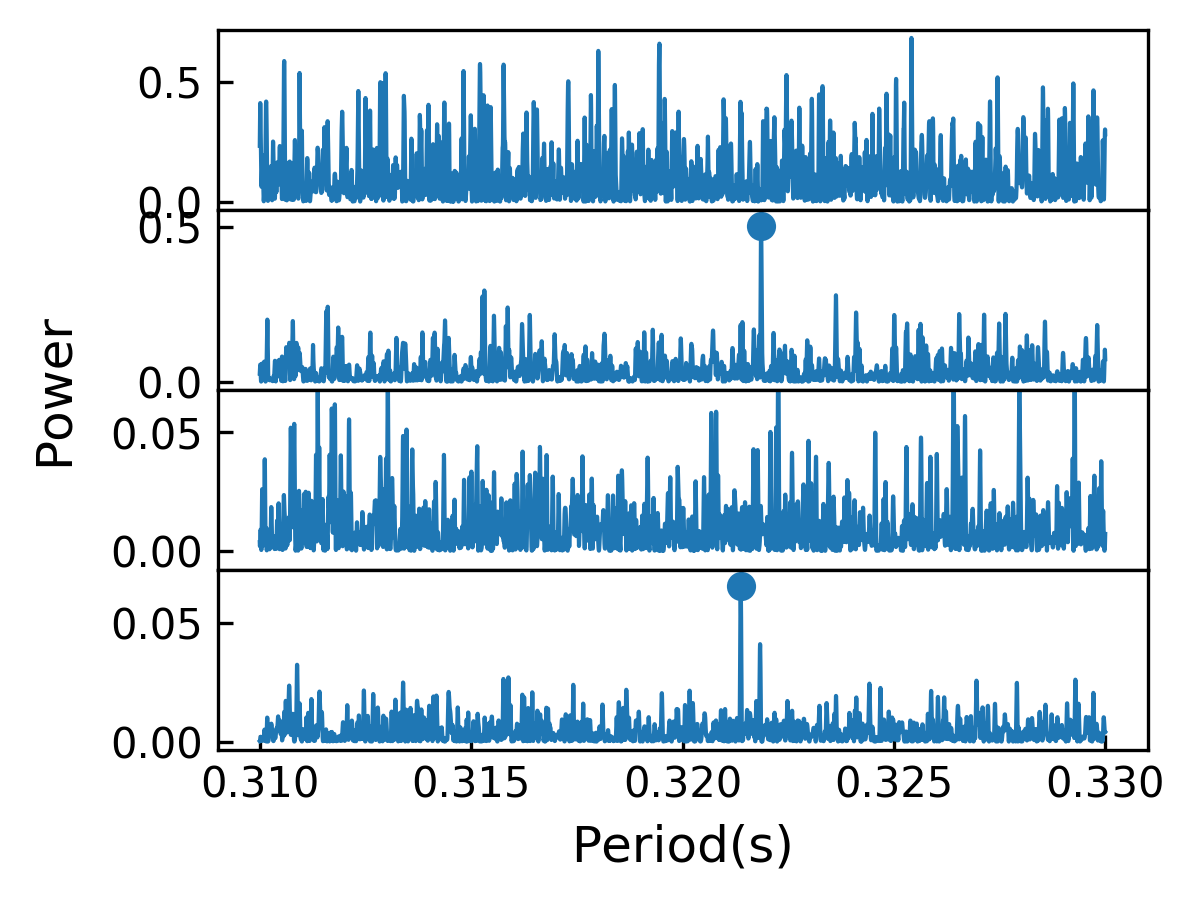}
    \caption{In the LSP, we find same periods showing significant peaks on the 26th and 28th in $P=0.3216(2)$\,s with ${\rm FAP}=0.1$.}
    \end{subfigure}
    \hfill
    \caption{Two periods showing significant peaks in LSP. }
\label{tab:LSPzoomin} 
\end{figure}

We search for periodicity by the LSP of the TOAs collected on September 25th-28th, shown in Figure \ref{tab:lssearch0928} and select significant signals by using \texttt{find\_peaks}. 
The significance of the LSP detections depend on the number of trial periods and the power of the signal. We use the False Alarm Probability (FAP) to quantify the significance of the peaks. For normalized periodogram, \cite{vanderplas2018understanding} shows that
$$
{\rm FAP}(z) \approx 1-\left[P_{\text {single }}(z)\right]^{N_{eff}}
$$
where,$P_{\text {single }}(Z)$ represents the cumulative probability of seeing a peak less than $Z$ at a single specific frequency assuming Gaussian noise. 
Taking into account the folding of $N$ number of data samples (TOAs), the degree of freedom for null hypothesis is ${\rm NH} = N - 1$, and $P_{\text {single }}(Z)=(1 - Z) ^{({\rm NH}/2)}$, where $Z$ is the normalized periodogram power value, $N_{eff}$ is the number of independent peaks and can be estimated as $N_{eff}=f_{max}T$. $T$ is the range of the time series and $f_{max}$ is the maximum frequency of the LSP search.
We adopt the \texttt{find\_peaks} method again and set the prominence parameter to the $Z$ value that leads to a certain FAP threshold, thereby reducing the influence of large differences in baselines in different period ranges. 
When using an FAP threshold of 0.01, we find no period recurs on multiple days. 
Considering a 10-day orbital period and a companion star with a mass of 100~M$_{\rm \odot}$, which leads to a $a/c$ of $1\times10^{-8}$~s$^{-1}$, we estimate the error for matching $F_0$ should be about 1/1000 of $F_0$.
With a FAP threshold of 0.1, we do not find a period that appears in all four days, but find two period candidates ($P=0.7924(3)$ and $P=0.3216(2)$) that appears in the search of two different days in Figure \ref{tab:lssearch0928}, we zoom in on the LSP results and show in the Figure \ref{tab:LSPzoomin}. In addition, we perform simulations to generate the same four sets of random numbers as the four-day TOAs, perform an LSP search with the same threshold of ${\rm FAP}=0.1$, and found that in 1000 simulations, the probability of peak matching is 1.4\%. Considering that the two periods do not appear in both the 27th and 28th data, and simulation shows a small chance of such match occurring in random signals. Thus, we do not consider them credible spin periods, but we list them in candidates  Table \ref{tab:lsp}. 

\subsection{Period and acceleration search with TOAs \label{sec:P-PdotTOA}}

A linear acceleration, described by a constant $\dot{P}$ is the first-order approximation to a binary orbit. A real orbit would normally require at least five Keplerian parameters to describe. But the simple linear approximation could apply if the duration of the observation $T_{\rm obs}$ is much shorter than ($<$1/10) the orbital period $P_{\rm b}$ \citep{ransom2003}. 
Due to the limitation of this approach, our search is only sensitive to binaries with a period much longer than our observing time, i.e., binaries with $P_b>10$~hours if we consider only the TOAs within one session and $P_b>40$~days when we combine the TOAs from all four days. 
As an example, \citet{wzdc22} proposed a model involving a wide orbit magnetar-Be star binary to explain FRB\,20201124A's RM variation \citep{xu2021fast}. 
In their model, the magnetar-Be binary could have an orbital period of 80~days and companion mass up to 30~M$_{\rm \odot}$.

Using a linear acceleration search would enable us to test models such as \cite{wzdc22} up to a certain range of the companion mass $M_c$. 
If the magnetar is subject to gravitational pull of its companion, then it would experience a gravitational acceleration of $a = (2\pi/P_b)^2\, R \, \eta = GM_c/R^2$, where R is the distance between the two objects and $\eta = M_c/(M_p+M_c)$ with $M_p$ being the pulsar mass (usually $\simeq 1.4$~M$_{\rm \odot}$). These equations give us $a = (2\pi/P_b)^2 \,\eta \,(GM P_b^2/(2\pi)^2/\eta)^{1/3}$. 

Such acceleration could lead to a $\dot{P}_b/P_b = a/c$, where $c$ is the speed of light. We choose an upper limit of $M_c = 100\,$M$_{\rm \odot}$, which is a large mass even for a Be star or a stellar black hole. 

We find that the upper limit of $a/c$ is $7\times10^{-7}$~s$^{-1}$, assuming $P_b = $ 10~hours, $a/c = 2\times10^{-9}$~s$^{-1}$ for $P_b =$ 4~days.

In the meantime, a negative linear acceleration, or a positive $\dot{P}$ could also come from a spinning down pulsar or magnetar. 
The spin down rate $F_1$ is determined by the surface magnetic field strength of the pulsar: 
$$B_s=3.2\times 10^{19} {\rm G} \sqrt{P \dot{P}}$$.
Most pulsars have magnetic field strength around $10^{12}$~G and most known magnetars have $B_s$ around $10^{13}$~G to $10^{15}$~G. 
Assuming that we are searching for a magnetar with $B_s = 10^{15}$~G, we could derive that the {pulsar would experience an $F_1$ of} $F_1 = - F_0^3 (B_s/3.2\times 10^{19}{\rm G})^2$.

Pulses experiencing a linear acceleration are affected by the parameters $P$ and $ \dot P $. We write the phase $\phi_i$ as a function of $F_0 = 1/P$ and spin frequency derivative $F_1 = -\dot{P}/P^2$ in the following way:
$\phi_i = F_0 \, t + 0.5 \, F_1 \, t_i^2$.
To complete our acceleration search, we need to search $F_0$ and $F_1$ up to the intended accelerations ($F_1/F_0 = -\dot{P}/P = a/c$).
We fold the TOAs into a pulse density profile using $\phi_i$ and use the $H$-test to determine the significance of the resulting pulse profile. 
The $H$-test provides a better sensitivity to low-significance pulsar candidates. So it is extensively used for periodicity search in X-ray and $\gamma$-ray pulsar research \citep{de2010h,bachetti2021extending}. 
The $H$-test statistic is defined as:
$$H=\max \left(\mathrm{Z}_{m}^{2}-4 \mathrm{~m}+4\right), 1 \leq m \leq 20, $$
where $Z_{n}^{2}$ is defined as:
$$Z_{n}^{2}=\frac{2}{N} \sum_{k=1}^{n}\left[\left(\sum_{j=1}^{N} \cos k \phi_{j}\right)^{2}+\left(\sum_{j=1}^{N} \sin k \phi_{j}\right)^{2}\right],$$
where N is the number of detected photons, and $\phi_i$ (i $= 1,...,N$) is the pulse phase.
\cite{de2010h} obtained the cumulative probability of $H$ statistic by simulations as
$$  \operatorname{Prob}(>H)=\exp (-0.4 H).  $$
To simplify the phase computation, we utilize the frequency $F_0$ and its derivative $F_1$. 

When conducting acceleration search for the one hour sessions,
we search $F_0$ in the range between 0.01\,Hz and 1000\,Hz. $F_1$ is selected to cover the range of $a/c$ in [\,-1$\times$10$^{-6}$, 1$\times$10$^{-6}$\,] s$^{-1}$. 
Such an acceleration exists if the neutron star is being pulled by a 200~$M_{\rm \odot}$ companion in a 10-hr orbit.
We also estimate the resolution needed for the search by calculating how much of a change in $F_0$ and $F_1$ would lead to a small change in phase $\delta\phi_{\rm last}$ for a $t = 3600$\,s. It is clear that the smallest change in $F_0$ and $F_1$ that we need to search for $\delta\phi_{\rm last}$ not to exceed 0.2 is
$\delta F_0 = 5.6\times10^{-5}$~s$^{-1}$ and $\delta F_1 = 3\times10^{-8}$~s$^{-2}$.
The number of trials of the $F_0$-$F_1$ test reaches 7.2$\times$10$^{11}$. 
This is a substantial computing challenge. 
We split the computation into smaller blocks of the size $\Delta F_0=1$~Hz by $\Delta F_1=1\times10^{-5}$~s$^{-2}$, and use GPUs to conduct the phase calculation and compute the $H$ value for the folded profiles.
For each of the smaller blocks, we evaluate the folded profile from a fine grid of 20480$\times$320 of $F_0$ and $F_1$ to achieve the aforementioned resolution.

\begin{figure*}[hbt]
\centering
\includegraphics[width=1\linewidth]{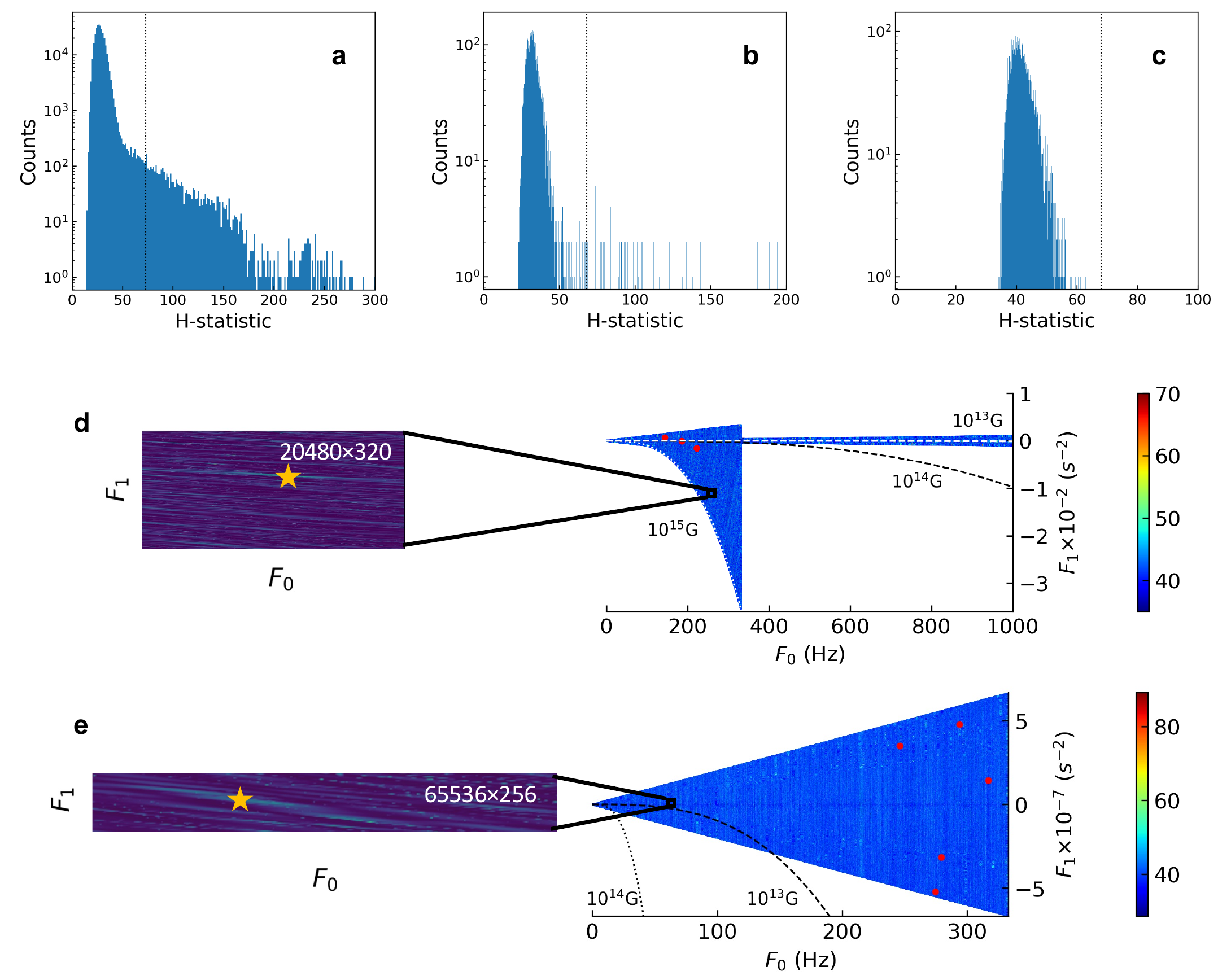}
\caption{
Panel {\bf a} is the $H$-test result of $P$-$\dot{P}$ search with 638 raw TOAs including all the micro-components. 
The gray line represents the threshold $H$ statistic that leads to a unity null-hypothesis probability when taking the number of trials into account. 
Panel {\bf b} shows the $H$ map of searching with 638 TOAs augmented with ms-level random noises. 
Panel {\bf c} shows the $H$ map of searching with 424 grouped and augmented TOAs that do not include all the micro-components. 
Panel {\bf d} right plot shows the $H$ map results of September 28th using the augmented grouped TOAs. We search the data of all four days, respectively, and present here the result of September 28th as an example. 
The search is conducted for $F_0$ in the range of 0.01$\sim$1000\,Hz and $\rm a/c$ in the range of [-1, 1] $\rm \times10^{-6} ~s^{-1}$.
Here we display $H$ greater than 74 as red circles. This threshold is determined by the probability of the $H$-test based on the number of trials.
The dash curve represent the $F_0$--$F_1$ value corresponds to a surface dipole magnetic field strength of $10^{13}$~G, $10^{14}$~G and $10^{15}$~G.
The left plot of panel {\bf d} shows an example of the grid search block, in which the highest $H$ value pixel is marked with star.
Panel {\bf e} shows the search results for the augmented and grouped TOAs combining all four days. The $H$-test result is on the folded profile for assumed $F_0$ in the range of 0.01$\sim$333\,Hz from the TOAs of September 25th-28th. The $\rm a/c$ is in the range of [-2, 2] $\rm \times10^{-9} ~s^{-1}$. The left plot is a zoomed-in display of the grid search for each pixel of the right plot.  
Panel {\bf e} right plot shows the $P$-$\dot{P}$ search results. Here we display $H$ greater than 79 as red circles. The dashed curve represent the $F_0$--$F_1$ values correspond to a surface dipole magnetic field strength of $10^{13}$~G and $10^{14}$~G.
}
\label{ppdot}
\end{figure*}

When conducting an acceleration search for TOAs, we conduct three initial test searches, each with a different setting of TOAs from September 28th, the highest event rate.
For the first test, we search all TOAs, including individual TOAs from the multi-components. 
The number of trials reaches $10^{12}$ with the search range and resolution described above. 
The theoretical false alarm threshold, i.e. null-hypothesis probability times number of trials ($10^{12}$), reaches unity when $H = 68$. 
However, we find many signals with $H$ values larger than 68 (as shown in Figure \ref{ppdot} panel {\bf a}). The folding results look like multiple frequencies of smaller periods. So we increase the search range to 1 -- 5500\, Hz with the number of trials of $10^{13}$. In particular, we find a strong signal that seems to be the harmonics of the fundamental frequency $F_0 = 5086.263$~Hz. This signal happens to correspond to a period of 196.6~us -- 4 times of our sample time. Therefore, we suspect that this is an artifact of our data's finite sampling frequency.
Figure \ref{tab:exampleTOAsearch} (a) shows a candidate with a period that is four times the sample time. 
We check the folded profile by eye for the candidates with $H>$ 200 and find most high $H$-test statistics are harmonically related to the sampling artifact.
Therefore, we conclude that we have to remove this significant artifact signal to find any real period.

In the second test, we augment the TOAs by adding a random, normally distributed noise to the TOAs with a Gaussian $\sigma$ of 1\, ms. We call these TOAs augmented TOAs hereafter.
Using these augmented TOAs, we run our search again, and this time the $5086.263$\,Hz signal, and its harmonics are all gone.
Surprisingly, there are still hundreds of periodic signals with $H>$ 68 (as shown in Figure \ref{ppdot} panel {\bf b}), some of which appeared both in the September 27th and 28th data.
We inspect these profiles and find that they are likely to be caused by the presence of a few clusters of pulses, i.e., the multi-components of the pulses.
These small clusters of TOAs tend to significantly boost the counts in a single profile bin, leading to many high $H$ signals in the low spin frequency end.
Such artificial signals also prevent us from correctly detecting any real spin period.

To minimize this micro-component effect, we conduct a third test.
This time, we group the TOAs from a cluster of multi-components into one. 
The one TOAs is derived from the weighted center of the multi-components group.
In this third test, only a few period candidates slightly exceed $H$ of 68 (as shown in Figure \ref{ppdot} panel {\bf c}).
Similarly, only a few detections come from the other three days and no matching signals from different days.
We show the $H$ distribution from the third test in Figure \ref{ppdot}.
So again, no credible spin period was detected after considering a large range of $P$ and $\dot{P}$.
In Figure \ref{ppdot} {\bf d}, we show the $H$-test result from September 28, 2021 for $P$ between ~1~ms and 100~s and the range of $\dot{P}$ limited by $|a/c|<1\times 10^{-6}$~s$^{-1}$. 
In the September 28th results presented here, we also {expanded}the search to a significant negative acceleration ($F_1$) that translate to surface magnetic fields $B_s$ up to $10^{15}$~G for an isolated magnetar with $P$ between 3~ms and 100~s.

Finally, we conduct an $P$-$\dot{P}$ acceleration search for the grouped and augmented TOAs from all four days.
We search $F_0$ between 0.01\,Hz and 333\,Hz. $F_1$ is selected to cover the range of $a/c$ in [\,-2$\times$10$^{-9}$, 2$\times$10$^{-9}$\,]~s$^{-1}$, which corresponds to companion mass up to 100\,M$_{\rm \odot}$ and orbital period longer than 40~days.
Due to the longer observing time when combining all 4 days, the resolution in $F_0$ and $F_1$ needed to be much higher: $\delta F_0 = 6\times10^{-7}$~s$^{-1}$ and $\delta F_1 = 3\times10^{-12}$~s$^{-2}$.
To complete this search, we split the computation into smaller blocks of the size $\Delta F_0=1$~Hz by $\Delta F_1=1\times10^{-5}$~s$^{-2}$, each containing 65536$\times$256 individual pairs of $F_0$ and $F_1$ that can be evaluated in a single GPU run.
This leads to a substantially higher number of trials and changes the threshold for $H$ to 79.
We show the search result in Figure \ref{ppdot} panel {\bf e}.

\subsection{Periodicity search with special pulses}

\begin{figure*}[!hbt]
\centering
\includegraphics[width=0.9\linewidth]{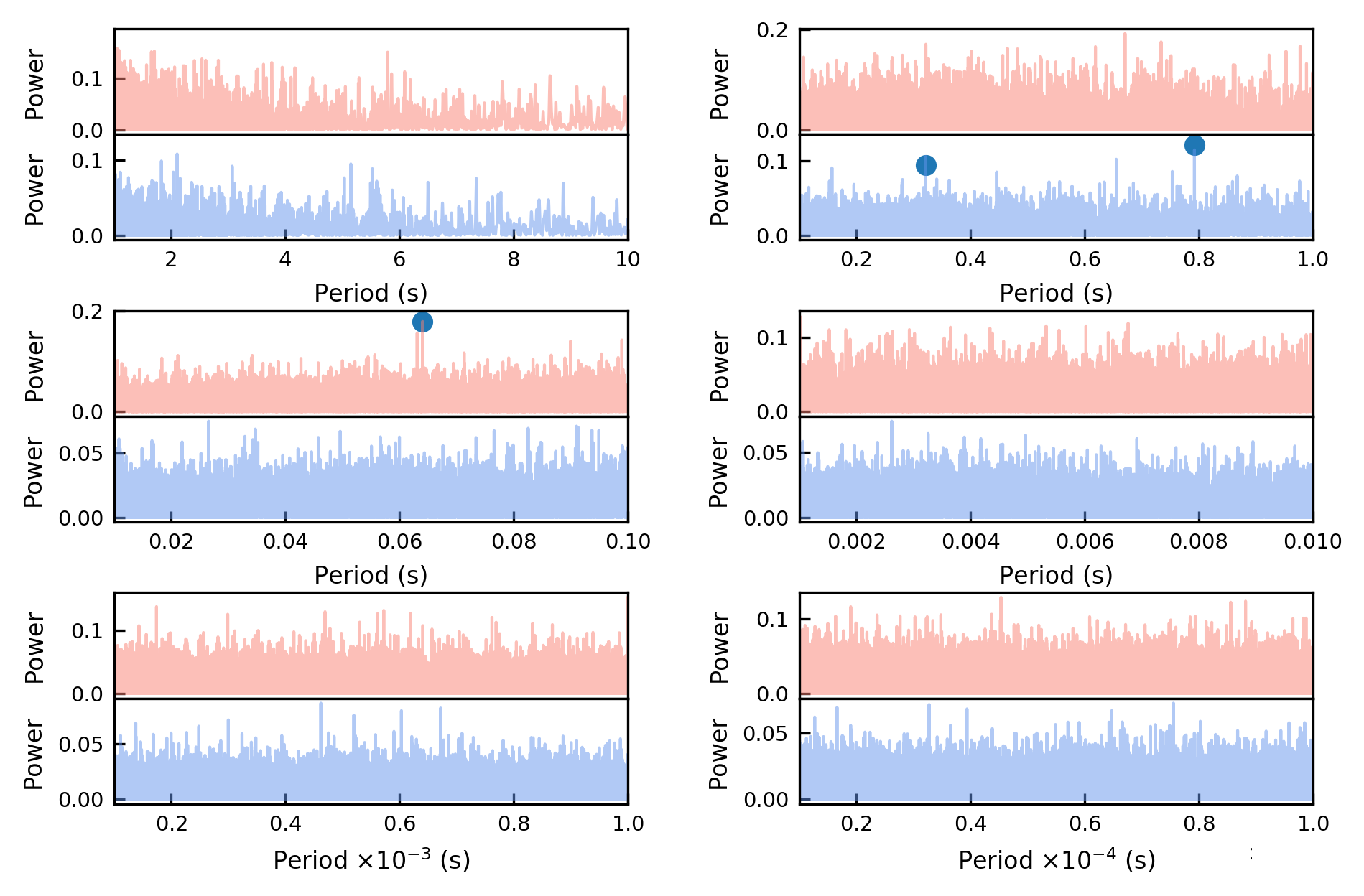}
\caption{Lomb-Scargle periodogram for the high frequency and low frequency pulses detected on September 28th. Red line represents high frequencies, blue line represents low frequencies, and circles represent significant peaks.}
\label{tab:fenleipsearch0}
\end{figure*}

This section summarizes the periodicity search in pulses selected based on frequency and polarization. 
We observed that some of the FRB pulses show up only in part of our observable band, as shown in Figure \ref{tab:mulall}, e.g. burst \#(29). The four pulses burst in \#(29) appear in the low frequency part with  periods of 54.2\,ms.
Pulses with different spectral shapes may originate from different rotational phases of the source object. 
In light of this possibility, we split pulses based on their frequency range: high frequency and low frequency. Pulses appearing in the full bandwidth are not considered here. We perform the Lomb-Scargle periodicity search in different frequency bands and list the top candidates. The periodicity search results of pulse categorized by frequency on September 28th are shown in Figures \ref{tab:fenleipsearch0}. No obvious period was found after folding.

\begin{figure}[!hbt]
\centering
\includegraphics[width=\linewidth]{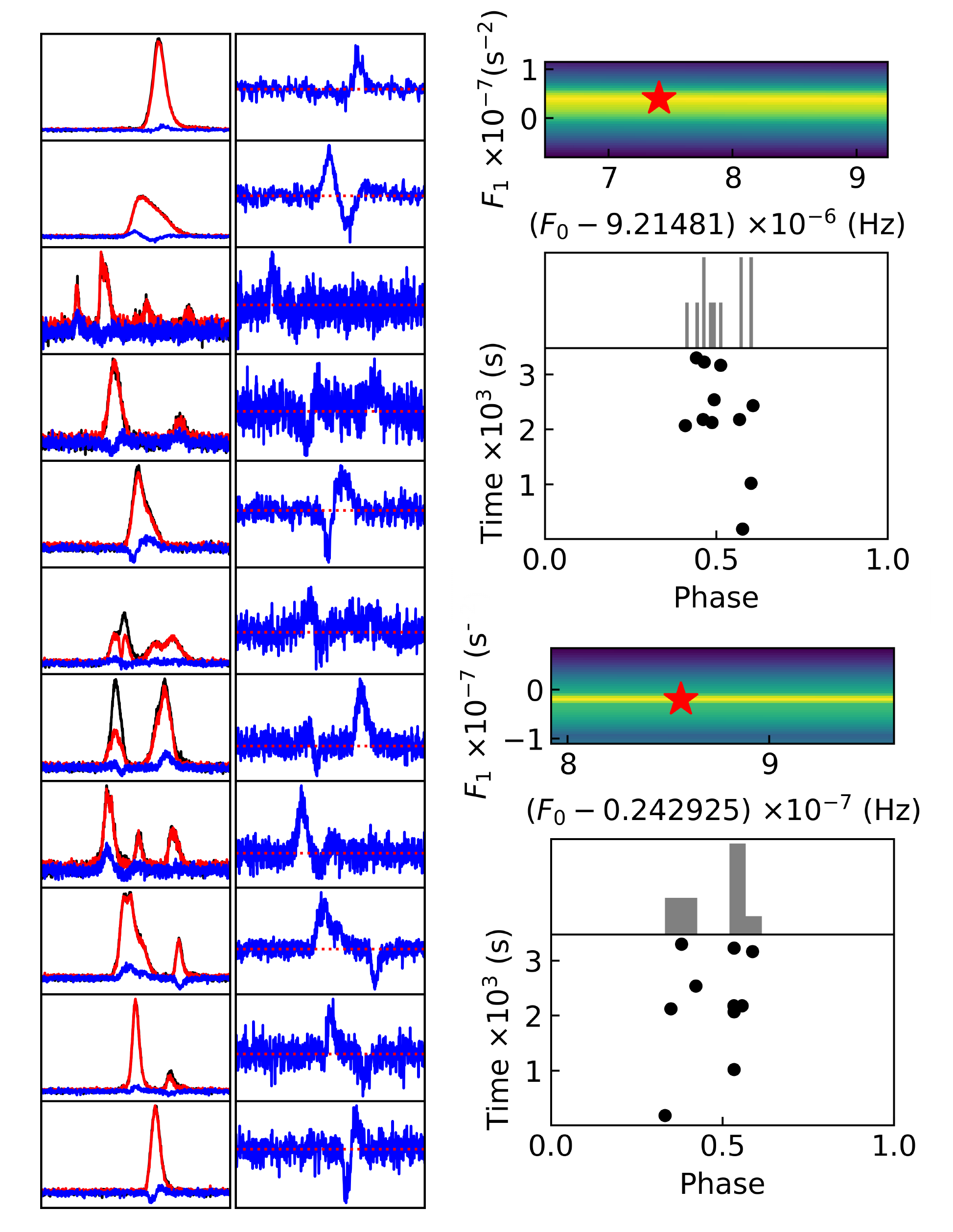}
\caption{
The left image shows the observed pulse with the sign changing of the circular polarization, where both the linearly and circular polarization are shown. The middle plot only shows the circular polarization when zoomed in. The right side shows the folded events based on two top period candidate the $F_0$-$F_1$ search graph.}
\label{tab:polpsearch}
\end{figure}

For pulsars, the phases where their circular polarization changes sign usually correspond to their polar cap's phase center.
\cite{xu2021fast} found considerable circular polarisation and quick position angle swing in FRB\,20201124A. \cite{wang2022,2021arXiv211206719W} and \cite{tong2022circular} discussed the geometric and non-geometric origin of FRB circular polarization.
If the FRB is caused by a pulsar emitting in all rotational phases, we might not be able to find its period through our methods. However, its pulses with unique polarisation features could still show periodicity. 
We identify the pulses with sign-changing circular polarization from the September episode. We then use the LSP search and perform  $P$-$\dot P$ search based on its barycentered pulse arriving TOAs. The results are shown in Figure \ref{tab:polpsearch}. 
Only 11 pulses are selected, and the period candidates are not statistically significant. We show the folding results of two possible top period candidates (top plot: 0.10852\,s and bottom plot: 4.11649\,s) in Figure \ref{tab:polpsearch}. 
To verify the significance of our detection, we randomly pick samples of 11 pulses from all TOAs and test the significance of the periodicity search. We compare the highest $H$ value from the randomly selected samples and the pulses that we found using sign-changing circular polarization. The highest $H$ of the random sample is larger than the $H$ of circular polarization search for 1455 times in 4000 trails.
Therefore, we conclude that the number of special circular polarization pulses is too small to obtain a significant detection. Still, this method may be worth trying after more such pulses are collected in the future.

\subsection{Periodicity analysis of multi-components}

\cite{hessels2019frb,gourdji2019sample,amiri2019second} discovered FRB multi-component phenomena from FRB\,20121102A, FRB\,
180814. Recently, \cite{2022arXiv220208002P} reported the FRB\,20201020A with five components and quasi-periodicity of 0.415\,ms.
But multi-components have not been systematically studied since bright repeated bursts are uncommon in FRBs.

The FRB multi-component phenomena are similar to the micro-structure already observed in pulsars \citep{craft1968submillisecond}. Micro-structures have been noticed in the observations of pulsars and millisecond pulsars~(MSPs) \citep{cordes1990quasiperiodic,de2016detection, liukuo2022}. Angular radiation and temporal modulation are some possible theoretical explanations \citep{benford1977model,rickett1981location,boriakoff1976pulsar,cheng1980particle,arons1979some}.

In this paper, {we analyze the periodicity of multi-component bursts if they are composed of at least 4 components with the longest waiting time less than 1 second}. Many fine structures are composed of three consecutive peaks, but we do not include them. This is because we can always find a common denominator period from three pulses. Thus the resulting period is not a {periodic} signal with statistical significance. 

We find a total of 53\,multi-components in the second active episode (Figure \ref{tab:mulall}). The bottom panels show the pulse profiles; the solid black line in the middle indicates the pulse profile, the solid red line represents linear polarization, and the solid blue line represents circular polarization. The top panel represents Position Angle (PA). As indicated in the Table \ref{tab:ms2}, we count the quasi-period information of all multi-components.
We define the residual $Res = \sqrt{\sum r_i^2}$ and Dimensionless residual $\overline{R}$ to characterize the quality of the quasi-period.
$$  \overline{R}=\frac{t_{\rm bin} \times \sqrt{\sum r_i^2}}  {(N-1)\,P} $$

\begin{figure}[!hbt]
\centering
\includegraphics[width=\linewidth]{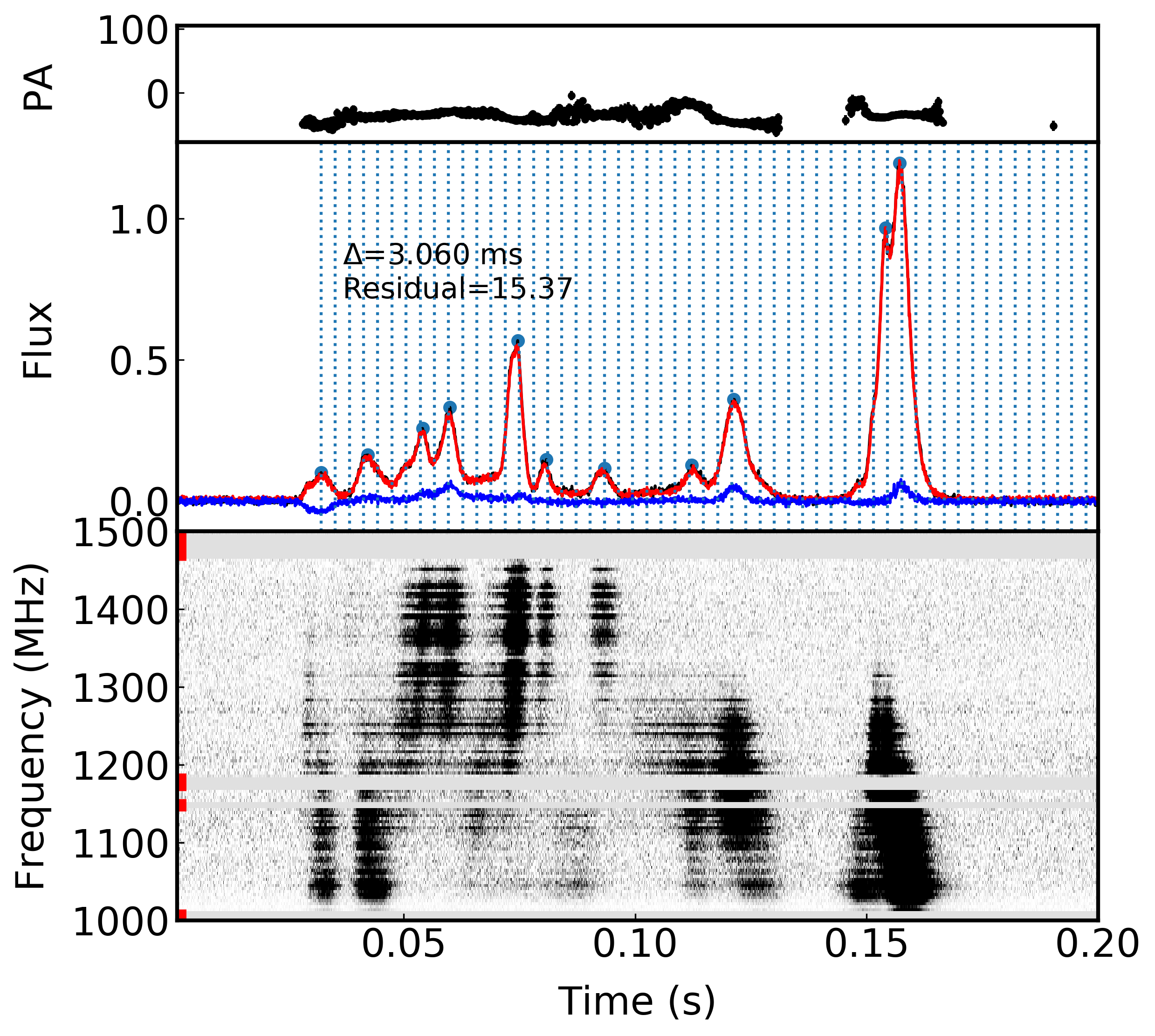}
\caption{An example of a typical pulse containing multi-components, the total pulse width reaches 150\,ms, there are 11 apparent structures, that can well meet the interval of 3.060\,ms. The red line in the figure represents linear polarization, the blue line represents circular polarization, and the sub-figure on the top is the position angle.}
\label{tab:0374}
\end{figure}

\begin{figure*}[!hbt]
    \begin{subfigure}[b]{0.5\textwidth}
        \centering
        \includegraphics[height=2.3in]{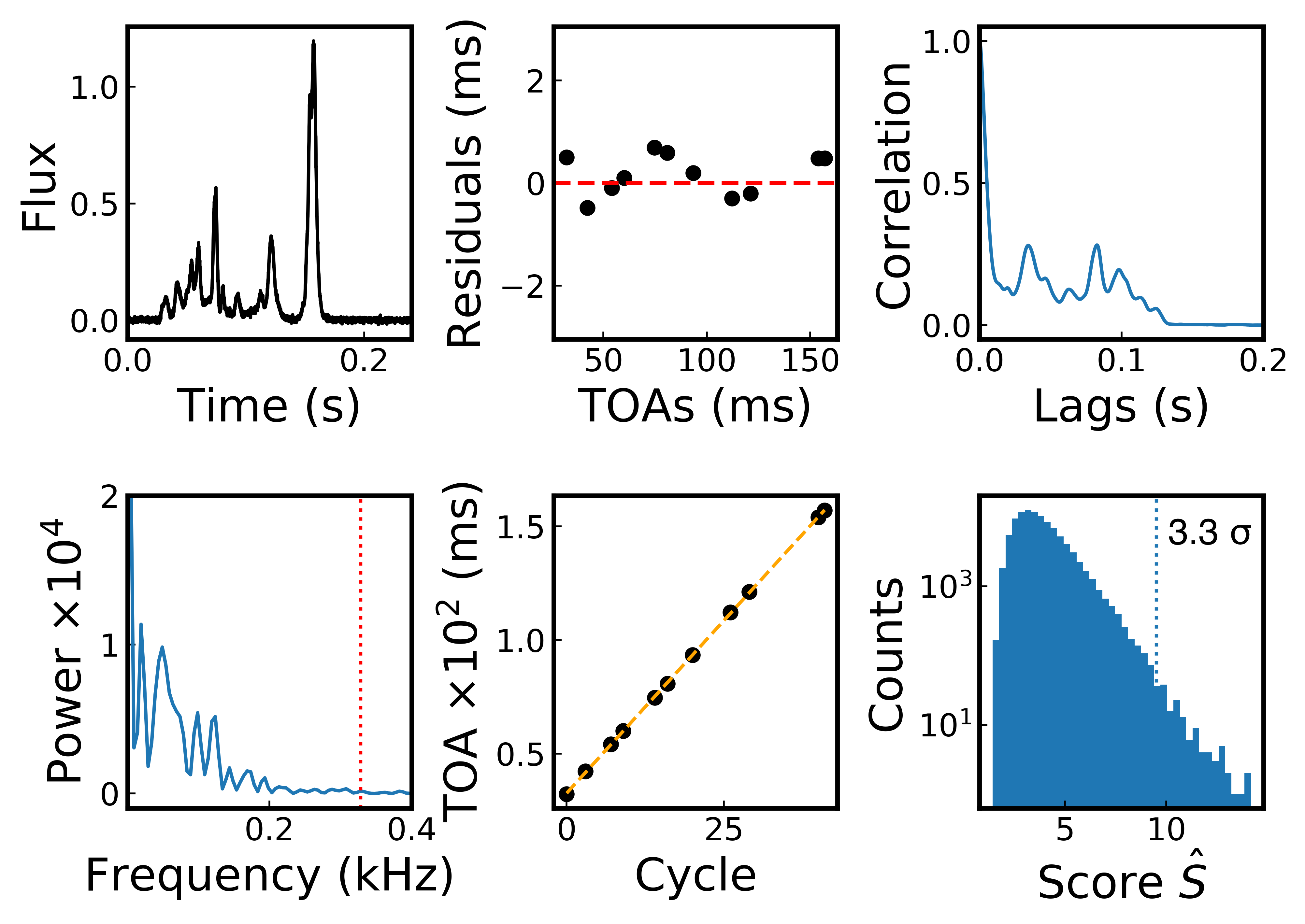}
        \caption{Detailed analysis of multi-components in Figure \ref{tab:0374}.}
        \label{tab:0374d}
    \end{subfigure}%
    \hfill
    \begin{subfigure}[b]{0.5\textwidth}
        \centering
        \includegraphics[height=2.3in]{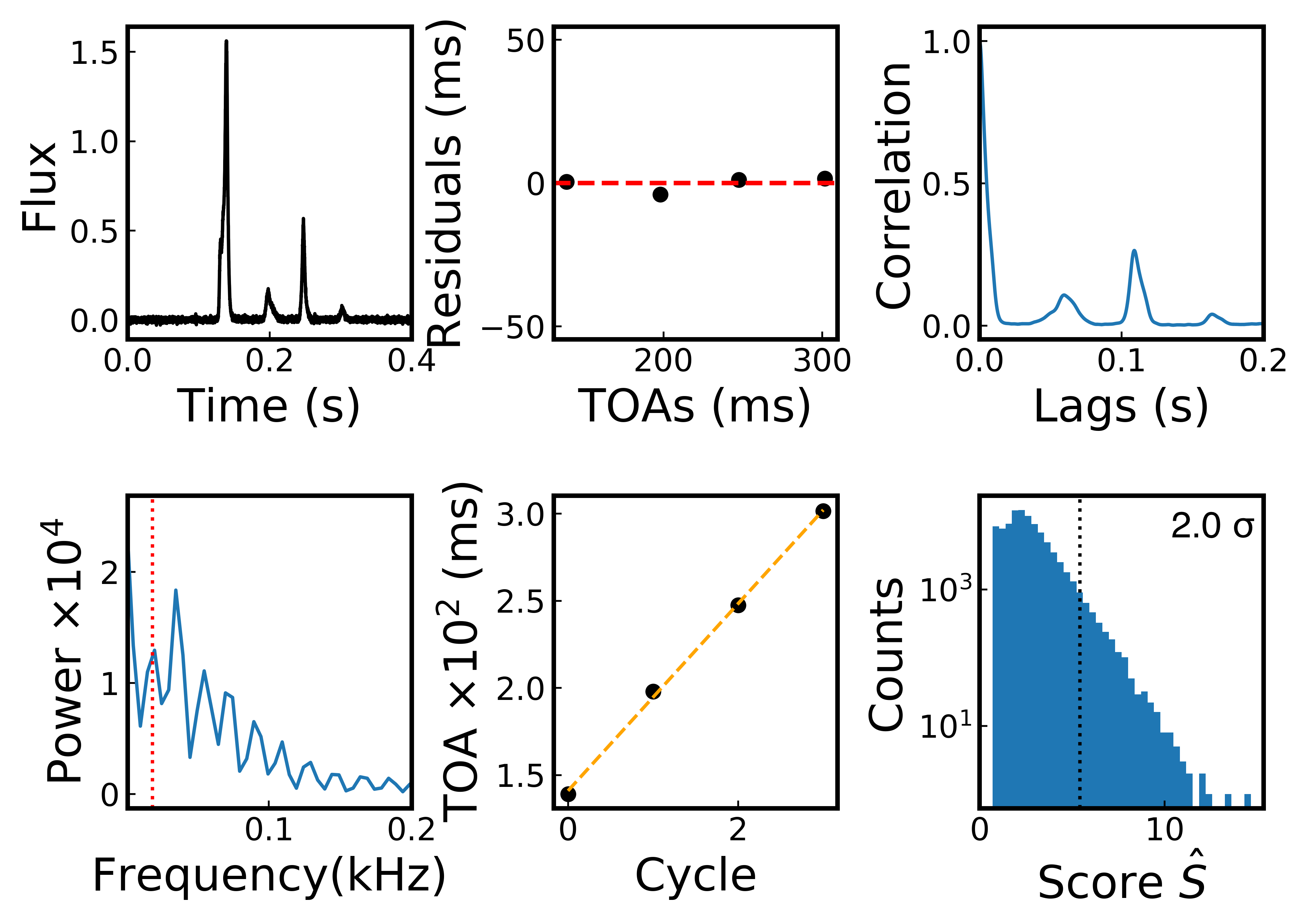}
        \caption{Detailed analysis of multi-components (29).} \label{tab:0243d}
    \end{subfigure}
    \hfill
    \begin{subfigure}[b]{0.5\textwidth}
        \centering
        \includegraphics[height=2.3in]{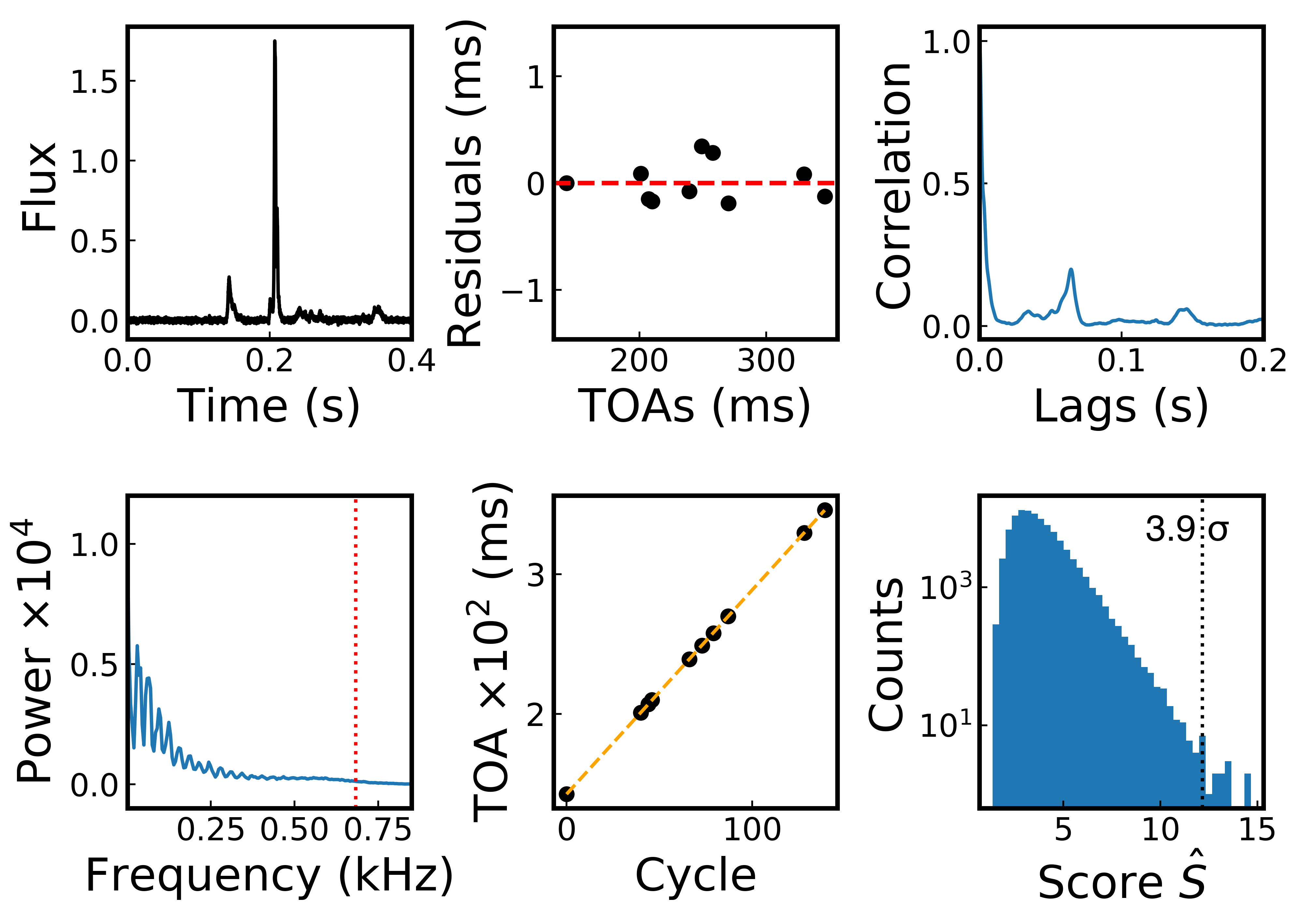}
        \caption{Detailed analysis of multi-components (25).} \label{tab:0190d}
    \end{subfigure}
    \hfill
    \begin{subfigure}[b]{0.5\textwidth}
        \centering
        \includegraphics[height=2.3in]{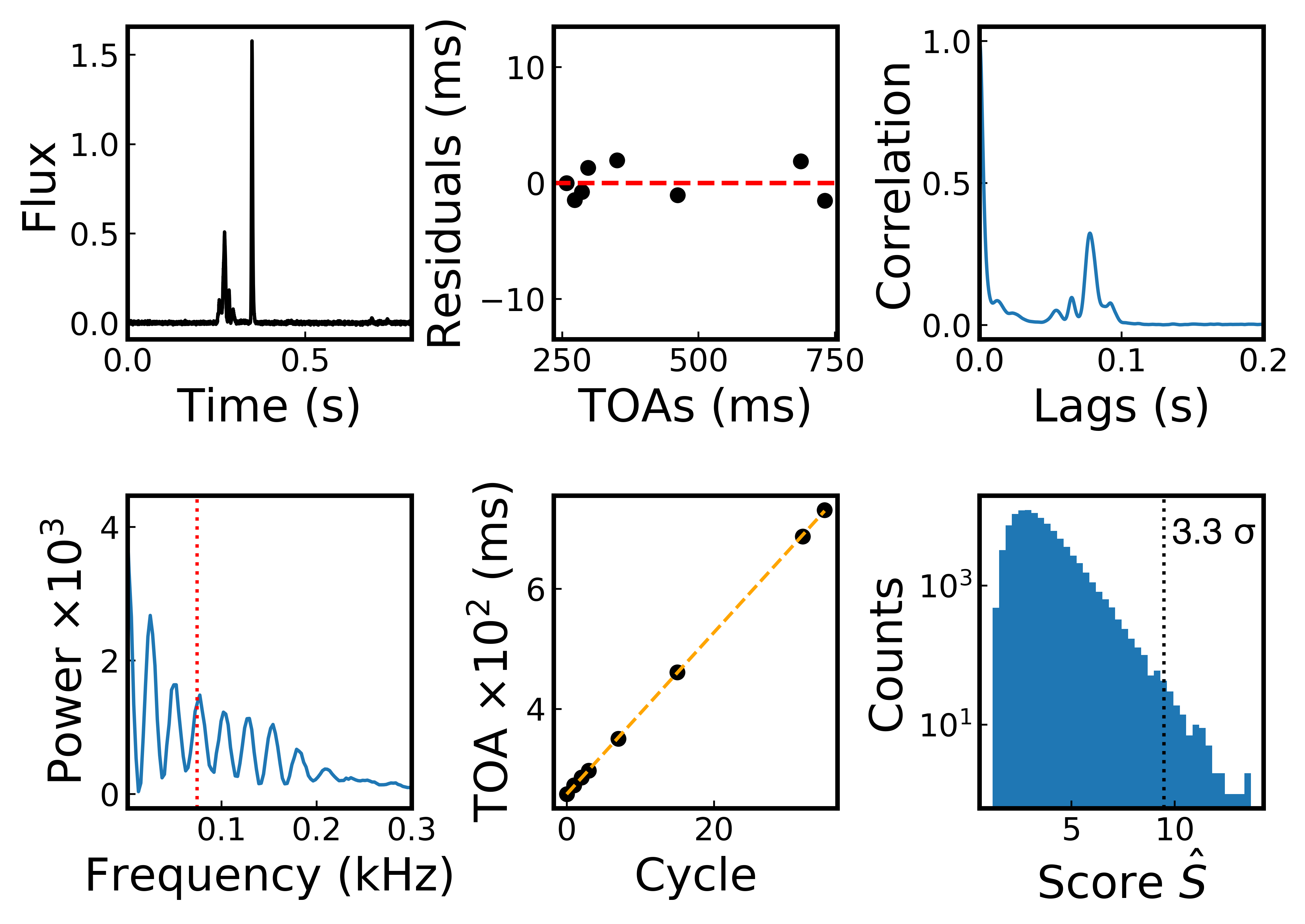}
        \caption{Detailed analysis of multi-components (10).} \label{tab:0407d}
    \end{subfigure}
\caption{In each subplot, the top left plot is the pulse profile. The lower left is the Fourier transform of the profile, and the scatter plot in the middle is the residual of TOAs and the straight line fitting of cycle and TOAs. The upper right is the auto-correlation analysis of the profile, and the lower right is the result of the quasi-periodic hypothesis testing.}
\end{figure*}

\begin{figure}[!hbt] \centering \includegraphics[width=\linewidth]{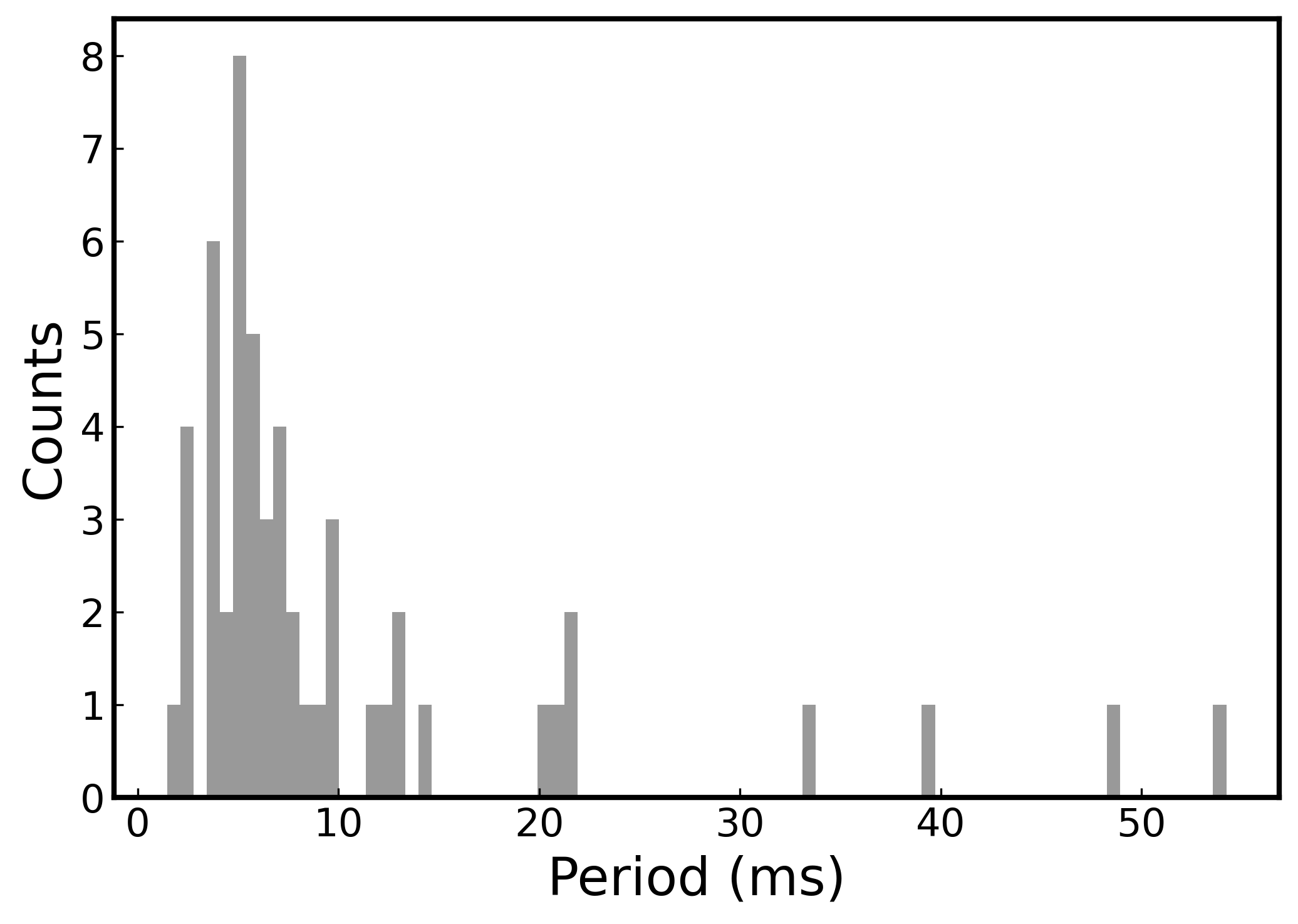} \caption{Multi-components quasi-period statistical histogram.} \label{tab:msst} \end{figure}

\cite{andersen2021sub} presented discoveries of multi-components in three CHIME FRBs. One of these FRB\,20191221A shows periodicity with a confidence level of $\rm 6.5\,\sigma$. The other two sources have lower confidences of 2.4\,$\sigma$, and 1.3\,$\sigma$, respectively. They pointed out that the sub-bursts of multi-components are often downward-drifting in the 400–800\,MHz. 
We adopt the same analysis method to check for the significance of periodicity found in FRB20201124A's multi-components.
We obtain the TOAs of the profile peaks and found an intriguing quasi-periodic pattern. Figure \ref{tab:0374} is one of the most complicated structures in our observations. Both the downward-drifting and upward-drifting components exist at this multi-component, and its period is 3.060\,ms. Detailed periodicity analysis of this multi-component is shown in Figure \ref{tab:0374d}. The picture in the upper left corner is the pulse profile. We separated the components to limit the impact of the baseline on the periodicity search. The Fourier transform of the separated profile is shown in the lower-left corner. The red dotted line is the frequency where the quasi-period is located. The picture in the upper-middle part is the residual analysis of the arrival time. Influenced by the sub-pulse strength, not all FFT results are obvious, {so we determine the best quasi-period where} the multi-component residuals are smallest. The black dots in the lower middle of the figure are pulses in the 11-component burst. X-axis is the integer-valued vector of the burst. $n_i$ $=$ (0, 3, 7, 9, 14, 16, 20, 26, 29, 40, 41), Y-axis is the TOA\, $t_i$. The period is given by the minimum TOA residual. In order to compare with \cite{andersen2021sub}, we use the same statistical analysis method, and fit the gap $n_i$ and TOA $t_i$ with a straight line, $t_{i} = f^{-1} n_{i}+T_{0}+r_{i}$. Define a statistic value $L$ that measures the difference to the straight line
$$
\hat{L}[n]=\frac{1}{2} \log \left(\frac{\sum_{i}\left(t_{i}-\bar{t}_{i}\right)^{2}}{r_{i}^{2}}\right),
$$
and define the periodicity-sensitive statistic $\hat{S}$:
$$
\hat{S}=\max _{n}(\hat{L}[n]).
$$
We sample the time series from the uniform distribution, and obtain the null-hypothesis probability by comparing the $\hat{S}$ of the simulated events with the observed data. We get a confidence of $\rm 3.3\,\sigma$ for the pulses in the bottom right corner Figure \ref{tab:0374d}. The figure in the upper right corner shows the results of auto-correlation analysis of the pulse profile. The definition of the auto-correlation function (ACF) is $ R$ = $\sum I(i) ~ I({i+\tau})$.
Other bursts and the same analysis are shown in Figures \ref{tab:0243d},\ref{tab:0190d},\ref{tab:0407d}. We show in Figure \ref{tab:msst} the distribution of multi-component periods and find a peak at 5\,ms with a wide distribution.

The highest confidence quasi-periodic structure that we found is $\rm 3.9\,\sigma$, less than the $\rm 6.5\,\sigma$ level found from FRB\,20191221A by CHIME. 
However, multiple multi-components of FRB\,20201124A are found to show periods with the confidence higher than 3\,$\sigma$. 
As shown in Figure \ref{tab:msst}, the multi-component periods do not seem to be related to each other. Since these multi-components come from the same source that lacks a global period, these period signals are false positives. Our result shows that we should be cautious when linking low-significance quasi-periodicity in FRB micro-components to the rotation of the source object at least for signals with significance below $\sim 4\,\sigma$.

\begin{figure*}[!htb]
\centering
\includegraphics[width=0.8\linewidth]{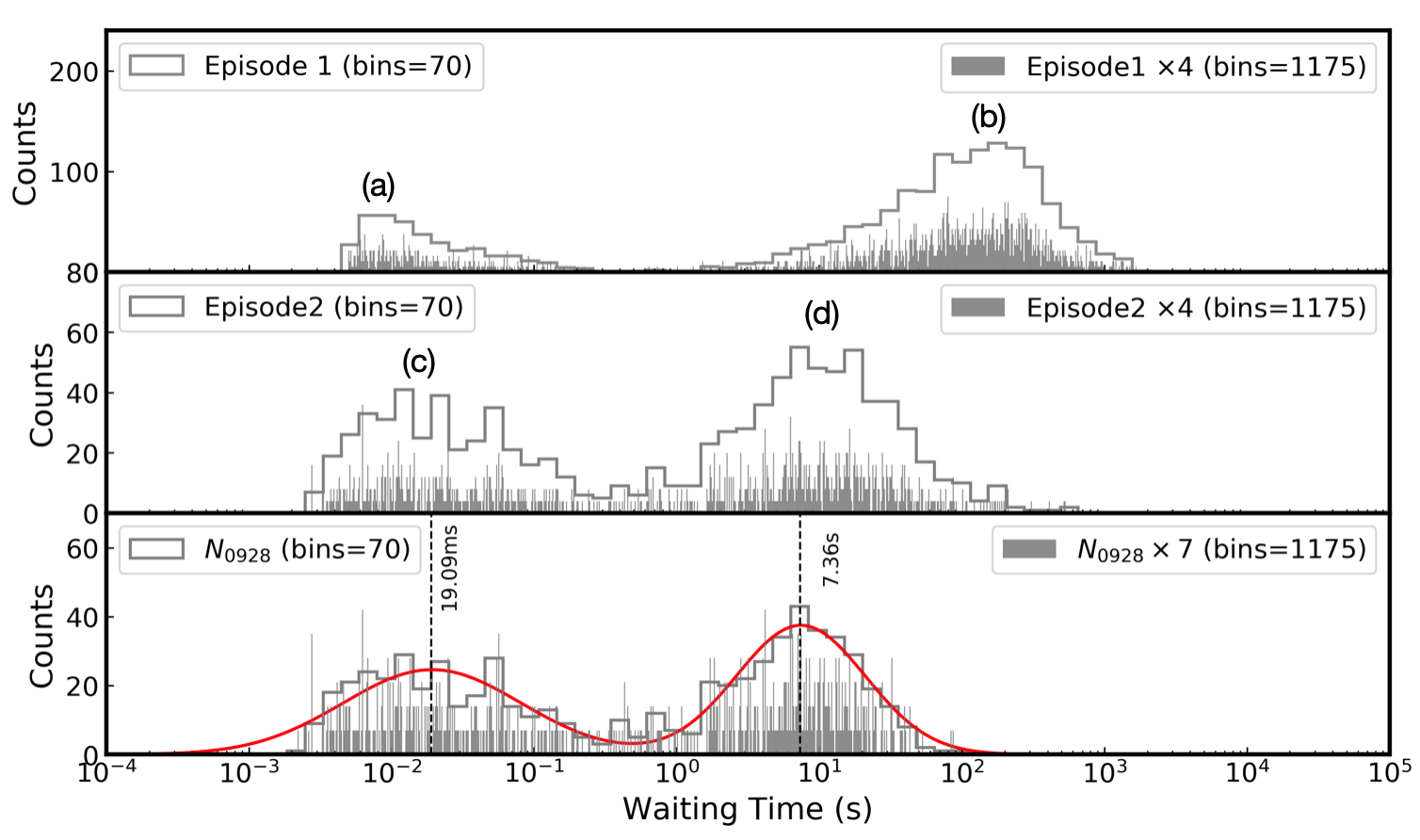}
\caption{The figure is divided into upper, middle and lower panels. The upper panel is the waiting time of the first active episode, the middle is  the second active episode, and the lower is the waiting time of September 28th with the highest burst rate. We show the results for different numbers of bins. For the 28th with the highest burst rate, we perform a Gaussian fit and plot the mean values of the double peaks.}
\label{tab:waitingtime}
\end{figure*}

\begin{figure*}[!hbt]
\centering
\includegraphics[width=0.9\linewidth]{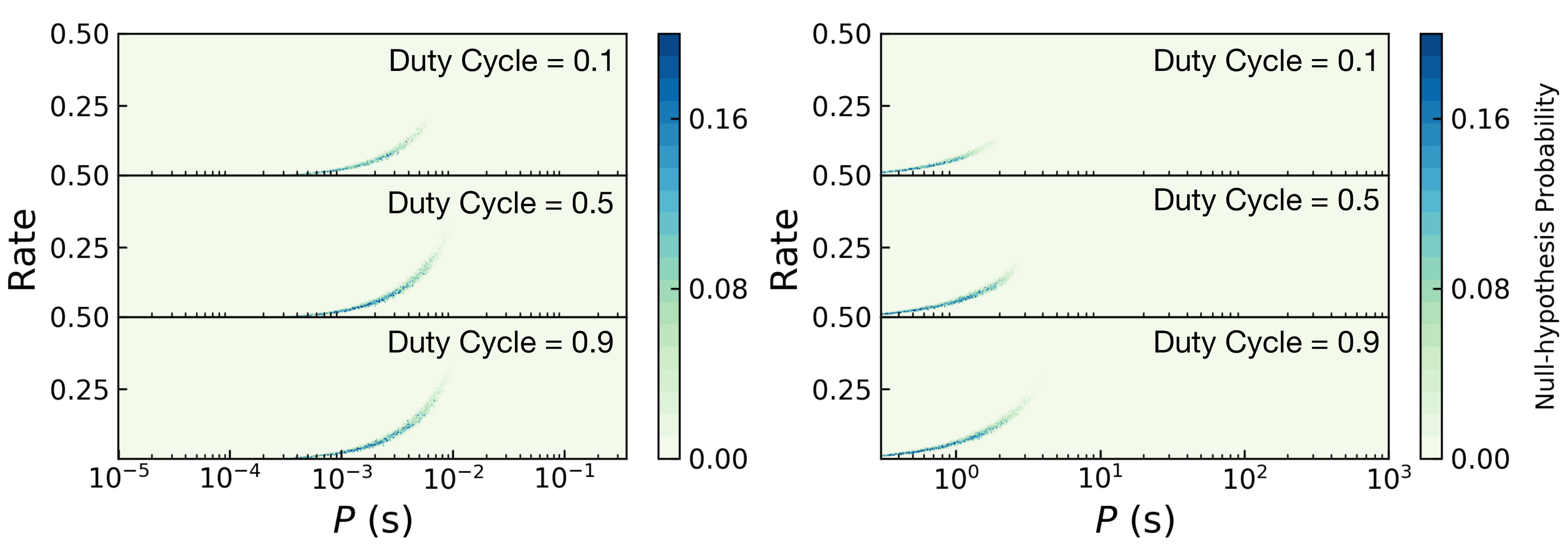}
\caption{The K-S test results of the simulated time series and the observed time series. The figure shows K-S test null-hypothesis probability of the waiting time distribution of the second active episodes. We simulate the two peaks independently since they may relate to different physics. The left panel corresponds to (c) in Figure \ref{tab:waitingtime}, while the right panel corresponds to (d).}
\label{tab:waitingtimekstest}
\end{figure*}

\subsection{Test putative periods using waiting times}
Waiting times are the time lapses between consecutive pulses. They often are strongly affected by the FRB repeating rate. 
\citet{cruces2021repeating, li2021bimodal} studied the waiting time distribution of FRB\,20121102A, focusing on the fitting of the distribution. They get the shape factor of the Weibull distribution to study their aggregation characteristics. In this section, we simulate {pulses from a rotating object of different spin periods and compare their waiting times} to the observed waiting times to gain insight into the range of the possible spin period for FRB\,20201124A.

This work focuses on episode 2 (September 25-28th, 2021) of the two active episodes from FRB\,20201124A, which reaches an order of magnitude higher peak event rate than episode 1 in April 2021 \citep{xu2021fast}.
The drastically different burst rates lead to significantly different waiting time distributions. 
We plot the waiting time distribution in Figure \ref{tab:waitingtime}. 
The first and second panels show the overall waiting times from the active episode 1 and 2, respectively, and the bottom panel shows those from the one hour observation on September 28th, 2021, from which the event rate is at its peak. 
The waiting time property has the following features: 1. two significant bumps, one peaking in milliseconds and the other peaking in seconds. 2. the left millisecond peak is almost identical in the two episodes, indicating that these small waiting times may be caused by intrinsic physics that is not affected by the burst rate. 3. the right, seconds-level peak {become smaller in episode 2}, driven by the change in the event rate.
The millisecond waiting time peak reported in this work appears to be less asymmetric than the same waiting time distribution reported for episode 1  \cite{xu2021fast}. This apparent change is likely because we split the multi-components into individual pulses when doing this paper's analysis, so we tend to get shorter waiting times than \cite{xu2021fast}.

\begin{figure*}[!hbt] 
\centering 
\includegraphics[width=0.83\linewidth]{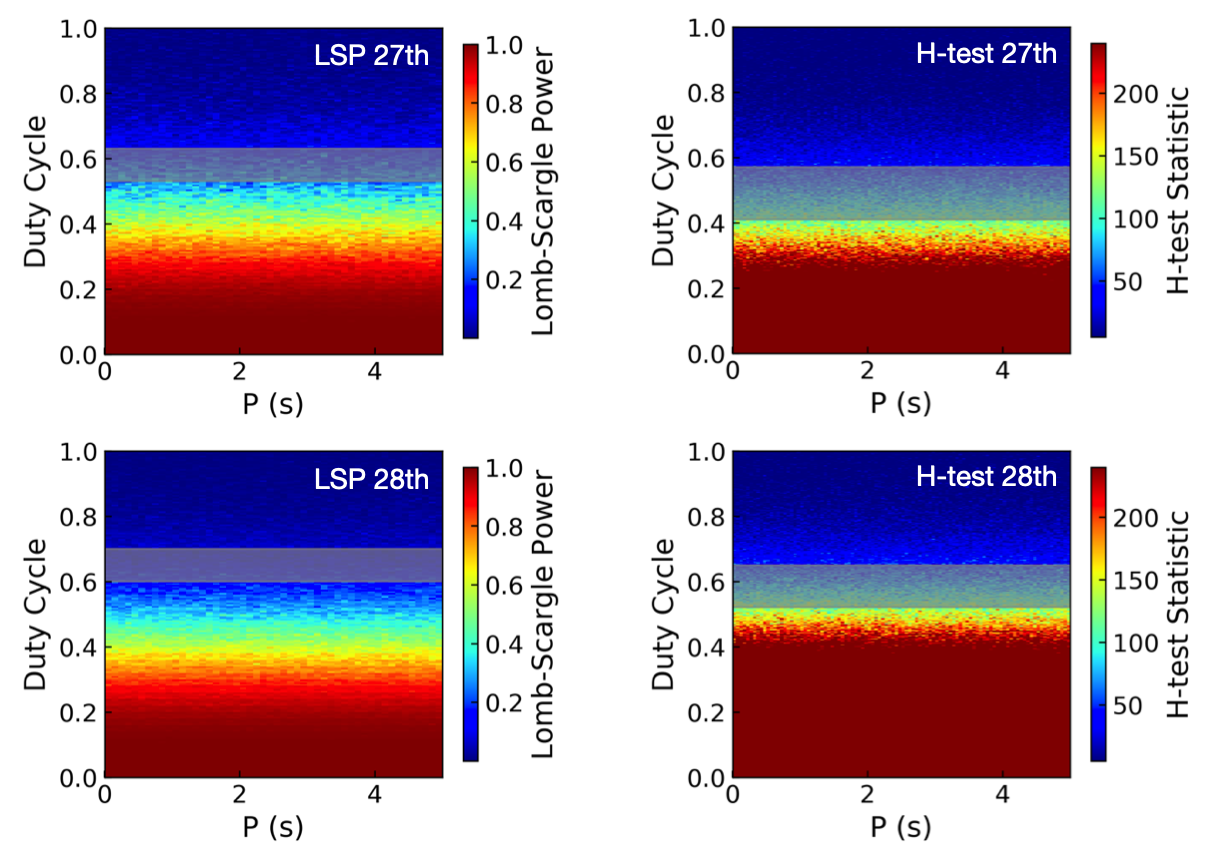} 
\caption{We generate periodic time series with different duty cycles and then use the previous methods $H$-test and LSP to search for periods, respectively. The figure shows the $H$ statistic and the power of the LSP, which becomes smaller when the duty cycle increases.} 
\label{tab:ssim}
\end{figure*}

The waiting time distribution itself {does not clearly show a spin} periodicity. {If a time series has a clear period, sharp peaks shall appear in the waiting time distribution. Considering that {for some pulsars with a significant duty cycle and nulling behavior} the pulses {could appear in a range of phases and may not appear in every period}, the peak of waiting time distribution may be broadened into the observed shape.}

{The double-peaked waiting time distribution} is likely the result of physics of two different time scales. 
{There are two possible scenarios:}
1. the millisecond-scale waiting time peak is caused by the spin of the source, while the second-scale waiting time peak is caused by the source turning on and off in timescales of many seconds, similar to the nulling phenomenon in pulsars.
2. the second-scale waiting time peak is caused by the rotation of the source, while the millisecond-scale peak comes from pulse micro-structures similar to those observed in some pulsars \citep{liukuo2022}.

For scenario 1, we only simulate a millisecond pulsar. This will only generate waiting time distribution analogous to the left waiting time peaks. 
In the scenario 2, we simulate an intermittent pulsar with period of seconds, and ignore the millisecond left peak, which is presumably related to the intrinsic radiation process. 
{We simulate pulses of various periods, burst rate and duty cycles.}
We omit $\dot P$ because it is a negligible effect in a timescale of seconds and should not affect the distribution of waiting times. 
We generate samples of {pulse arrival times} {following the von-Mises distribution} for a given duty cycle, {burst rate and 1 hour integration time.}  {Whether a pulse appears in the current period depends on a Poisson chance based on the burst rate.} 
Then we employ a Kolmogorov–Smirnov (K-S) test to compare the simulated pulses' waiting time distributions with those observed from the  active episode 2. The result is shown in Figure \ref{tab:waitingtimekstest}. 
The three plots in the left panel of the Figure correspond to the K-S test results from fitting the left waiting time peak for episode 2 marked with (c) in Figure \ref{tab:waitingtime}, and the right panel corresponds to the result of fitting the right waiting time peak of episode 2 marked with (d) in Figure \ref{tab:waitingtime}.

The larger the null-hypothesis probability, the higher the degree of agreement between the observation and the simulation. 
One can see that only a narrow range of periods are compatible to the observed sample. We find that possible compatible period range of the left and right peaks in [0.29\,ms, 10.36\,ms] and [0.31s, 3.04\,s ], assuming a minimum null-hypothesis probability of 0.05.

\subsection{Large duty cycle and period detectability}
To investigate the potential reasons for a non-detection of the spin period, we simulate periodic time series of different duty cycles and then perform the aforementioned searches on these time series using the $H$-test and LSP. We keep the {simulated observing} time at one hour and the total number of pulses {close} to the September 28th burst number (424 after grouping together the micro-components). 
We fold and compute the $H$ and LSP power of the mock time series. The resulting statistics of different periods and duty cycles are displayed in Figure \ref{tab:ssim}. We employ the $H$ statics threshold of 72 in 27th and 74 in 28th, and LSP power of 0.17 in 27th and 0.1 in 29th and mark the region where the $H$ and LSP power of the time-series start to drop below thresholds with the transparent gray band in Figure \ref{tab:ssim}. The selection of the $H$ statistical threshold is based on the cumulative probability and the number of trials. The LSP threshold is chosen to be the highest spectral power from the LSP search of that day. These simulations show that the spin would not have been detectable if the pulse duty cycle is larger than $0.49\pm0.08$ if we use the $H$ statics and $0.58\pm0.05$ if using the LSP.

Like the \texttt{PRESTO} S/N, the von-Mises duty cycle is an indication of the signal significance. 
A smaller duty cycle corresponds to a sharper and more significant pulse shape, while a larger duty cycle corresponds to a wider and less significant pulse shape. 
Since there are only 424 independent TOAs (from September 28th), {Figure \ref{tab:ssim} shows that} we can only detect the {period} signal if {its pulse distribution is} sharp enough.

\section{Summary and Discussion}
\label{sect:summary}
In this study we report a systematic search of pulsar spin periodicity from an extremely active episode of FRB20201124A.
We attempted different periodicity search strategies for the data of FRB\,20201124A taken during its active episode in September 2021 by the FAST telescope: 1. searching the time series dedispersed from the original data; 2. searching for periodicity based on the pulse TOAs; 3. searching for periodicity in the multi-component fine pulse structures; 4. Periodicity Search on selected pulses. 
Despite various attempts, we did not find a credible period with high confidence. Here we consider a period credible if
1. it is statistically significant;
2. it appears in the search result for multiple days; and
3. it is not caused by obvious artificial reasons such as the sampling frequency. 

Based on all the pulse TOAs from four days, we have folded and searched for a period in the range of 0.01\,Hz to 333\,Hz and a linear period derivative equal to the effect of a Doppler acceleration of $a/c$ in the range of [-2, 2]$\times 10^{-9}$~s$^{-1}$, with sufficient resolutions to resolve a 0.2 phase shift for the last TOA. This is the maximum acceleration produced by the pull of a 100~M$_{\rm \odot}$ mass star or a black hole companion in an orbit of 40\,days around the neutron star.
Any lighter companion star or longer orbit would lead to a smaller maximum acceleration.
We also folded and searched the TOAs from individual observing sessions, for a period in the range 0.01\,Hz to 1000\,Hz and period derivative for Doppler acceleration $a/c$ in [-1,1]$\times$10$^{-6}$s$^{-1}$.
This acceleration is equal to the pull of a 200~M$_{\rm \odot}$ mass star or a black hole companion in an orbit of 10\,hours.
Both searches yield no credible spin period.
We suspect that the possible reasons for a non-detection, assuming that the FRB is from a rotating neutron star, include a potentially large pulse duty cycle ($>0.49$) or that the pulsar has a heavier companion or a shorter orbit than we have assumed.

Although no credible period was identified, we provide in Table \ref{tab:candlist} a list of top period candidates that we could neither positively identify nor rule out. We present the periods discovered by various approaches as well as the $H$ value of the fold result.

From the fine structures of pulses, we have identified 53 bursts with fine multi-components through which tentative periods could be derived from more than four components.
The multi-components of FRBs share many similarities with the micro-structures of pulsars.
They are observed in different frequencies \citep{lange1998radio} and both exhibit quasi-periodic properties. However, there are also some slight differences. First, the characteristic timescale of pulsar micro-structure ranges from a few microseconds to hundreds of microseconds \citep{Johnston1996PulsarsP}. Its typical quasi-period ranges from several microseconds to several thousand microseconds \citep{popov2002pulsar,popov1987microstructure}. The quasi-periods observed in FRB\,20201124A are on the order of milliseconds, albeit only a sample of FRBs exhibit micro-components. Perhaps microsecond-level fine structures exist in FRBs but are not yet observable due to DM and scatter smearing in most observations.
Additionally, the polarization properties are also different. \cite{kramer2002high} found a strong correlation between Stokes parameters $V$ and $I$ in the micro-structure of Vela, and they did not find the handedness of the circular intensity changes within a micro-pulse. In FRB\,20201124A, there are circular polarization sign changes in multi-components, and the circular polarization has no obvious correlation with $I$. 

In this paper, we find that the periods of multi-components vary in a wide range and do not seem to correlate with each other. Therefore, we conclude that the low-significance periodicity observed in the pulse micro-structures is likely due to radiation mechanisms, not to the spin of the source object. Some Low-significance periods have also been found in some FRBs. We advocate caution when linking such quasi-periodic phenomenon to the spin period of the source.

We searched for spin period systematically between 1~ms and 100~s and linear spin acceleration up to a $F_1=3.33\times10^{-4}$~s$^{-2}$ from the second active episode of FRB\,20201124A and found no convincing spin period that appeared in two separate observations.
From the 424 bursts detected in the one hour FAST observation taken on 2020 September 28th and 186 bursts detected on 2020 September 27th, our periodicity search methods should have found a significant pulsar signal if it had a duty cycle less than 0.49 and with $P$ and $\dot{P}$ in our search ranges.
Most pulsars show a duty cycle much smaller than this \citep{Rankin90,Kramer98}, but
large duty cycles do exist for few special cases where strong bridge emission exist or the pulsar is an aligned rotator \citep{KouFF21,WangZL22,HanJL21}.
Some millisecond pulsars also show relatively wide pulses \cite{Johnston93,Kramer98,Stairs99}.
Single pulses from the magnetar PSR~J1622-4950 were found to spread in a wide range of phases \citep{Levin12}.
Our result contests the FRB models invoking a pulsar with typical polar cap emission or a spinning down magnetar with $B_s < 10^{15}$~G.
For magnetars with $P>1$~s, we also rule out detection with duty cycle $< 0.49$ and $B_s < 10^{16}$~G.
Note that all magnetars known to date have $P>1$~s and $B_s < 10^{16}$~G \footnote{\url{https://www.physics.mcgill.ca/~pulsar/magnetar/main.html}\citep{Olausen14}}.
The acceleration we searched also allows us to contest binary models of which the orbital period is longer than 10 hours and companion below 100~M$_{\rm \odot}$.
In summary, our results contest many of the contemporary models of repeating FRBs in a wide range of parameter space.

It is worth mentioning that the above statements apply to pulsar-like emission models of FRBs that invoke emission of FRBs from the open field line regions of magnetars \citep[e.g.][]{kumar17,yangzhang18,wang19,lu20}. However, since FRBs are much more energetic than radio pulsar emission, an FRB originating from the inner magnetosphere of a magnetar would likely greatly distort the field lines \citep{qu22}, making the well-defined open field line region fuzzier and enlarging the observed duty cycles. Alternatively, if the FRB emission region is outside of the magnetopshere as some models suggest \citep[e.g.][]{metzger19,beloborodov20}, then the emission phase could be more random. As a result, the non-detection of periods from this source does not necessarily rule out a magnetar as the FRB source, even though other possibilities may be also considered \citep{katz2022absence}.

\clearpage
\begin{table}[]
\scriptsize
\renewcommand\arraystretch{1.2}
\center 
\caption{Presto search Candidate list} 
\label{tab:candlistpresto}
\resizebox{0.99\linewidth}{!}{
\begin{tabular}{llllll}
\hline
Index   & Harm & $P_{bary}$ (ms)   & $\dot{P}_{bary}$ (s/s)        & Date & Sigma     \\
  \hline
1  & 32 & 126.06423(11)     & $-5.6(2.3)\times10^{-10}$ & 59484 & 6.2  \\
2  & 32 & 162.078505(96) & $-4.04(21)\times10^{-9}$  & 59484 & 17.1 \\
3  & 16 & 50.025310(27)     & $-1.13(59)\times10^{-10}$ & 59484 & 5    \\
4  & 32 & 71.575233(38)     & $-4.74(84)\times10^{-10}$ & 59484 & 15.9 \\
5  & 32 & 99.047556(69)     & $-5.0(1.5)\times10^{-10}$ & 59484 & 11.2 \\
6  & 32 & 88.044581(40)     & $-2.029(90)\times10^{-9}$ & 59484 & 14.4 \\
7  & 32 & 76.511895(42)     & $-1.807(93)\times10^{-9}$ & 59484 & 11.5 \\
8  & 32 & 61.905820(42)     & $1.185(93)\times10^{-9}$  & 59484 & 11.1 \\
9  & 32 & 117.055470(68)    & $-2.19(15)\times10^{-9}$  & 59484 & 17.6 \\
10 & 32 & 91.625691(41)     & $1.299(92)\times10^{-9}$  & 59484 & 12.2 \\
11 & 32 & 82.147963(36)     & $3.86(80)\times10^{-10}$  & 59484 & 13.4 \\
12 & 32 & 114.055239(60)    & $-4.67(13)\times10^{-9}$  & 59484 & 13.9 \\
13 & 32 & 90.044511(60)     & $-2.51(13)\times10^{-9}$  & 59485 & 13.2 \\
14 & 2  & 499.89481(70)     & $4.24(16)\times10^{-8}$   & 59485 & 60.9 \\
\hline
\end{tabular}
}
\end{table}

\begin{table}[]
\scriptsize
\renewcommand\arraystretch{1.2}
\center
\caption{FFA Candidate list} 
\label{tab:candlist}
\resizebox{0.99\linewidth}{!}{
\begin{threeparttable}
\begin{tabular}{clccclcc}
\hline
Index & Period(s) & Date  & SNR   & Index & Period(s) & Date  & SNR   \\
\hline
1     & 26.51(2)     & 59484 & 70.62 & 13    & 106.07(2)    & 59484 & 67.94 \\
2     & 31.84(1)     & 59484 & 63.31 & 14    & 123.1(1)    & 59484 & 67.00 \\
3     & 34.02(1)     & 59484 & 61.38 & 15    & 212.1(3)    & 59484 & 65.96 \\
4     & 37.86(1)     & 59484 & 68.61 & 16    & 265.2(6)    & 59484 & 77.31 \\
5     & 38.40(1)     & 59484 & 66.28 & 17    & 344.1(3)    & 59484 & 66.81 \\
6     & 52.18(1)     & 59484 & 63.78 & 18    & 424.4(8)    & 59484 & 67.45 \\
7     & 53.02(9)     & 59484 & 74.36 & 19    & 591.5(5)    & 59484 & 71.23 \\
8     & 67.50(2)     & 59484 & 63.69 & 20    & 36.714(3)     & 59485 & 62.91 \\
9     & 72.25(3)     & 59484 & 86.92 & 21    & 70.0(1)     & 59485 & 71.69 \\
10    & 75.76(4)     & 59484 & 64.99 & 22    & 136.41(8)    & 59485 & 66.50 \\
11    & 88.4(1)     & 59484 & 65.61 & 23    & 227.3(4)    & 59485 & 67.59 \\
12    & 98.56(3)     & 59484 & 62.60 & 24    & 301.2(7)    & 59485 & 66.44\\

\hline
\end{tabular}
      \end{threeparttable} 
}
\end{table}
\begin{table}[]
\tiny
\renewcommand\arraystretch{1.2}
\center
\caption{LSP Candidate list} \label{tab:lsp}
\resizebox{0.7\linewidth}{!}{
\begin{threeparttable}

\begin{tabular}{ccccc}
\hline
Index & Period(s)    & Date1 & Date2 & FAP \\
\hline
1     & 0.7924(3) & 59482 & 59485 & 0.1 \\
2     & 0.3216(2) & 59483 & 59485 & 0.1 \\
\hline
\end{tabular}

      \end{threeparttable} 
}
\end{table}
\begin{table}[]
\tiny
\renewcommand\arraystretch{1.2}
\center 
\caption{$P-\dot P$ search Candidate list} 
\label{tab:candlistppdot}
\resizebox{0.99\linewidth}{!}{
\begin{threeparttable}
\begin{tabular}{cccccc}
\hline
Index & $F_0$(Hz)                           & $F_1$($s^-2$)                           & Date  & $H$ value                       \\
\hline
1    & 185.729821  & -92.037618$\times10^{-6}$                             & 59485 & 76.894955                     \\
2    & 144.325944 & 735.642633$\times10^{-6}$                                & 59485 & 78.167674                     \\
3    & 222.950632  & -1540.698504$\times10^{-6}$                                  & 59485 & 78.905402                     \\
4    & {316.941093} & {143.882353}$\times10^{-9}$              & 4day  & {79.78374}  \\
5    & {293.883889} & {478.658824}$\times10^{-9}$             & 4day  & {83.259686} \\
6    & {279.292725} & {-315.964706}$\times10^{-9}$             & 4day  & {80.035248} \\
7    & {274.766622} & {-522.439216}$\times10^{-9}$             & 4day  & {83.077999} \\
8    & {246.126386} & {352.545098}$\times10^{-9}$             & 4day  & {89.229852} \\

\hline
\end{tabular}
      \begin{tablenotes}  		       
      \item The errors of $F_0$ and $F_1$ are the precision of our search.  
      \end{tablenotes}        
      \end{threeparttable} 
}
\end{table}
\begin{table}[]
\scriptsize
\centering
\caption{Multi-components of FRB\,20201124A} \label{tab:ms2}
\begin{threeparttable}
\begin{tabular}{lcccc}
\hline \\
Index & N$\ast$ & $P$  (ms) & $\overline{R}$  & MJD \\
\hline \\
1     & 5               & 7.164  & 0.0240 & 59484.81606 \\
2     & 7               & 2.638  & 0.0411 & 59484.81650 \\
3     & 5               & 2.471  & 0.0685 & 59484.82173 \\
4     & 4               & 11.996 & 0.0368 & 59484.82257 \\
5     & 6               & 5.587  & 0.0415 & 59484.82357 \\
6     & 5               & 3.801  & 0.0347 & 59484.82633 \\
7     & 6               & 3.504  & 0.0716 & 59484.83085 \\
8     & 4               & 5.637  & 0.0077 & 59484.83419 \\
9     & 5               & 5.182  & 0.0337 & 59484.84094 \\
10    & 8               & 13.521 & 0.0272 & 59484.84349 \\
11    & 5               & 6.409  & 0.0308 & 59484.84719 \\
12    & 4               & 8.077  & 0.0609 & 59484.84938 \\
13    & 4               & 12.532 & 0.0532 & 59484.85148 \\
14    & 4               & 9.736  & 0.0748 & 59485.78325 \\
15    & 5               & 2.766  & 0.0198 & 59485.78338 \\
16    & 5               & 4.013  & 0.0761 & 59485.78488 \\
17    & 8               & 14.655 & 0.0238 & 59485.78561 \\
18    & 4               & 9.308  & 0.0381 & 59485.78653 \\
19    & 6               & 48.734 & 0.0058 & 59485.78912 \\
20    & 5               & 4.869  & 0.0436 & 59485.78933 \\
21    & 4               & 6.446  & 0.0931 & 59485.79068 \\
22    & 7               & 21.804 & 0.0355 & 59485.79374 \\
23    & 5               & 9.623  & 0.0456 & 59485.79432 \\
24    & 5               & 9.678  & 0.0552 & 59485.79515 \\
25    & 10              & 1.465  & 0.0471 & 59485.79612 \\
26    & 5               & 4.255  & 0.0748 & 59485.79922 \\
27    & 5               & 5.115  & 0.0209 & 59485.79949 \\
28    & 4               & 20.35  & 0.0087 & 59485.79996 \\
29    & 4               & 54.227 & 0.0292 & 59485.80002 \\
30    & 6               & 4.915  & 0.0120 & 59485.80020 \\
31    & 4               & 5.999  & 0.0629 & 59485.80131 \\
32    & 6               & 2.311  & 0.0463 & 59485.80133 \\
33    & 4               & 6.764  & 0.0444 & 59485.80161 \\
34    & 4               & 21.211 & 0.0045 & 59485.80171 \\
35    & 4               & 39.408 & 0.0172 & 59485.80309 \\
36    & 5               & 12.762 & 0.0208 & 59485.80323 \\
37    & 7               & 5.069  & 0.0078 & 59485.80379 \\
38    & 5               & 7.867  & 0.0674 & 59485.80521 \\
39    & 5               & 6.075  & 0.0679 & 59485.80717 \\
40    & 6               & 21.629 & 0.0080 & 59485.80729 \\
41    & 5               & 3.457  & 0.0429 & 59485.80764 \\
42    & 4               & 4.967  & 0.0888 & 59485.80808 \\
43    & 5               & 6.835  & 0.0148 & 59485.80927 \\
44    & 4               & 8.007  & 0.0439 & 59485.81131 \\
45    & 4               & 6.675  & 0.0213 & 59485.81193 \\
46    & 4               & 4.045  & 0.0574 & 59485.81267 \\
47    & 6               & 7.354  & 0.0171 & 59485.81270 \\
48    & 4               & 3.547  & 0.0763 & 59485.81270 \\
49    & 6               & 5.155  & 0.0141 & 59485.81702 \\
50    & 5               & 33.551 & 0.005  & 59485.81785 \\
51    & 4               & 4.956  & 0.0169 & 59485.81890 \\
52    & 7               & 5.918  & 0.0061 & 59485.82070 \\
53    & 4               & 4.677  & 0.0258 & 59485.82138
   \\   \\ \hline
\end{tabular}
      \begin{tablenotes}  		       
      \item $\ast$ Number of pulses.         
      \end{tablenotes}        
      \end{threeparttable} 
\end{table}

\begin{figure*}[!hbt]
    \begin{subfigure}[b]{0.5\textwidth}
        \centering
        \includegraphics[height=2.3in]{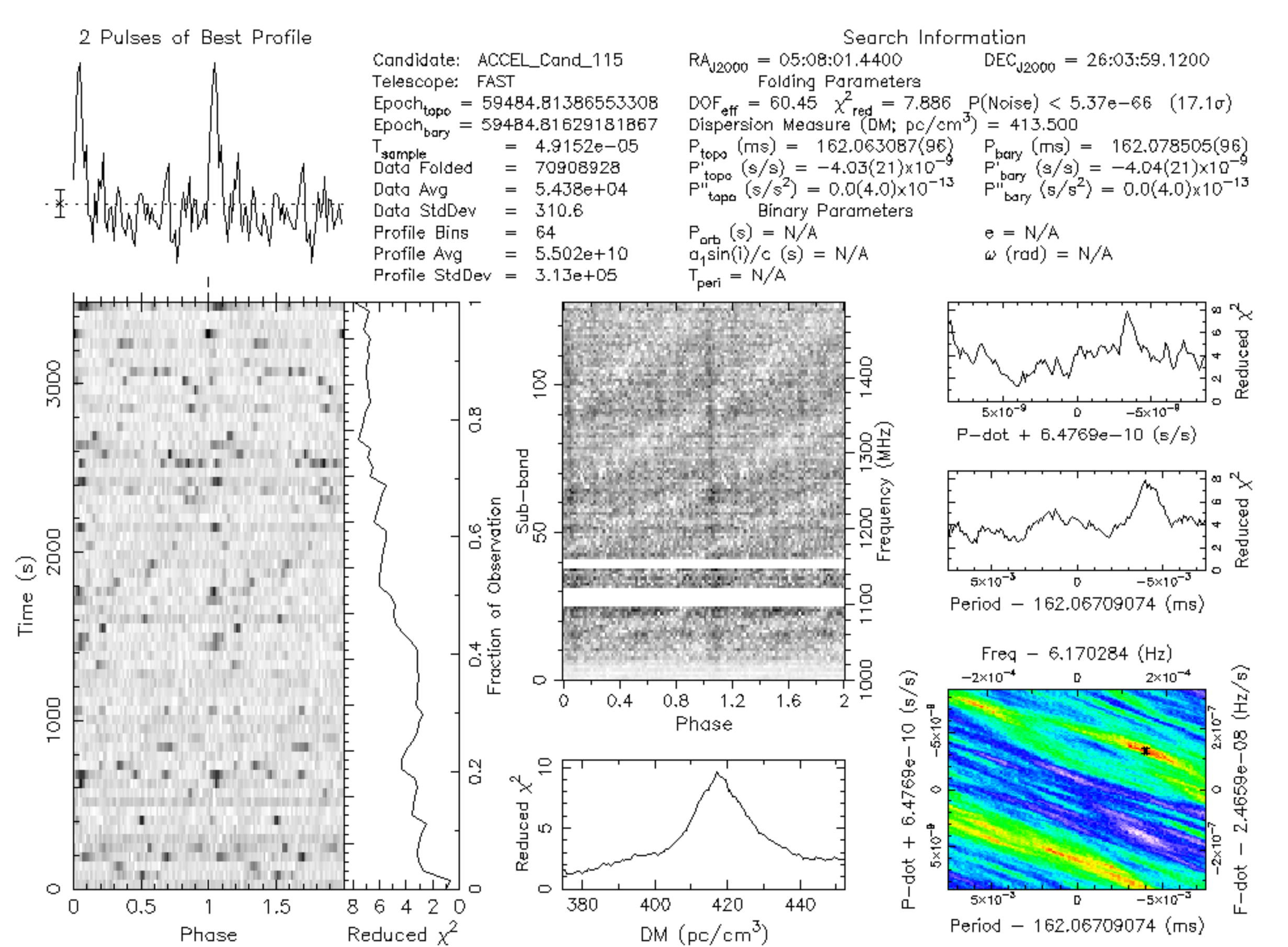}
        \caption{}
    \end{subfigure}%
    \hfill
    \begin{subfigure}[b]{0.5\textwidth}
        \centering
        \includegraphics[height=2.3in]{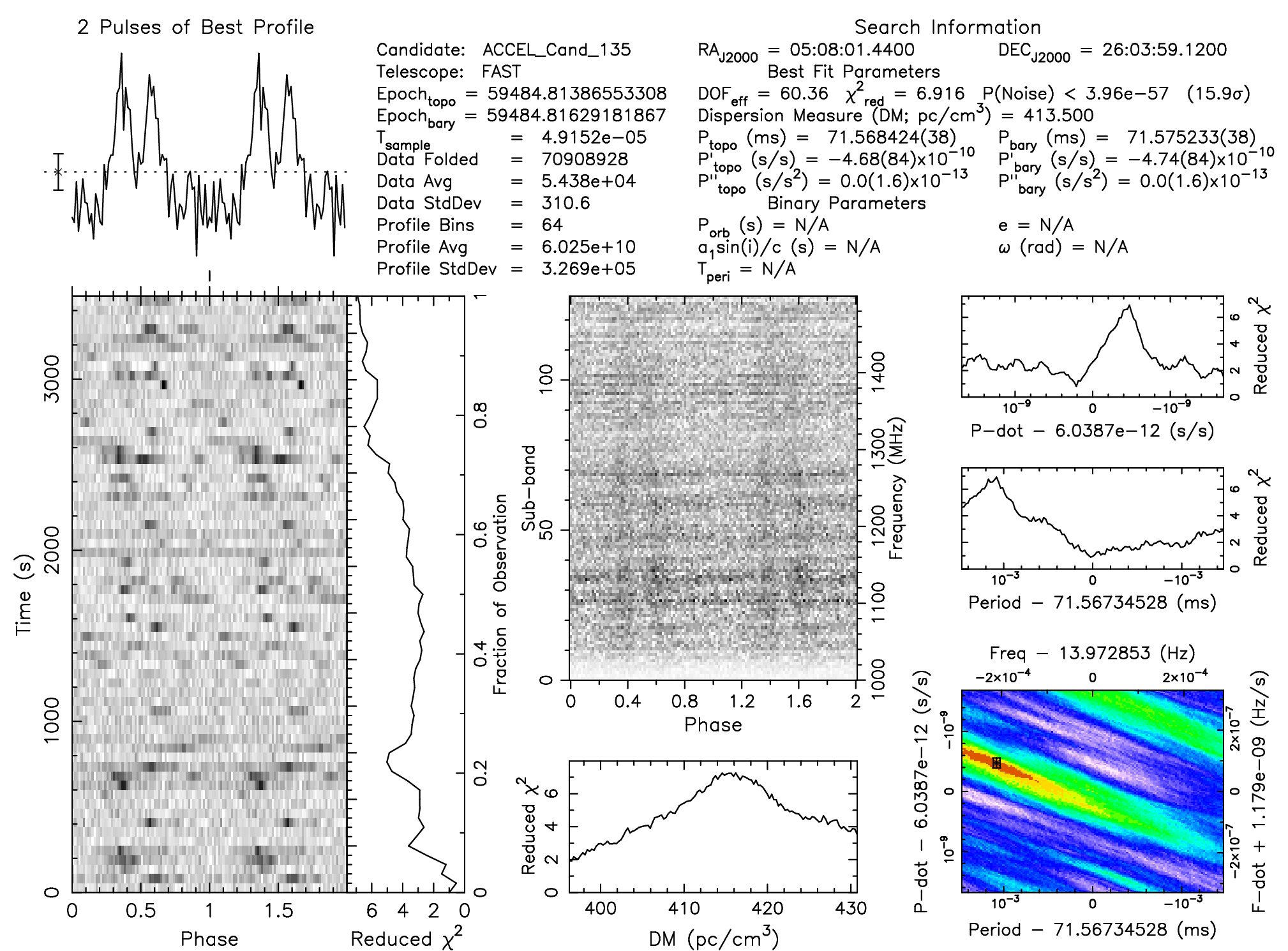}
        \caption{}
    \end{subfigure}%
    \hfill
    \begin{subfigure}[b]{0.5\textwidth}
        \centering
        \includegraphics[height=2.3in]{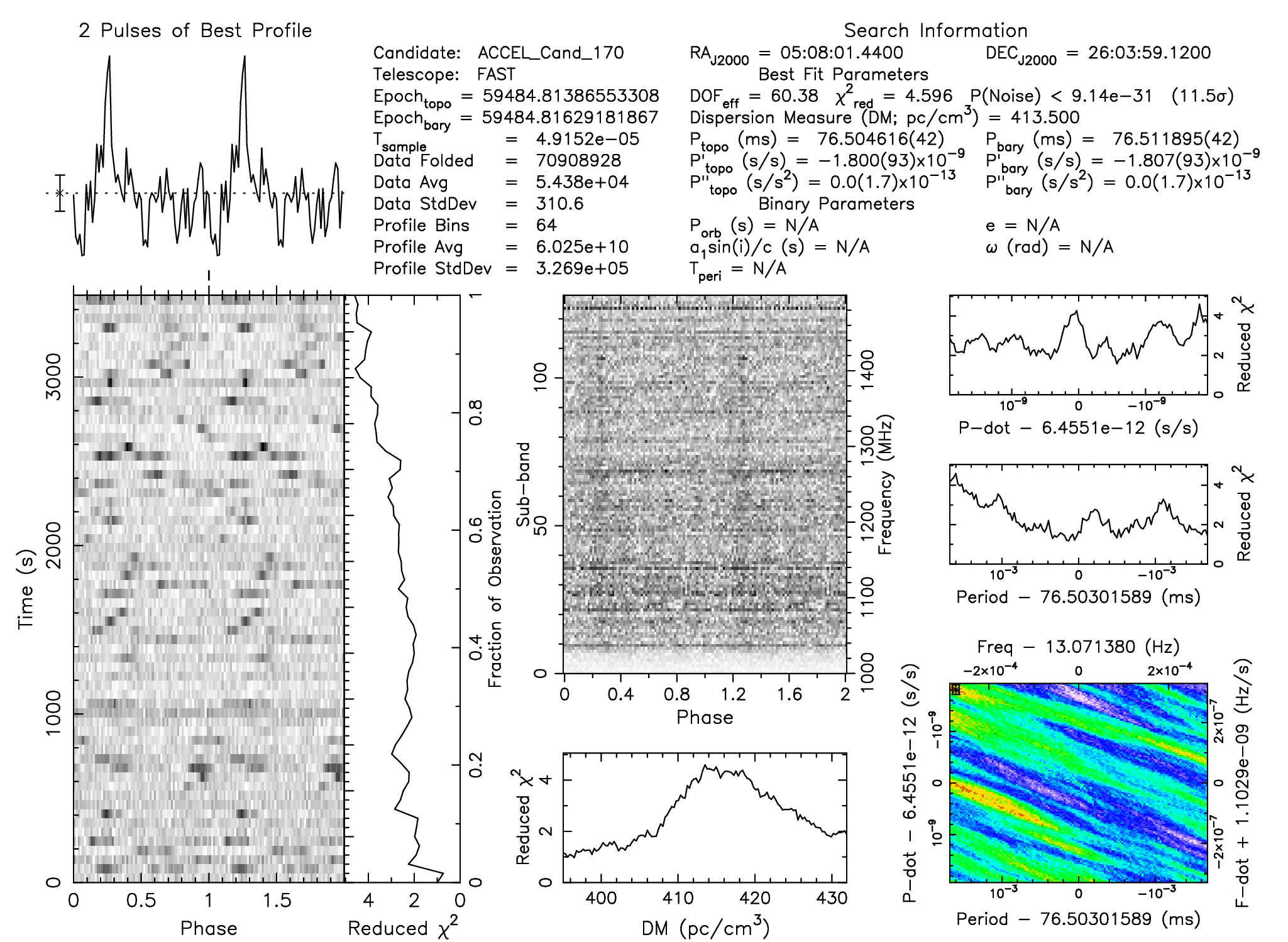}
        \caption{}
    \end{subfigure}%
    \hfill
    \begin{subfigure}[b]{0.5\textwidth}
        \centering
        \includegraphics[height=2.3in]{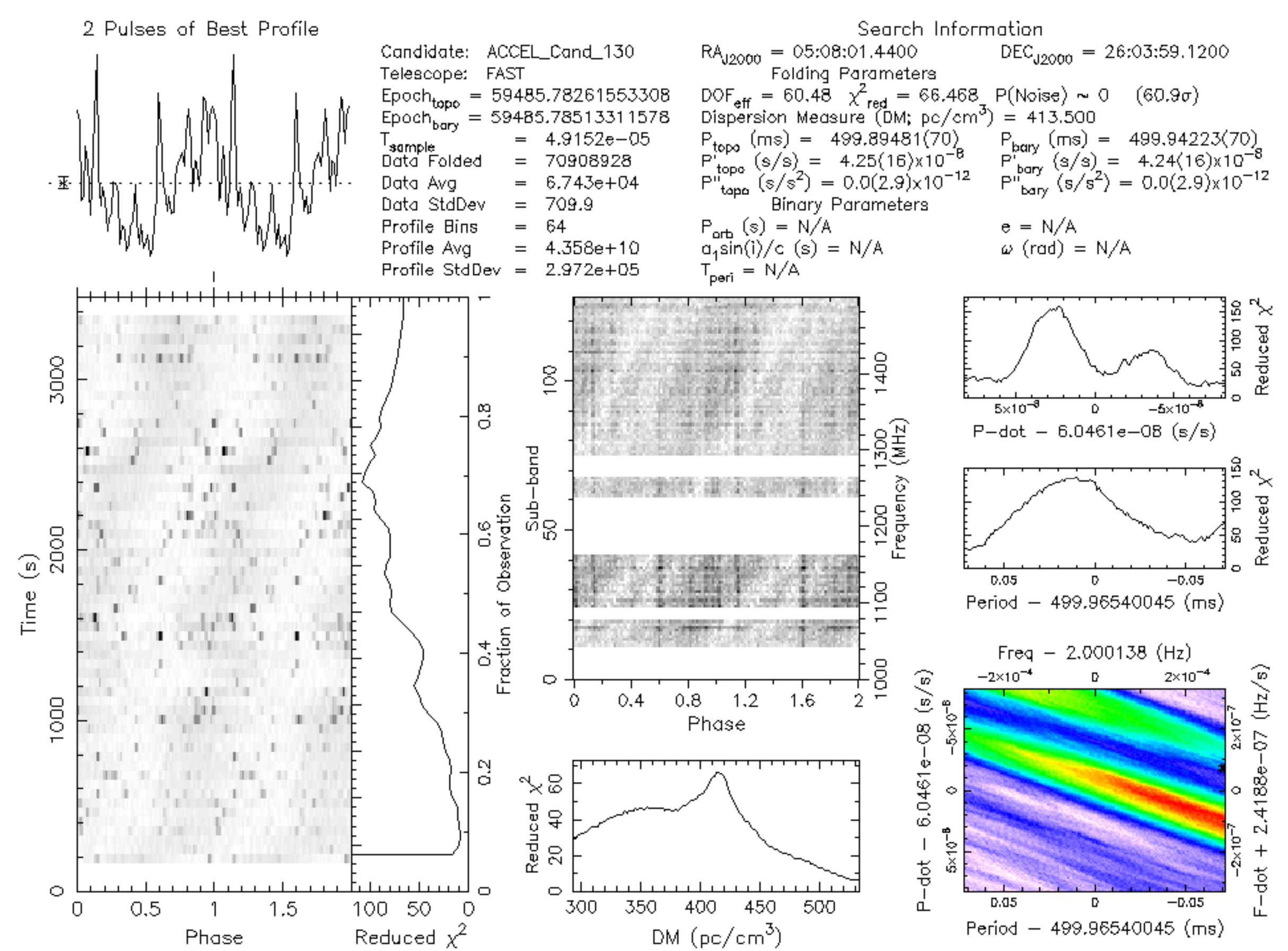}
         \caption{}
    \end{subfigure}%
     \caption{Presto candidates fold results.}
     \label{prestofig}
\end{figure*}

\clearpage

\clearpage
 \begin{figure*}[!hbt]
     \begin{subfigure}[b]{0.3\textwidth}
         \centering
         \includegraphics[height=2.3in]{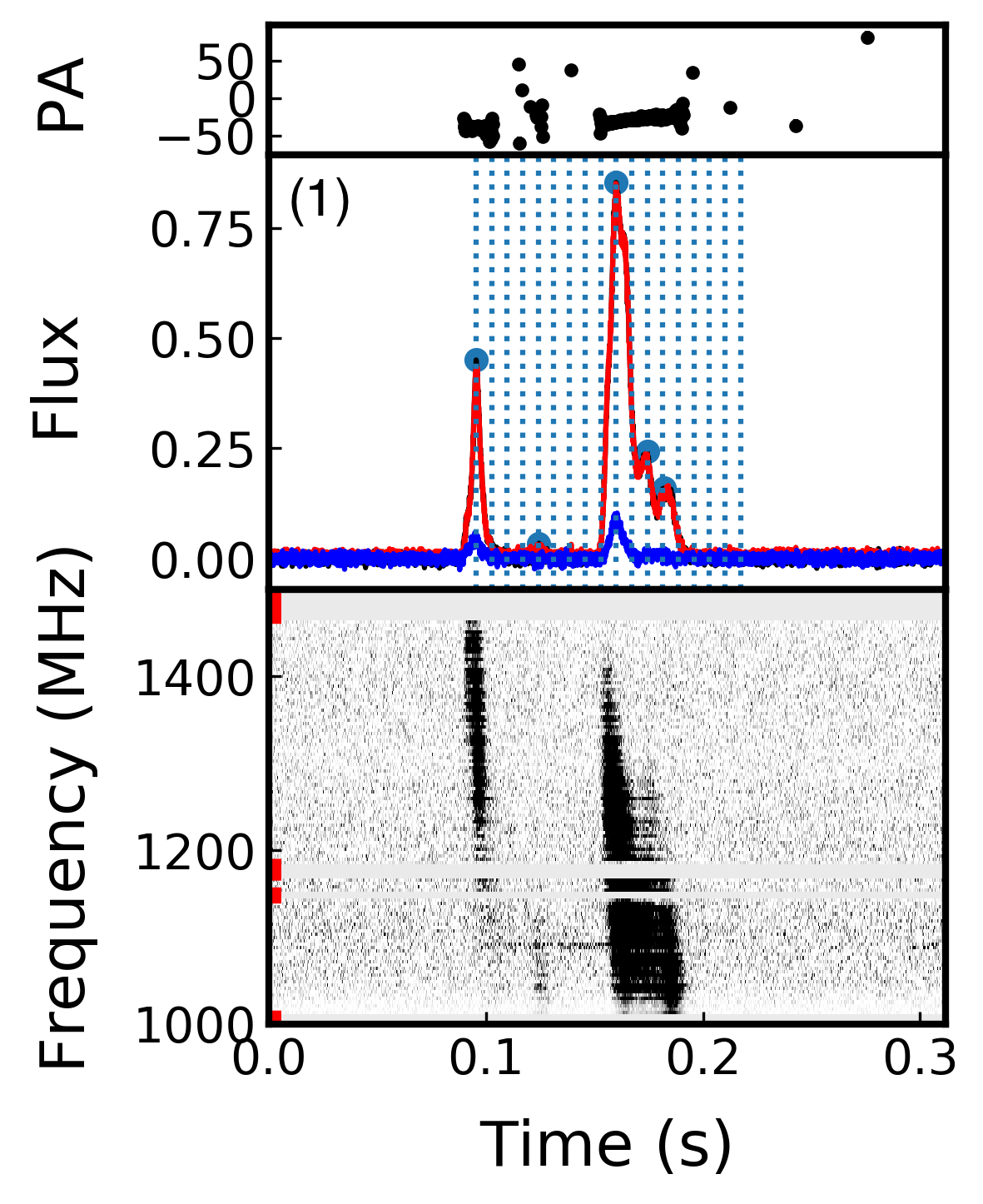}
     \end{subfigure}%
     \hfill
     \begin{subfigure}[b]{0.3\textwidth}
         \centering
         \includegraphics[height=2.3in]{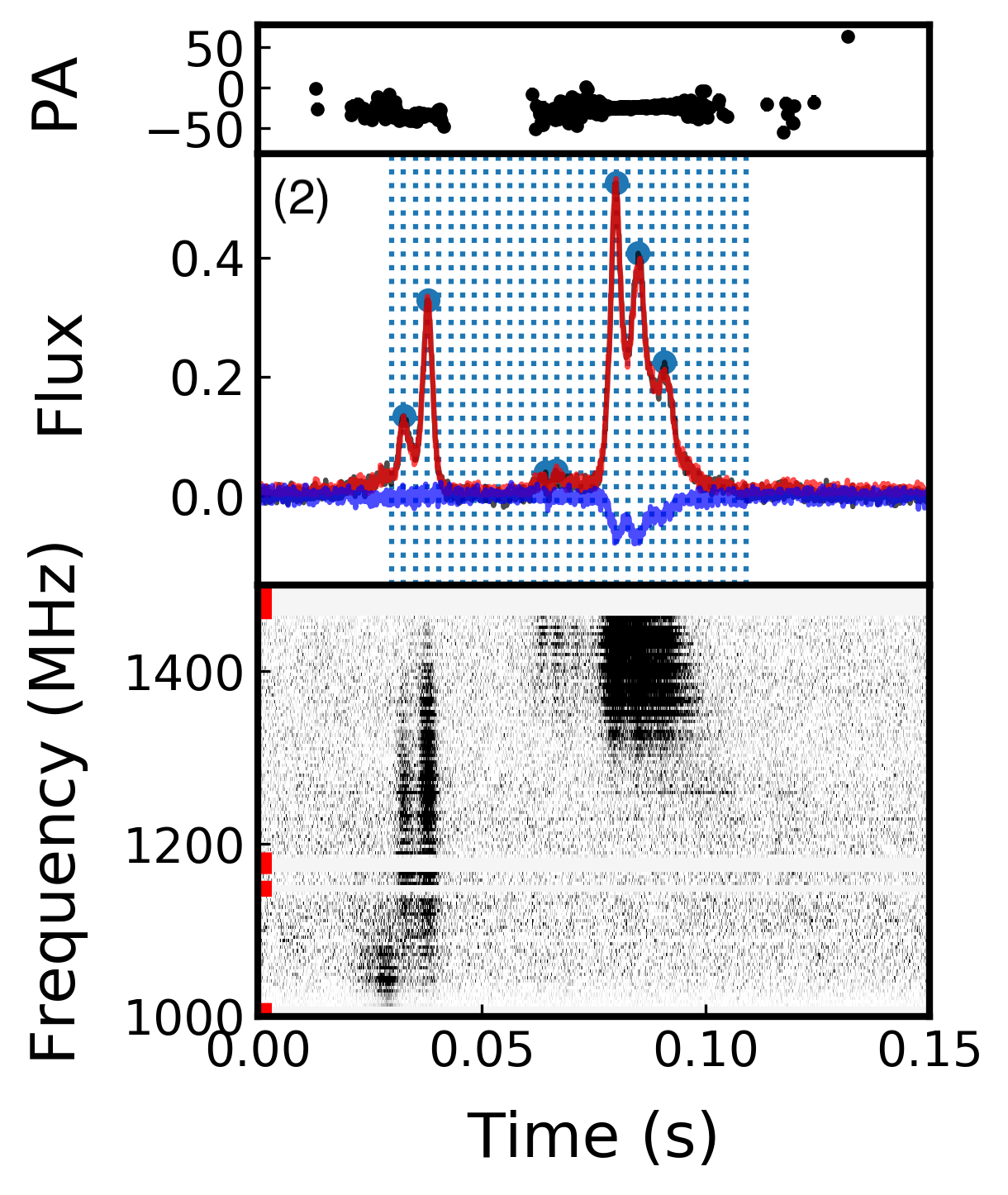}
     \end{subfigure}
     \hfill
     \begin{subfigure}[b]{0.3\textwidth}
         \centering
         \includegraphics[height=2.3in]{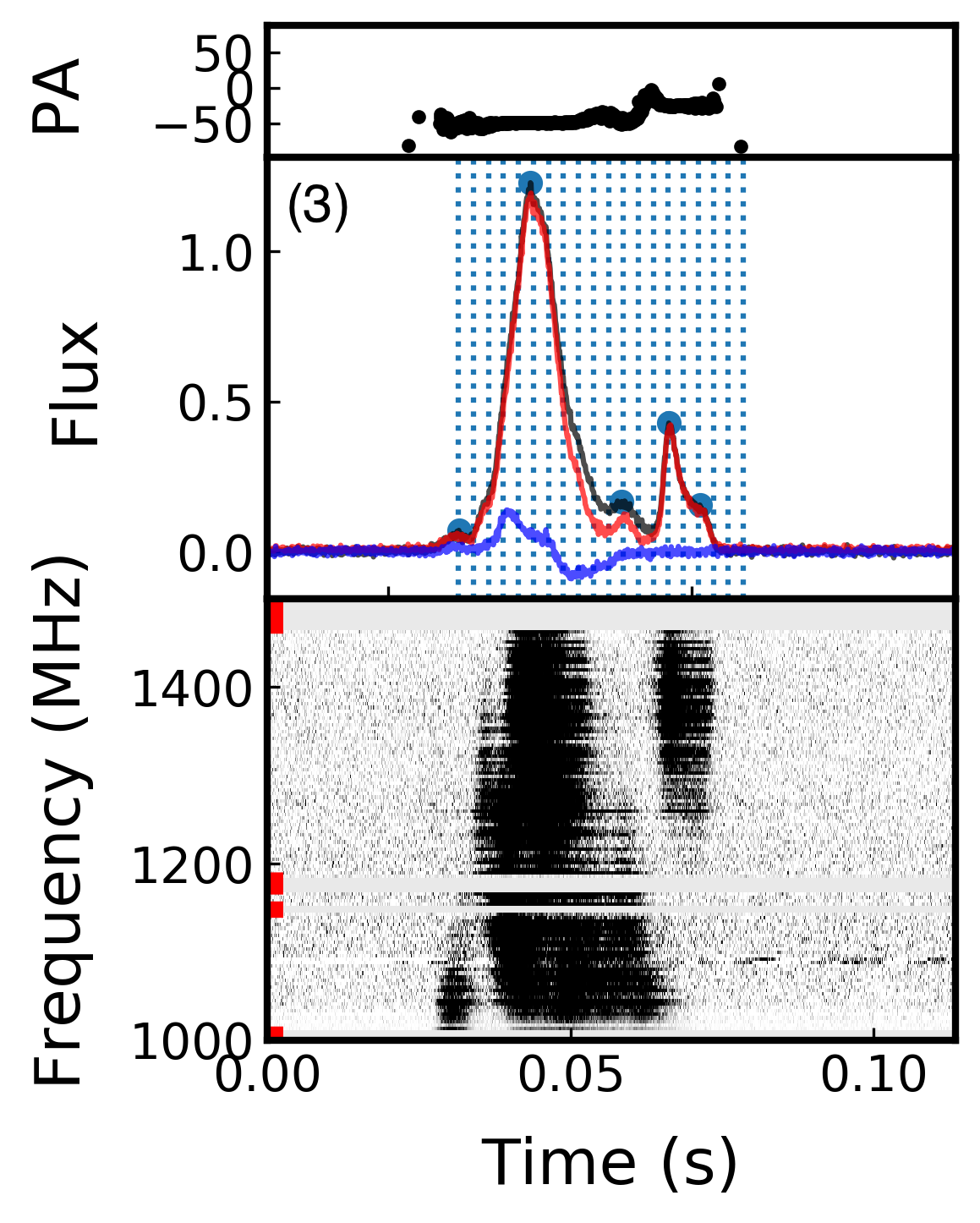}
     \end{subfigure}
     \hfill
     \begin{subfigure}[b]{0.3\textwidth}
         \centering
         \includegraphics[height=2.3in]{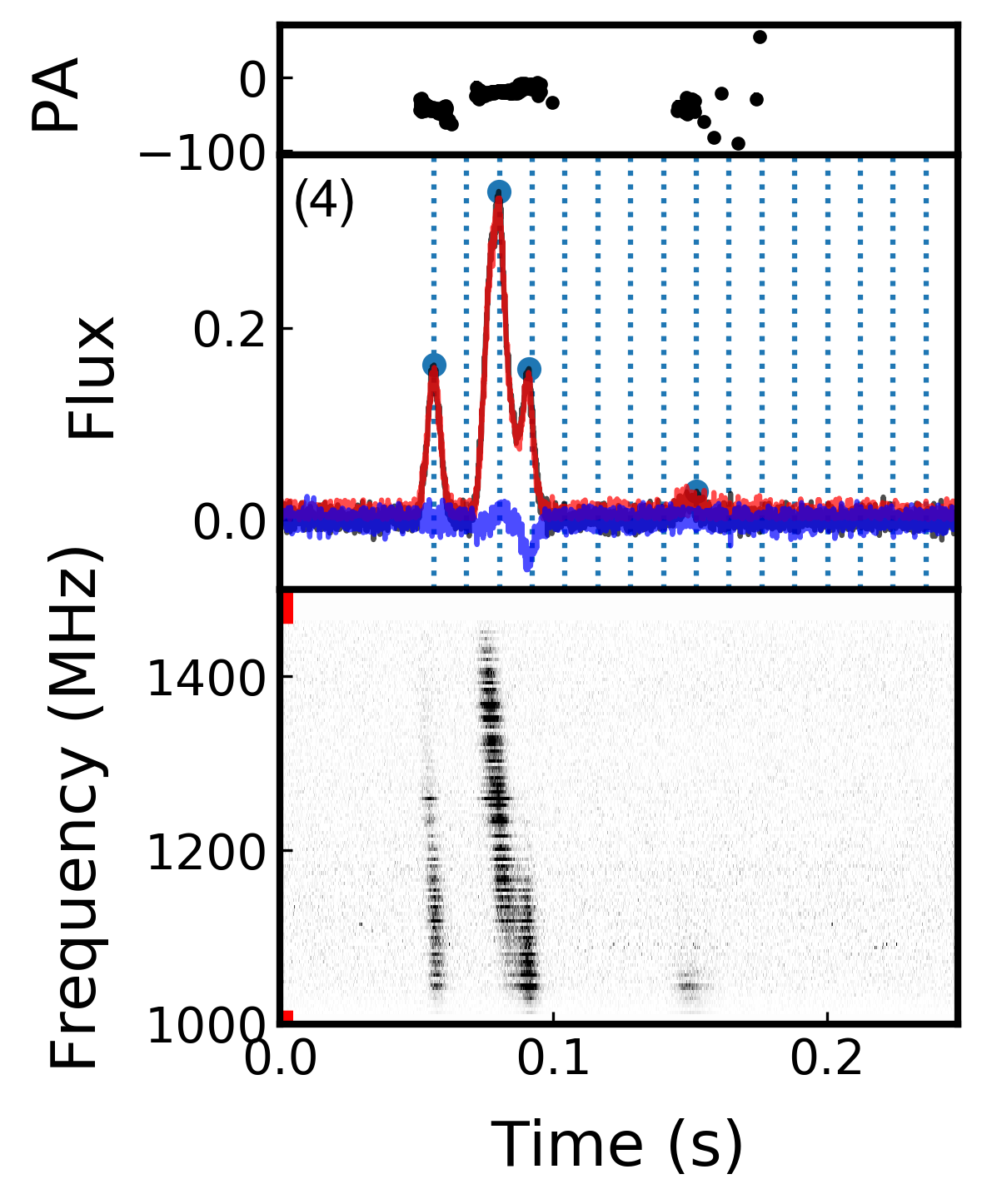}
     \end{subfigure}
     \hfill
     \begin{subfigure}[b]{0.3\textwidth}
         \centering
         \includegraphics[height=2.3in]{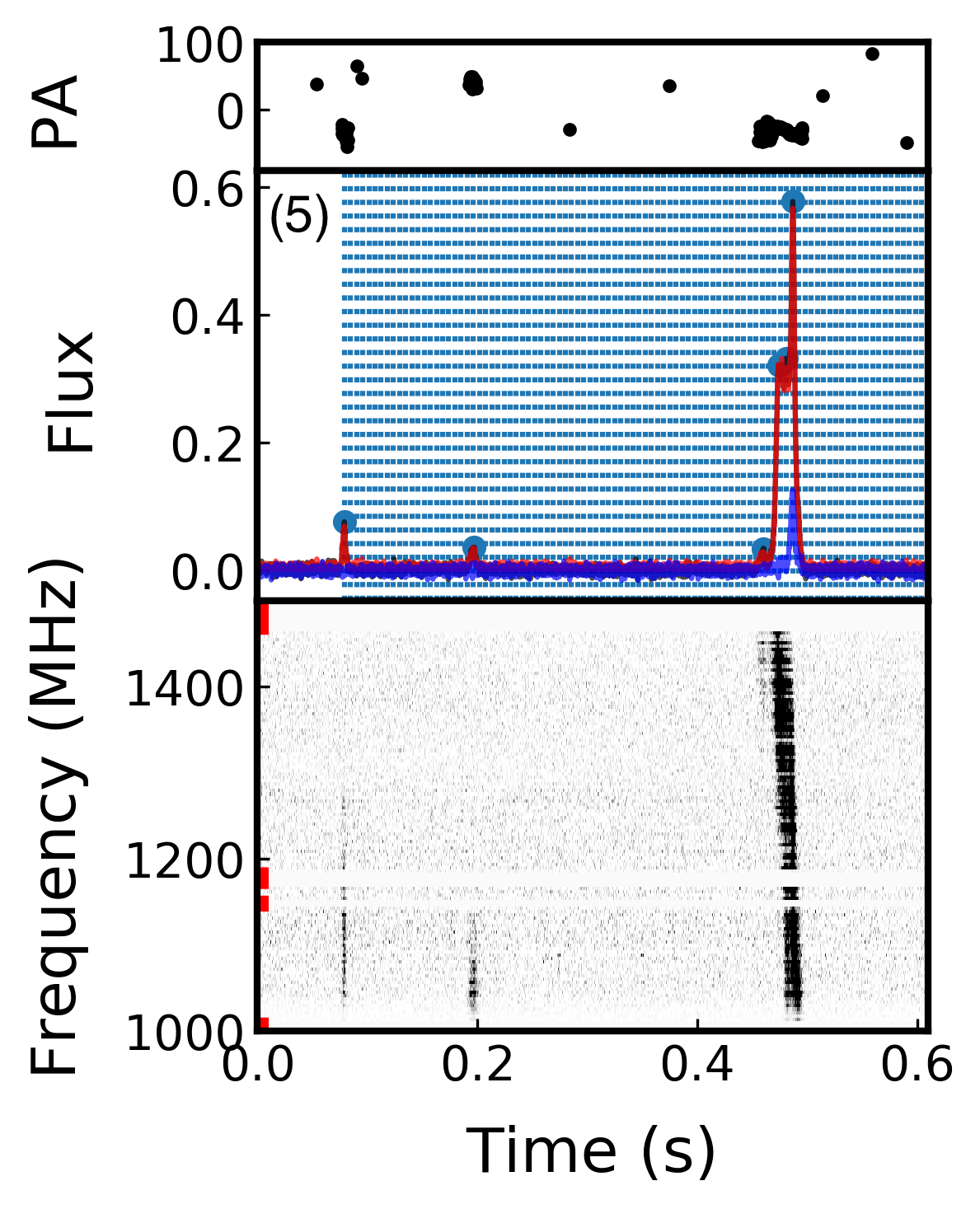}
     \end{subfigure}
     \hfill
     \begin{subfigure}[b]{0.3\textwidth}
         \centering
         \includegraphics[height=2.3in]{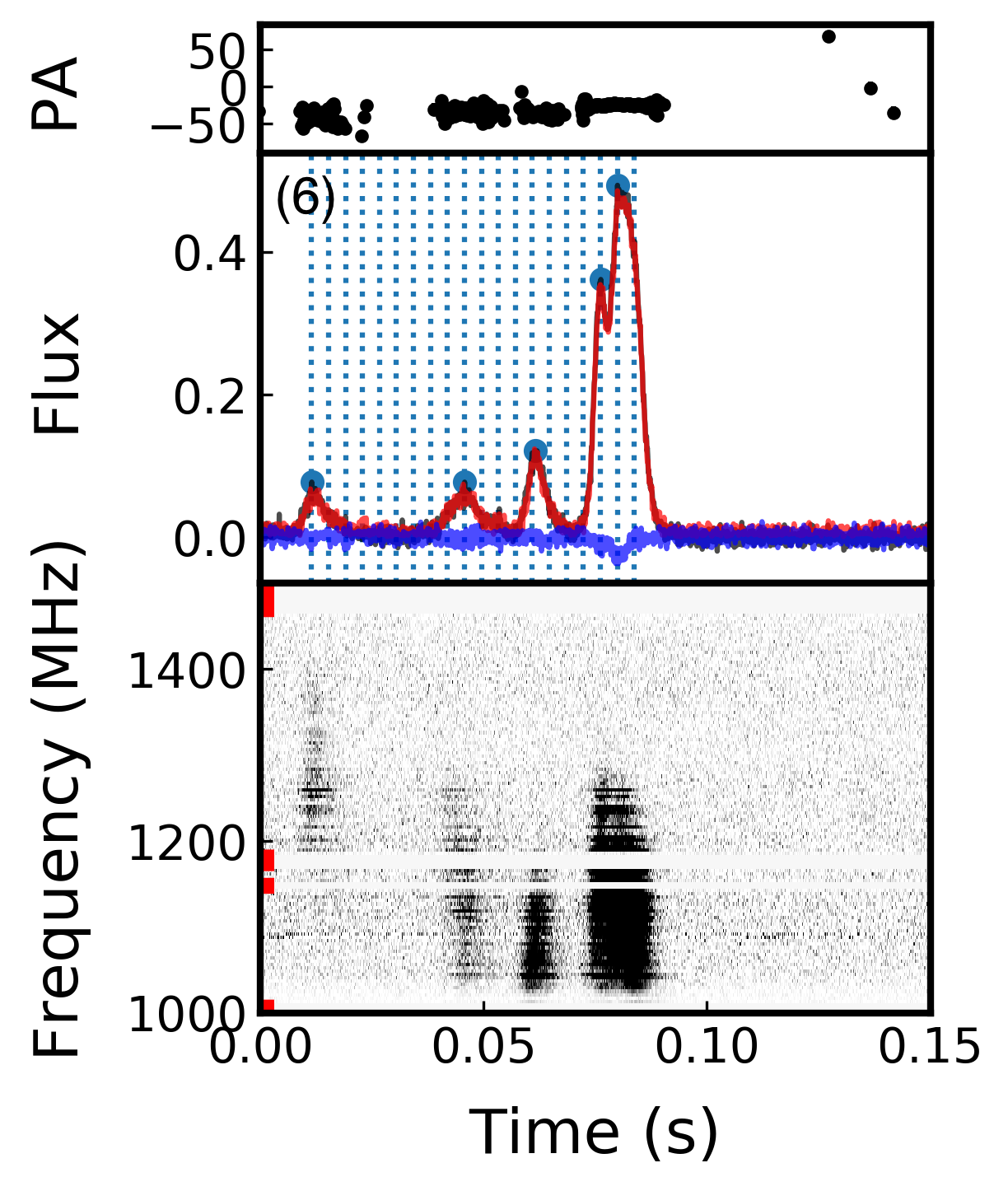}
     \end{subfigure}
     \hfill
     \begin{subfigure}[b]{0.3\textwidth}
         \centering
         \includegraphics[height=2.3in]{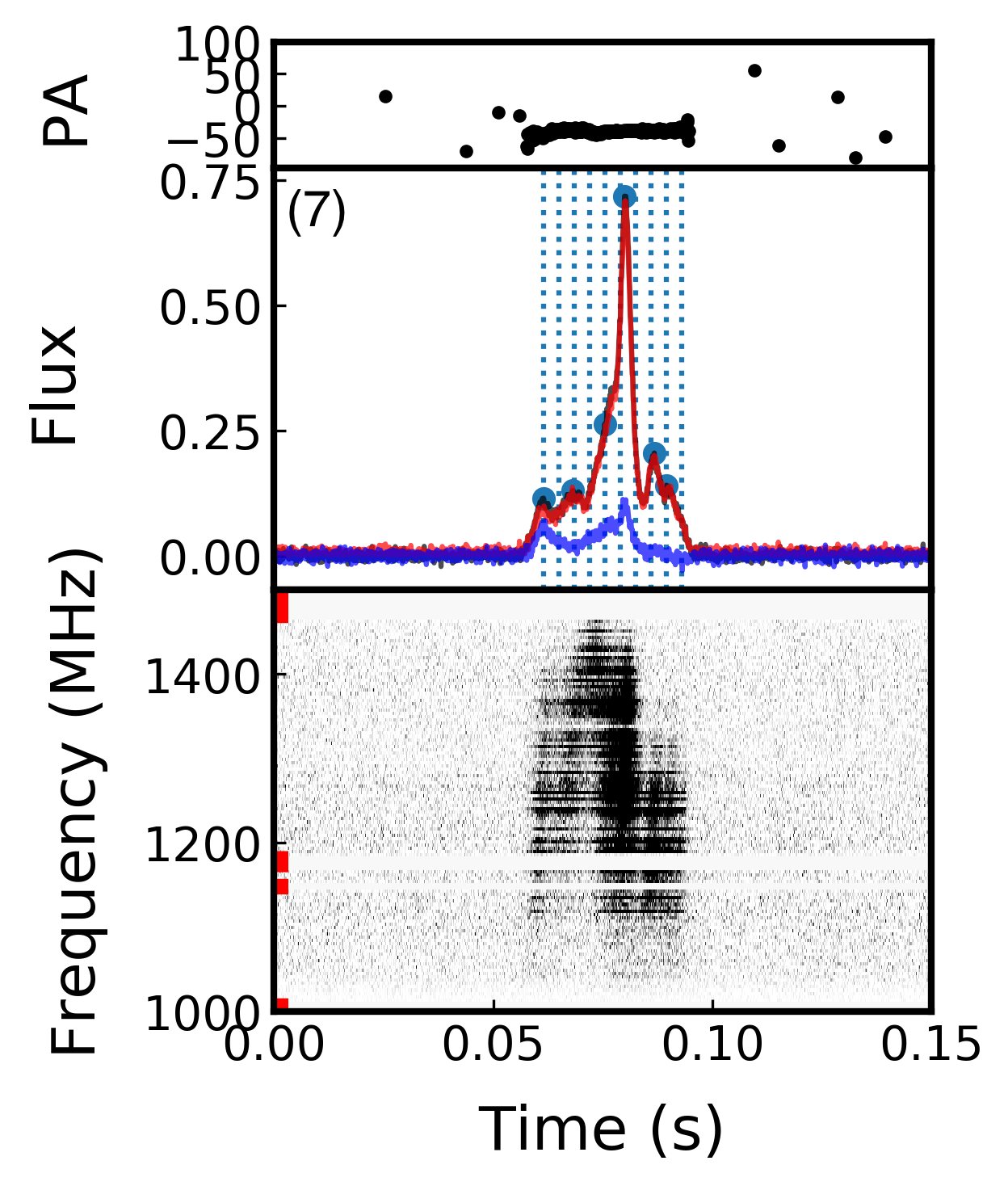}
     \end{subfigure}
     \hfill
     \begin{subfigure}[b]{0.3\textwidth}
         \centering
         \includegraphics[height=2.3in]{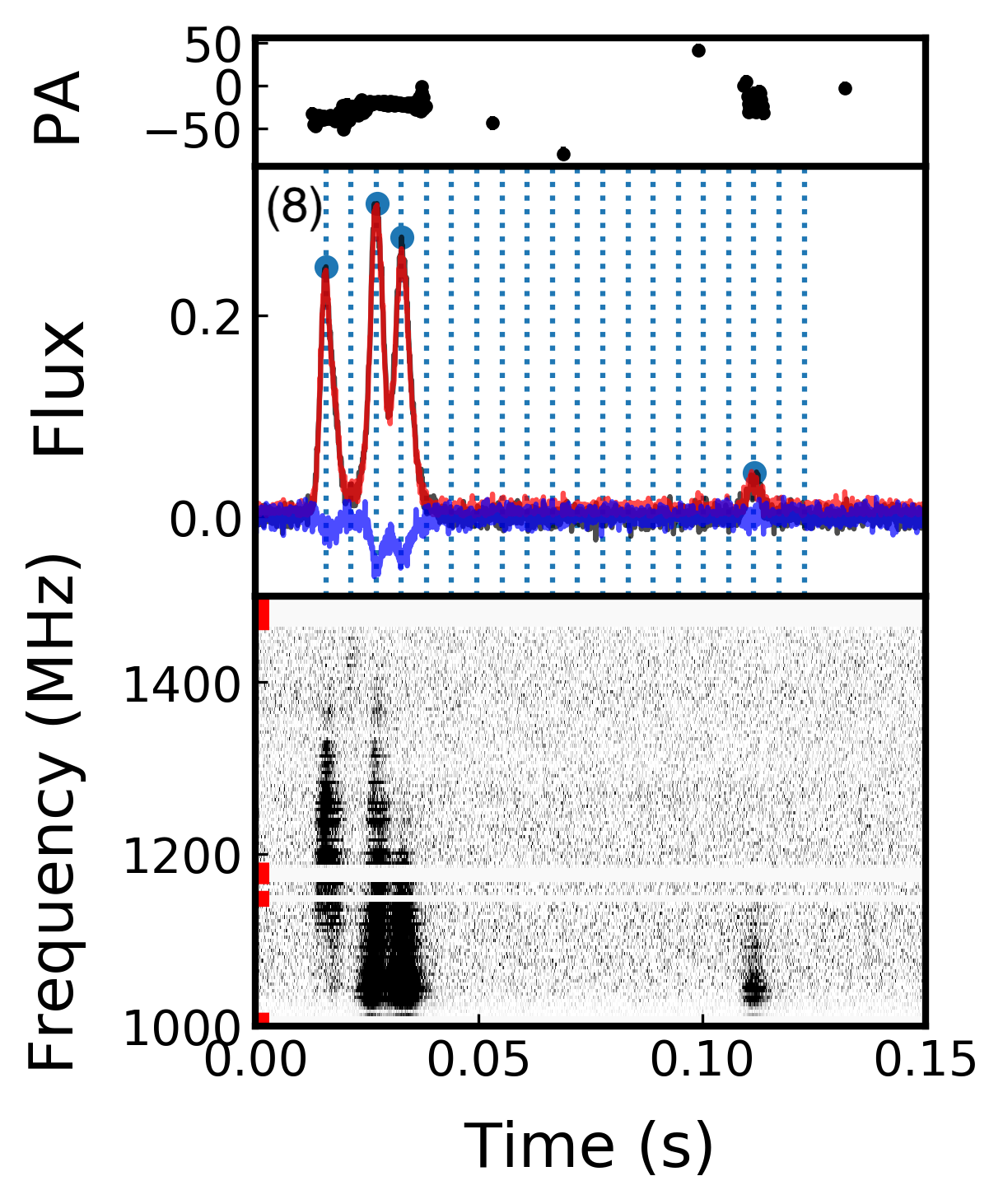}
     \end{subfigure}
     \hfill
     \begin{subfigure}[b]{0.3\textwidth}
         \centering
         \includegraphics[height=2.3in]{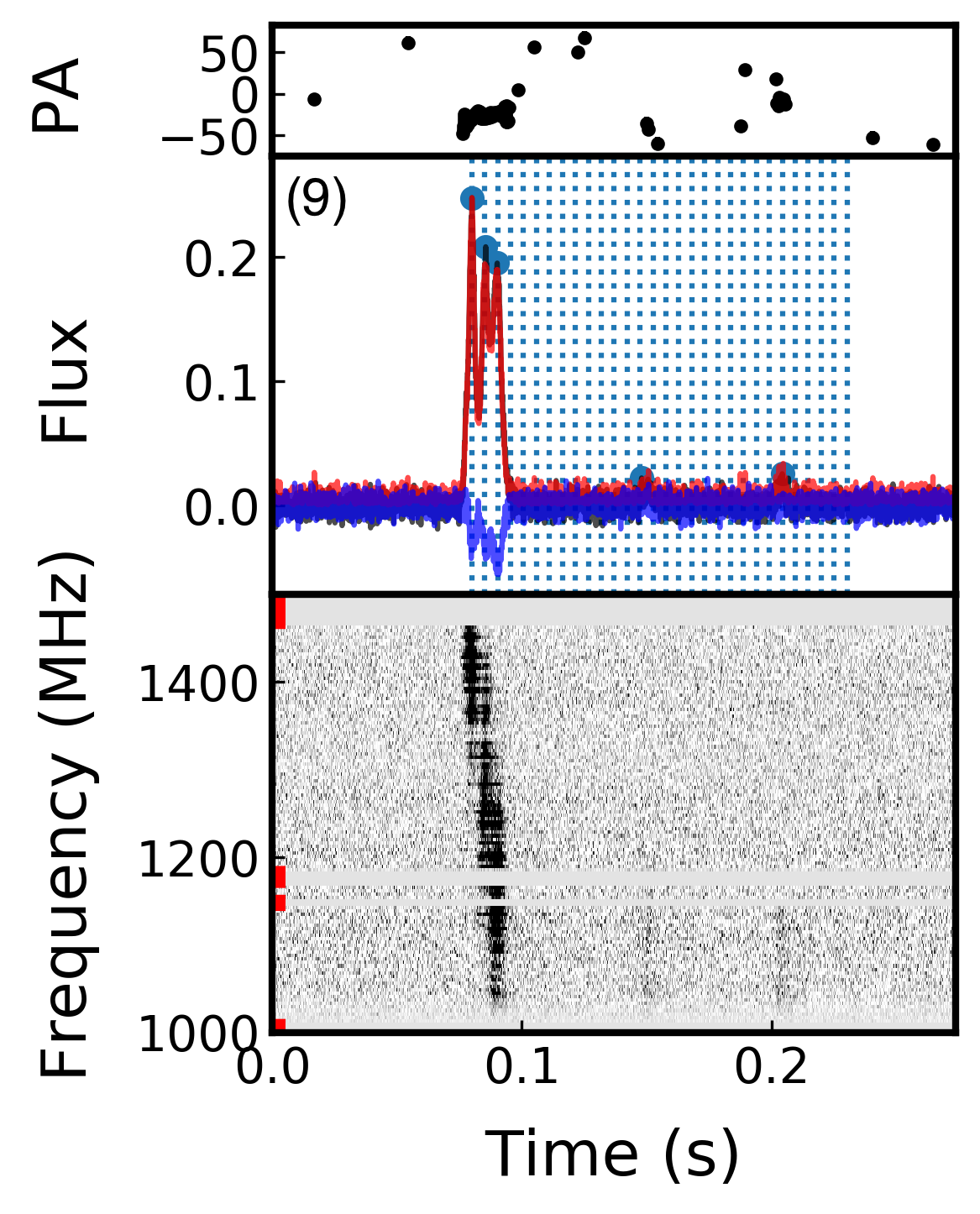}
     \end{subfigure}
     \begin{subfigure}[b]{0.3\textwidth}
         \centering
         \includegraphics[height=2.3in]{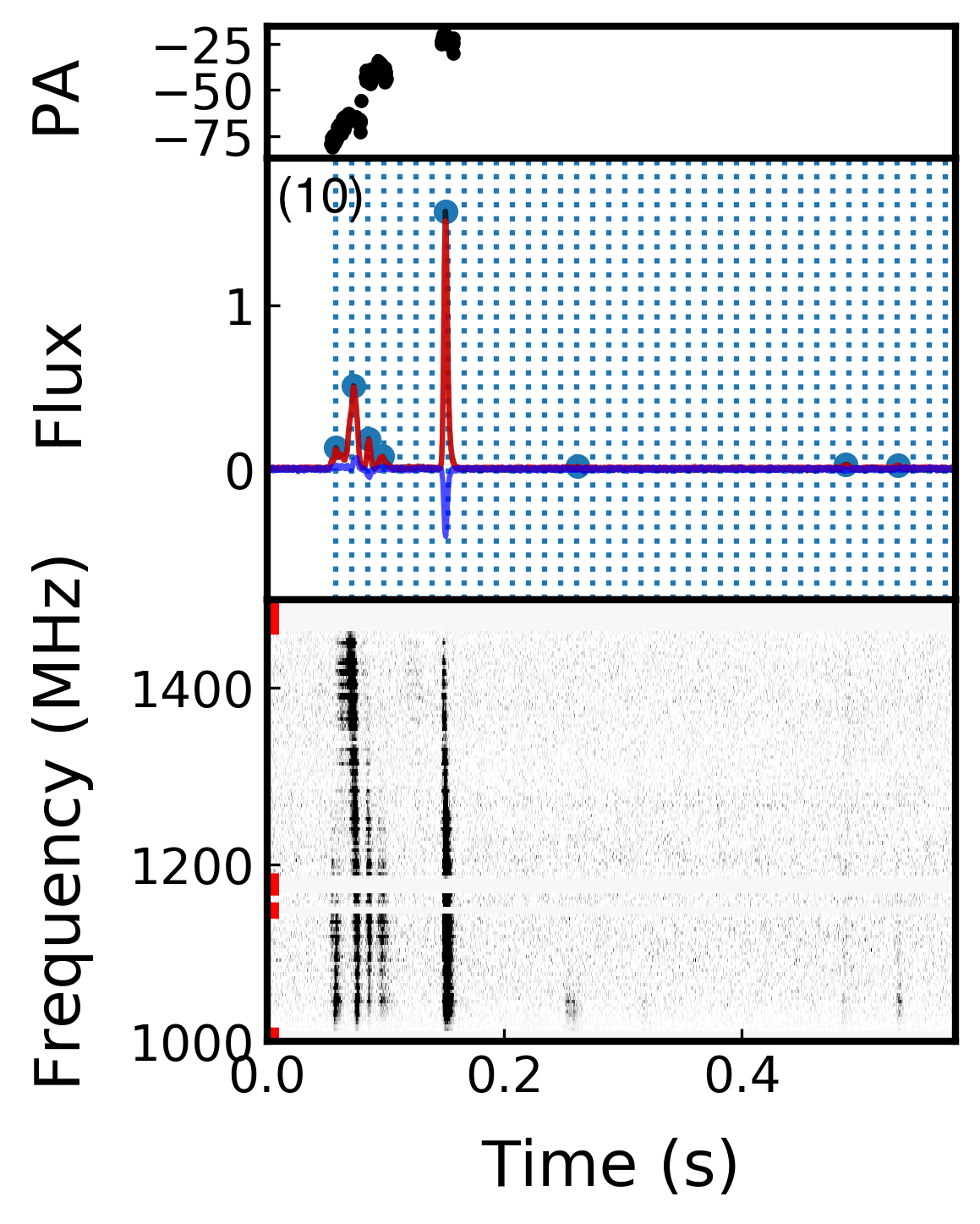}
     \end{subfigure}
     \hfill
     \begin{subfigure}[b]{0.3\textwidth}
         \centering
         \includegraphics[height=2.3in]{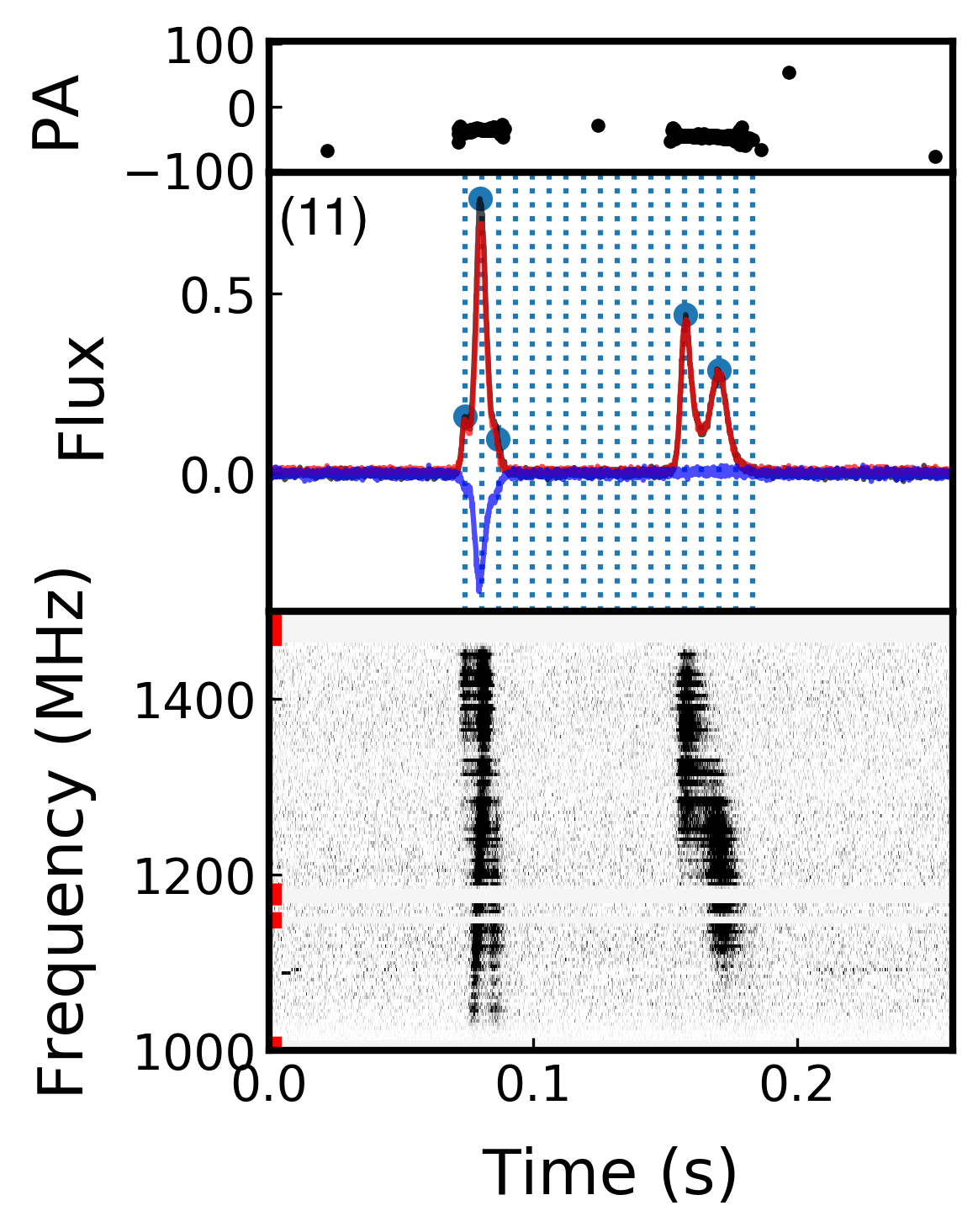}
     \end{subfigure}
     \hfill
     \begin{subfigure}[b]{0.3\textwidth}
         \centering
         \includegraphics[height=2.3in]{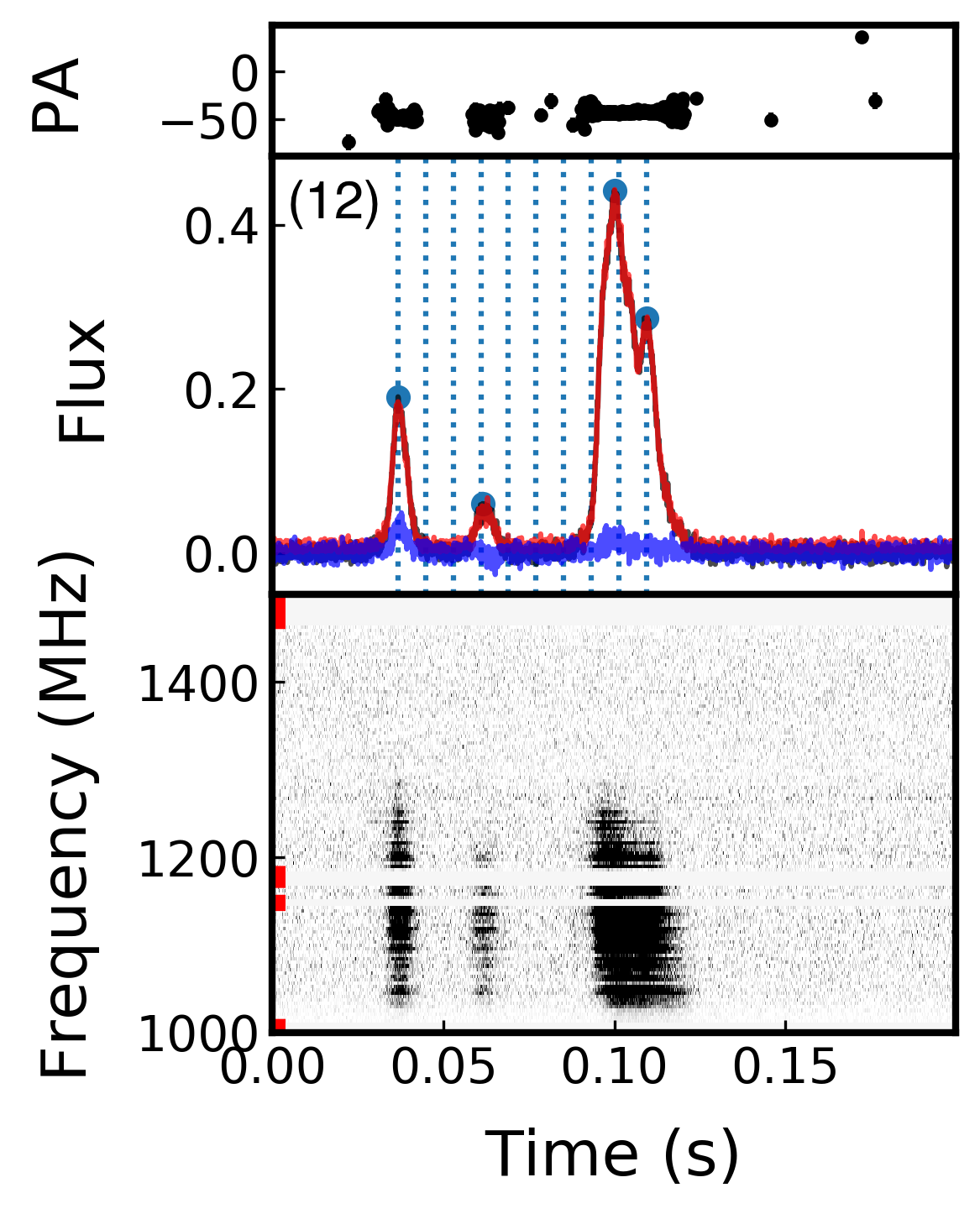}
     \end{subfigure}
  \end{figure*}

 \newpage
 \begin{figure*}[t!]
 \ContinuedFloat
     \begin{subfigure}[b]{0.3\textwidth}
         \centering
         \includegraphics[height=2.3in]{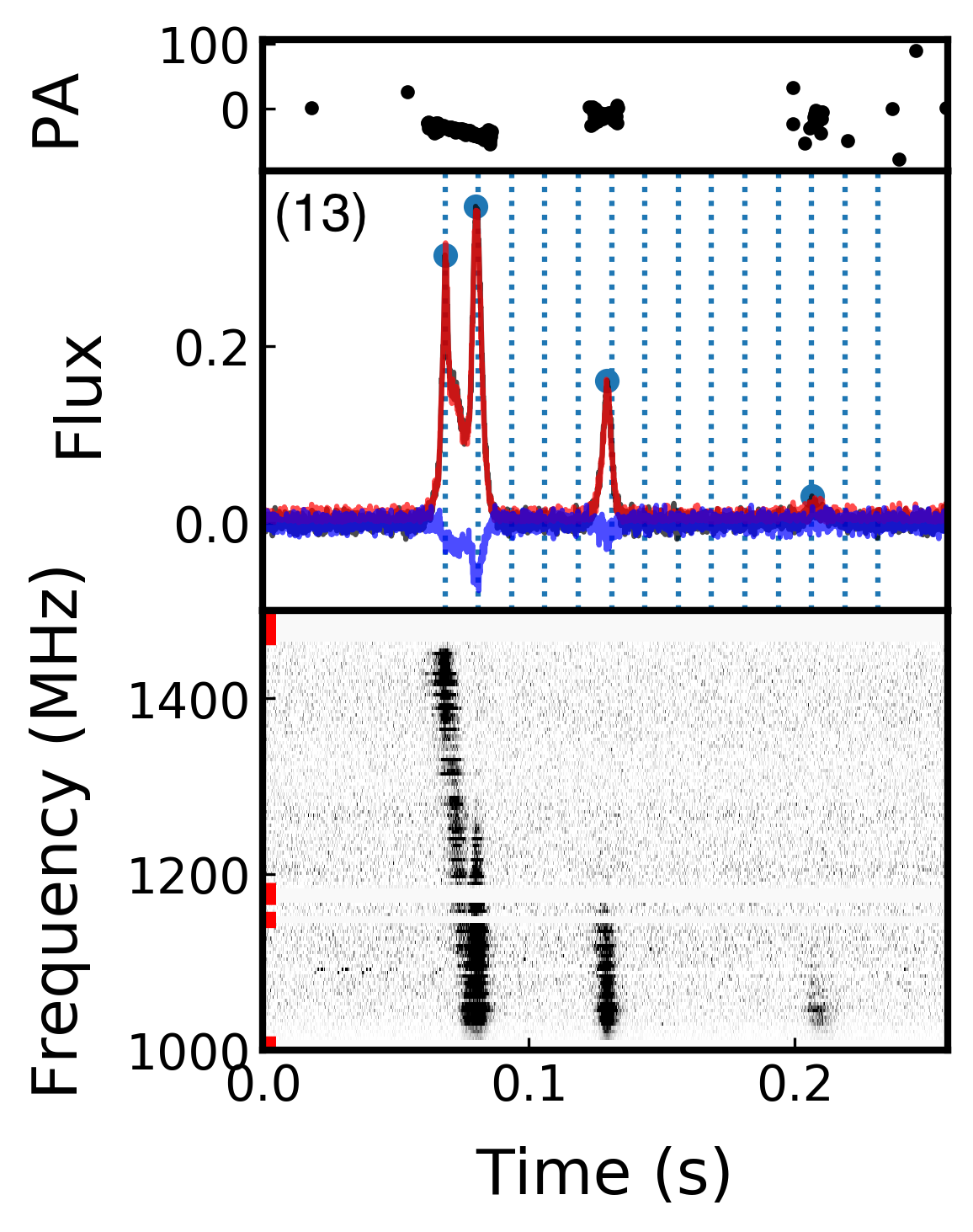}
     \end{subfigure}
     \hfill
     \begin{subfigure}[b]{0.3\textwidth}
         \centering
         \includegraphics[height=2.3in]{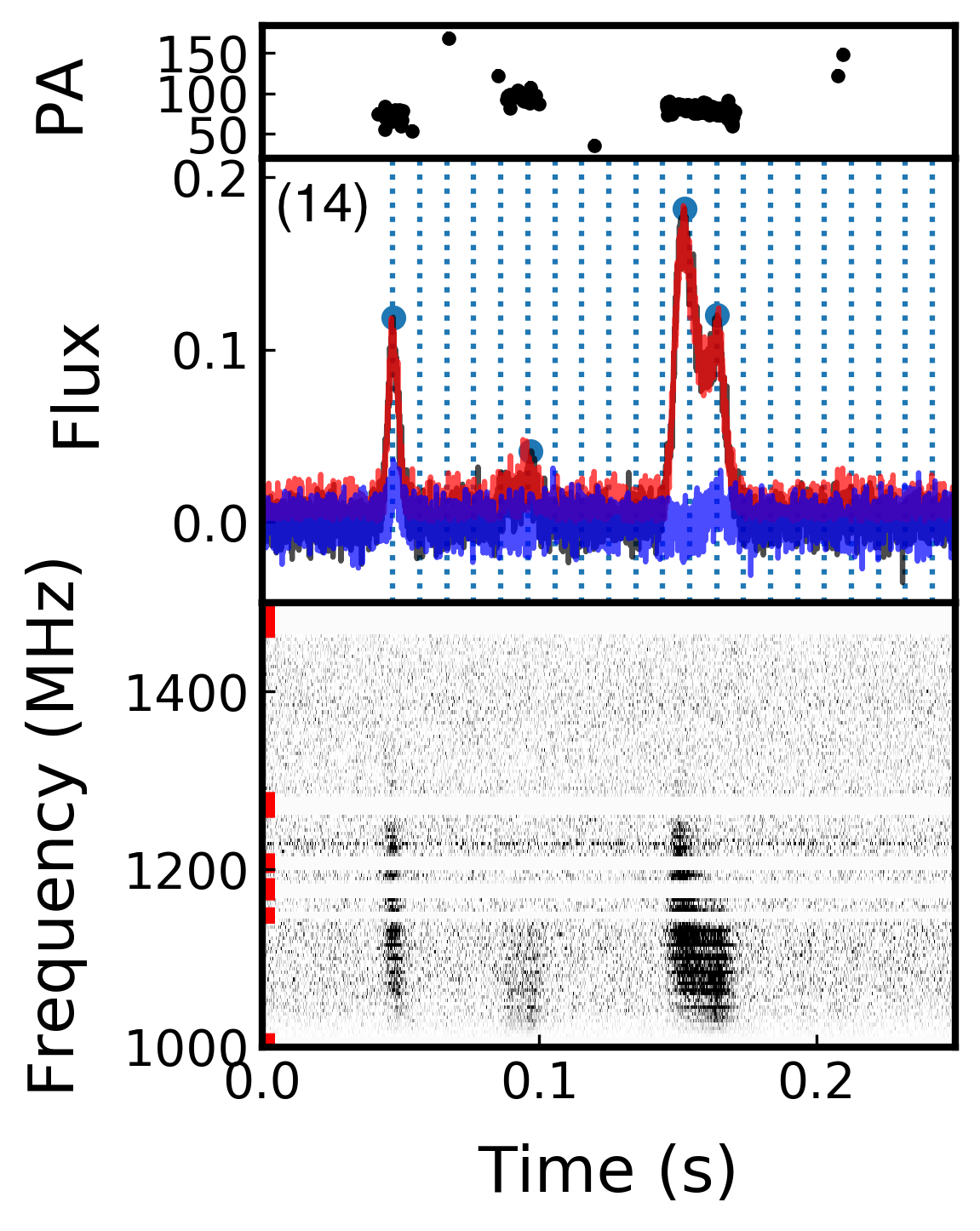}
     \end{subfigure}
     \hfill
     \begin{subfigure}[b]{0.3\textwidth}
         \centering
         \includegraphics[height=2.3in]{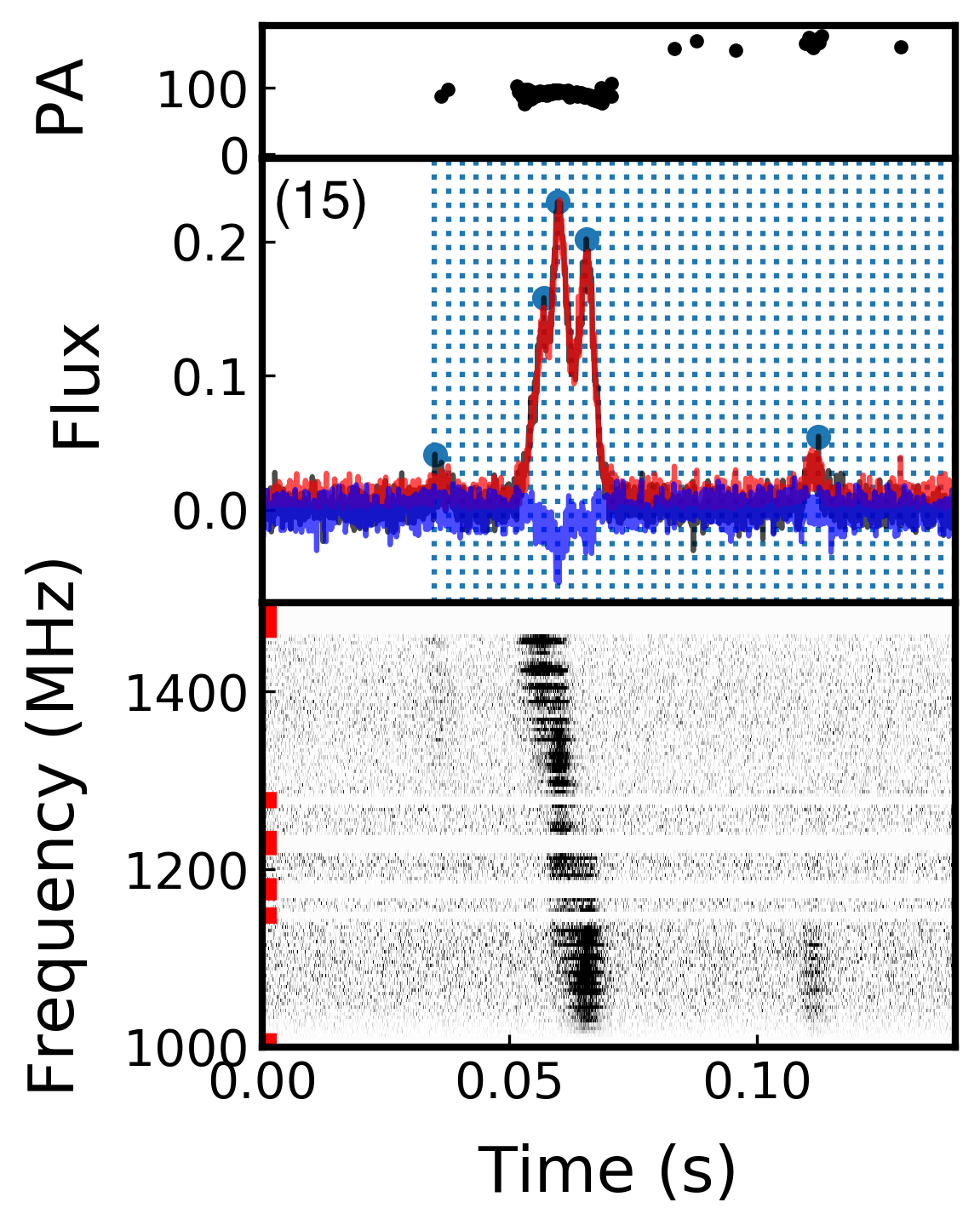}
     \end{subfigure}
     \hfill
     \begin{subfigure}[b]{0.3\textwidth}
         \centering
         \includegraphics[height=2.3in]{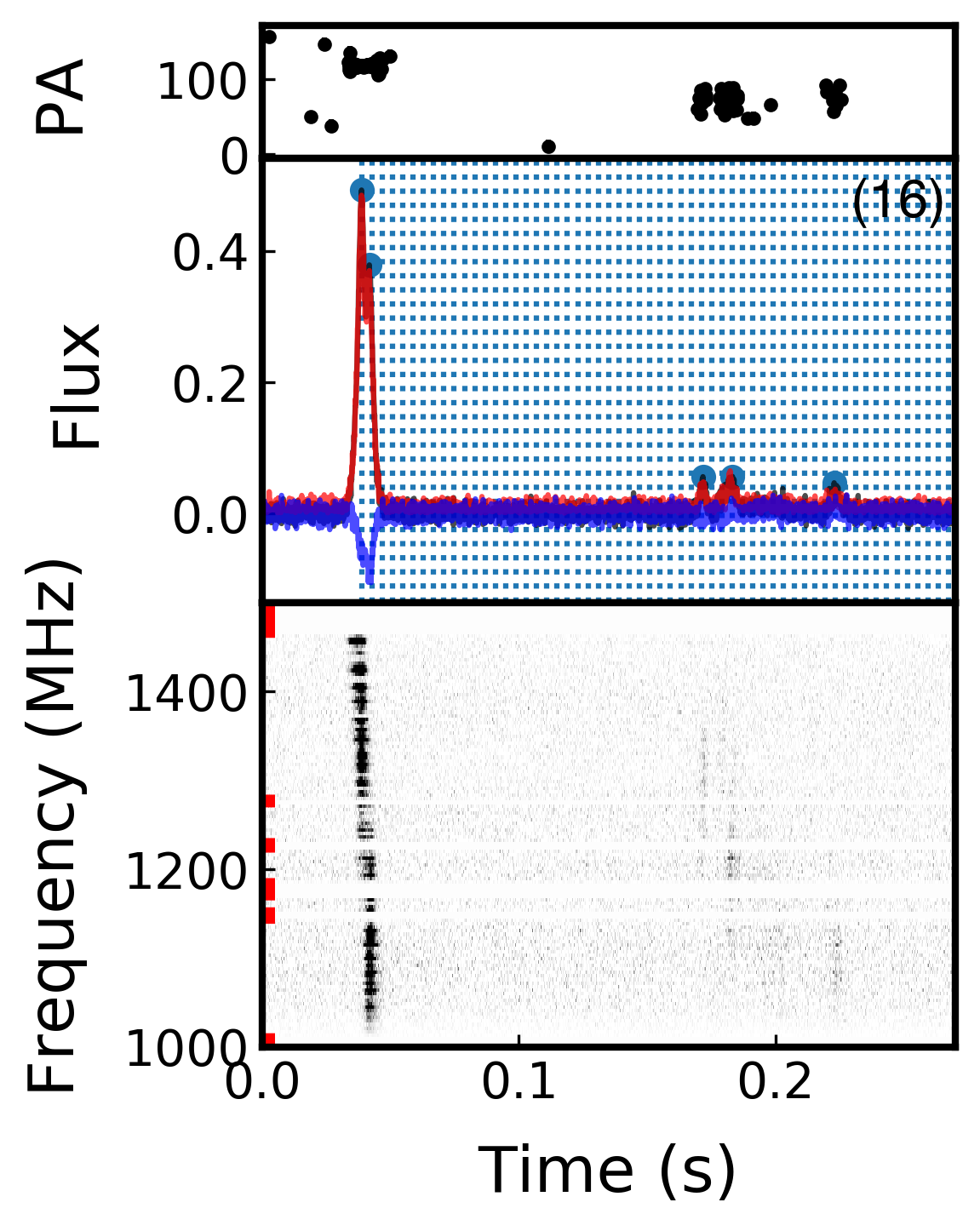}
     \end{subfigure}
     \hfill
     \begin{subfigure}[b]{0.3\textwidth}
         \centering
         \includegraphics[height=2.3in]{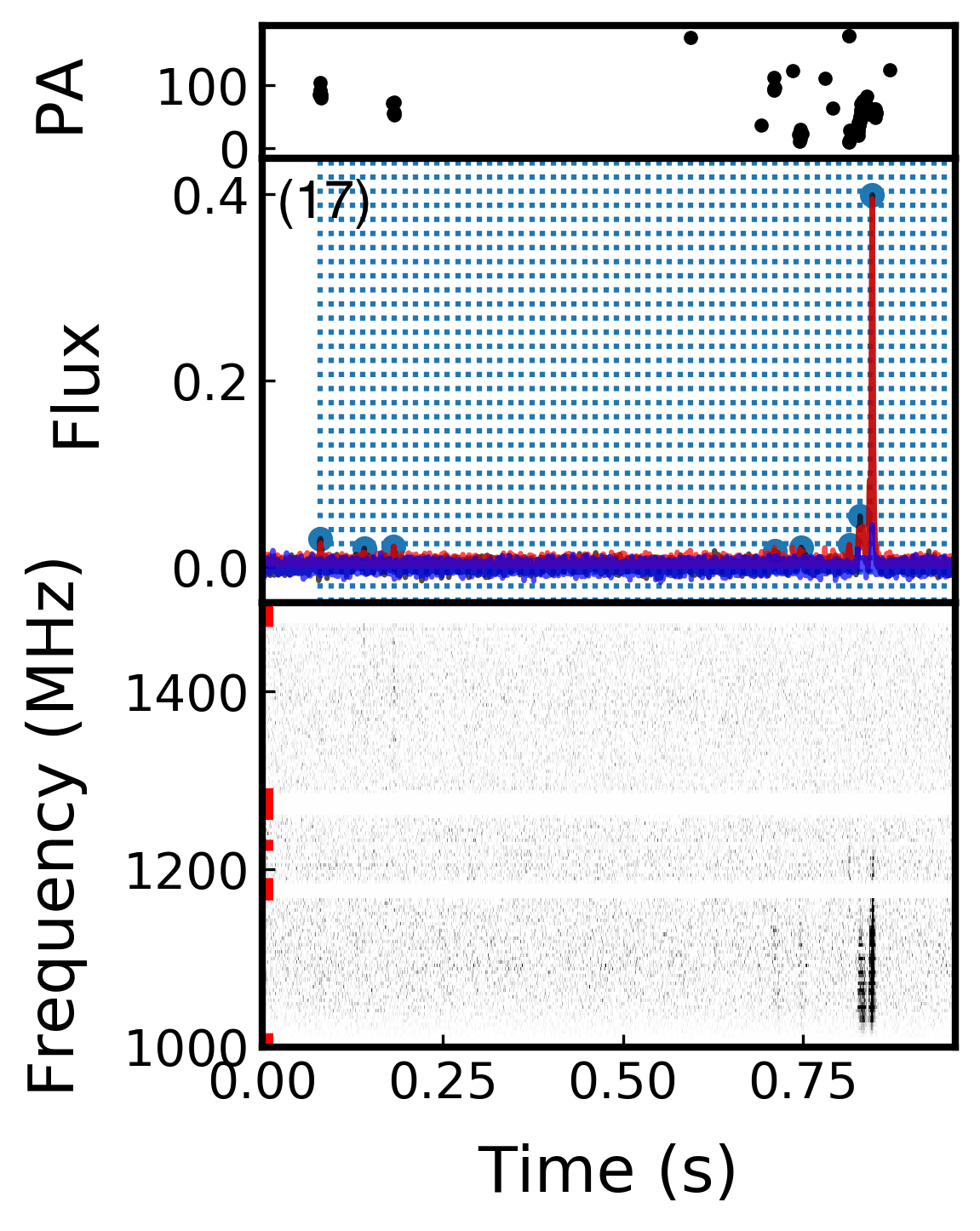}
     \end{subfigure}
     \hfill
     \begin{subfigure}[b]{0.3\textwidth}
         \centering
         \includegraphics[height=2.3in]{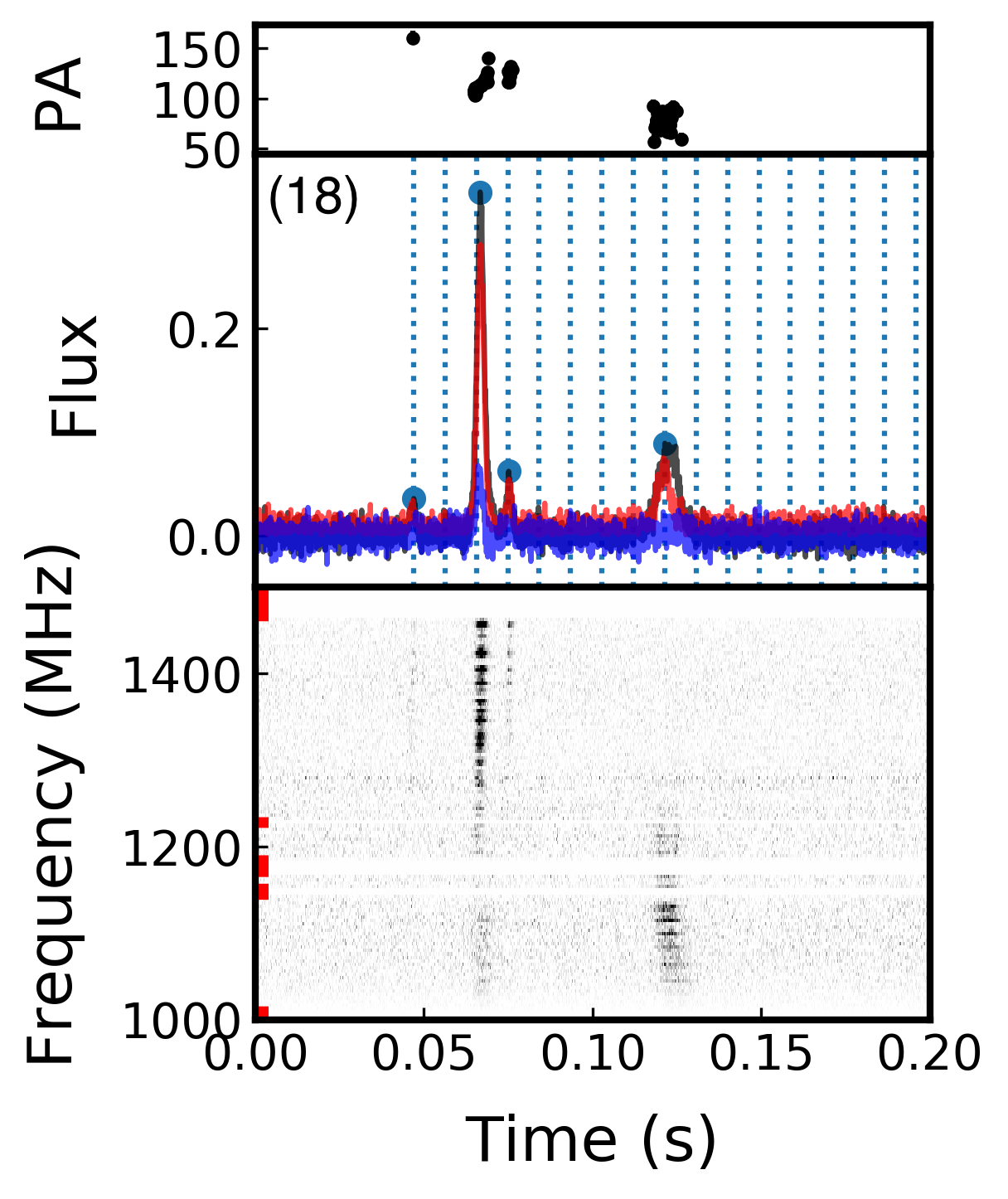}
     \end{subfigure}
     \hfill
     \begin{subfigure}[b]{0.3\textwidth}
         \centering
         \includegraphics[height=2.3in]{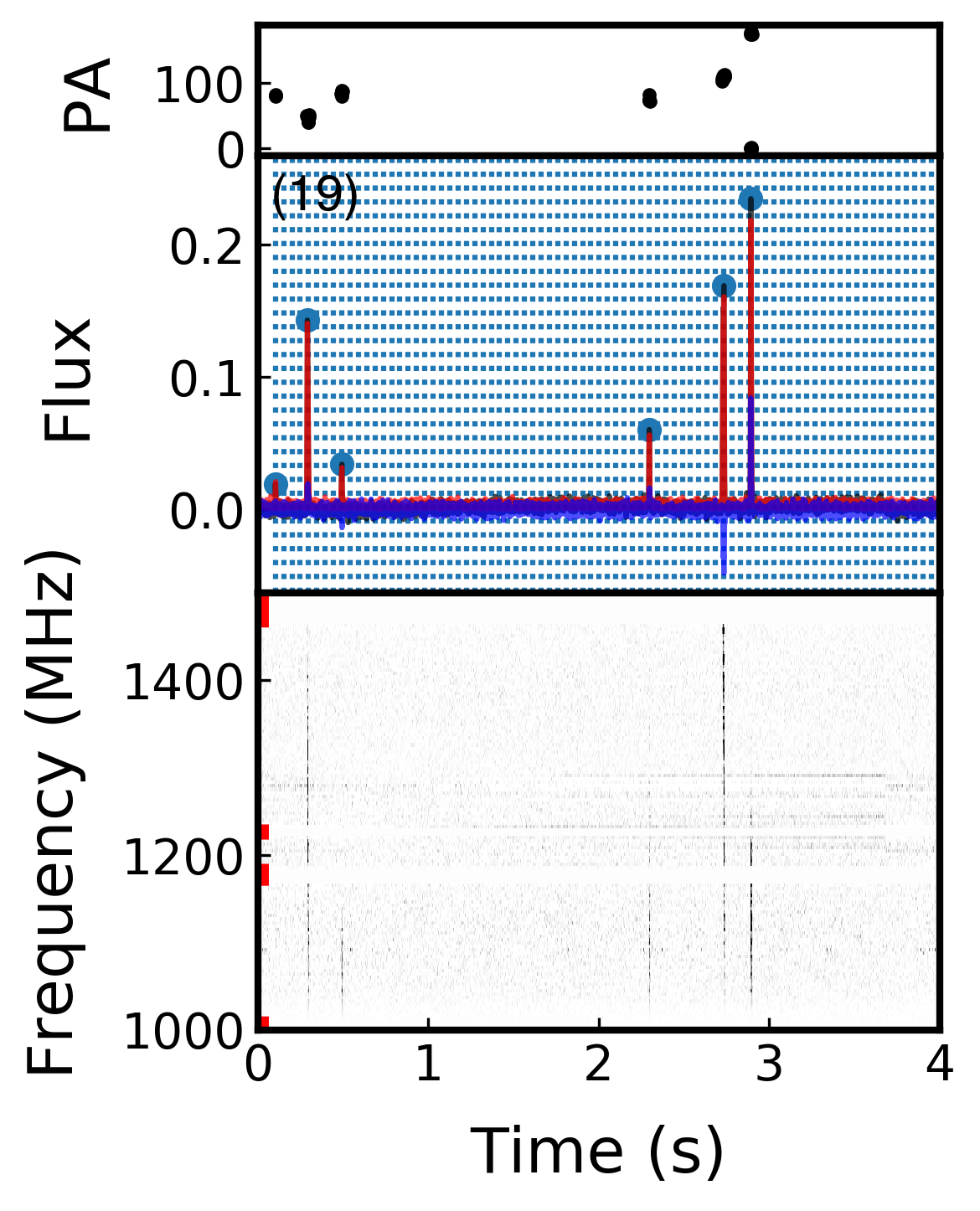}
     \end{subfigure}
     \hfill
     \begin{subfigure}[b]{0.3\textwidth}
         \centering
         \includegraphics[height=2.3in]{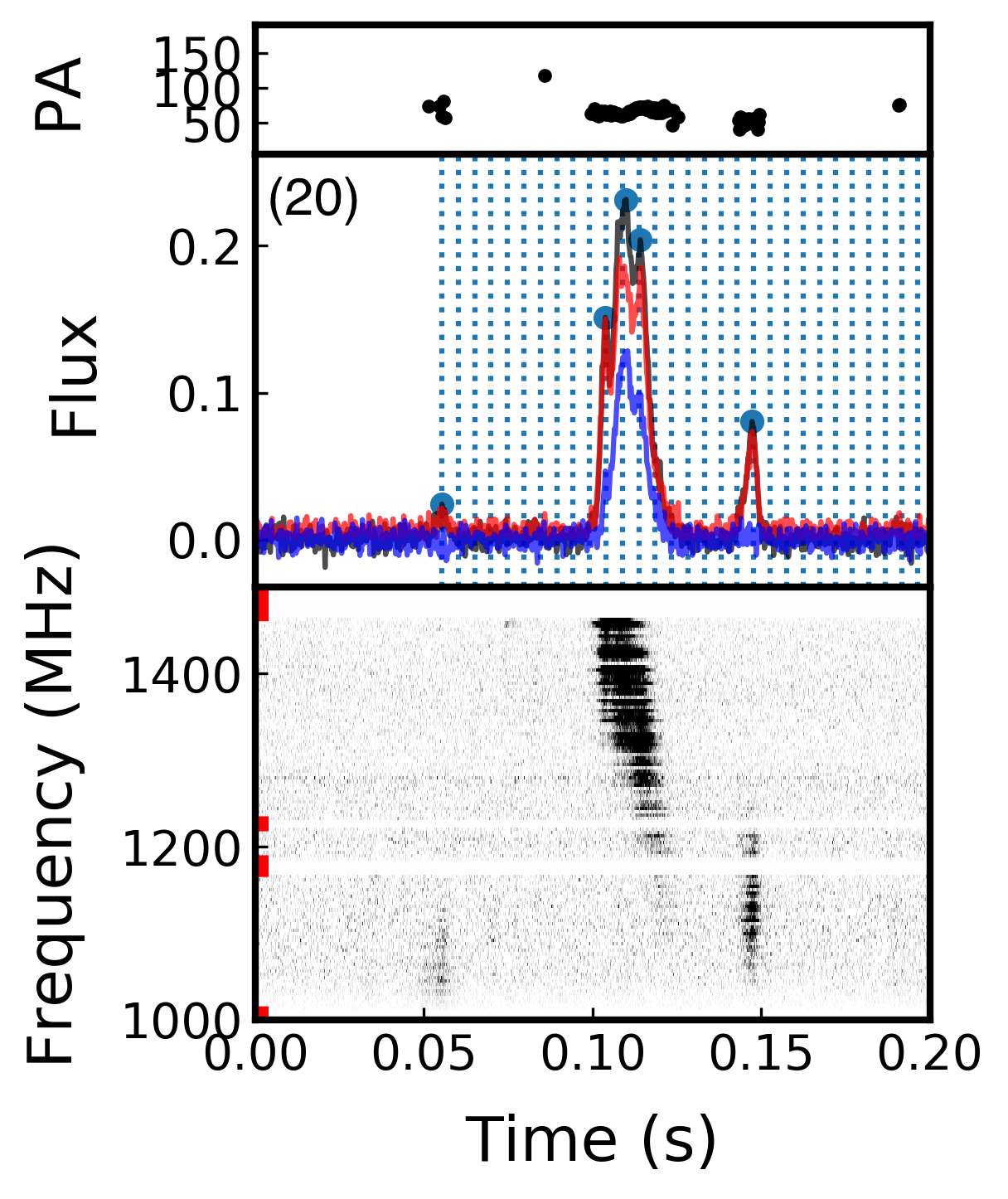}
     \end{subfigure}
     \hfill
     \begin{subfigure}[b]{0.3\textwidth}
         \centering
         \includegraphics[height=2.3in]{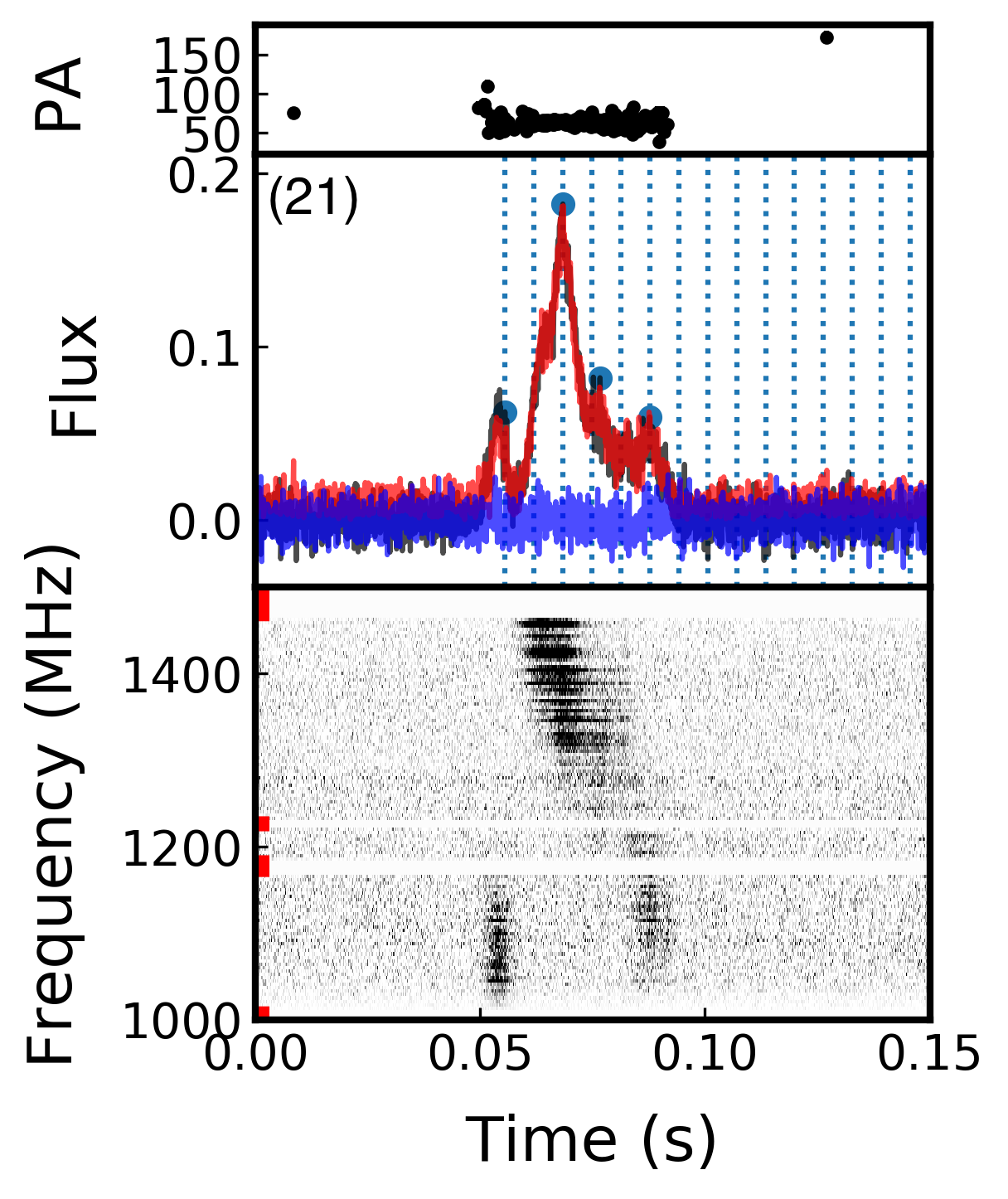}
     \end{subfigure}
     \hfill
     \begin{subfigure}[b]{0.3\textwidth}
         \centering
         \includegraphics[height=2.3in]{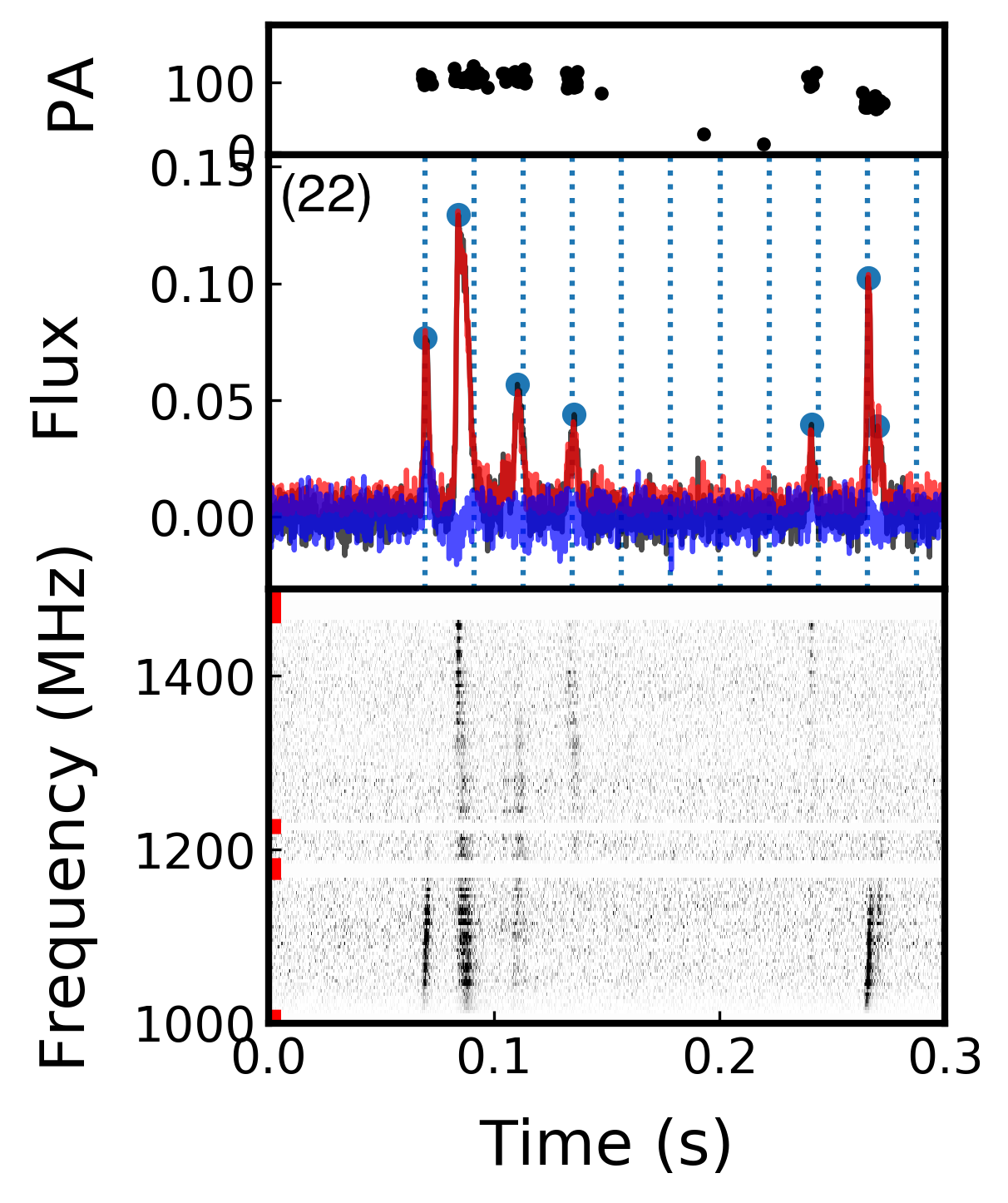}
     \end{subfigure}
     \hfill
     \begin{subfigure}[b]{0.3\textwidth}
         \centering
         \includegraphics[height=2.3in]{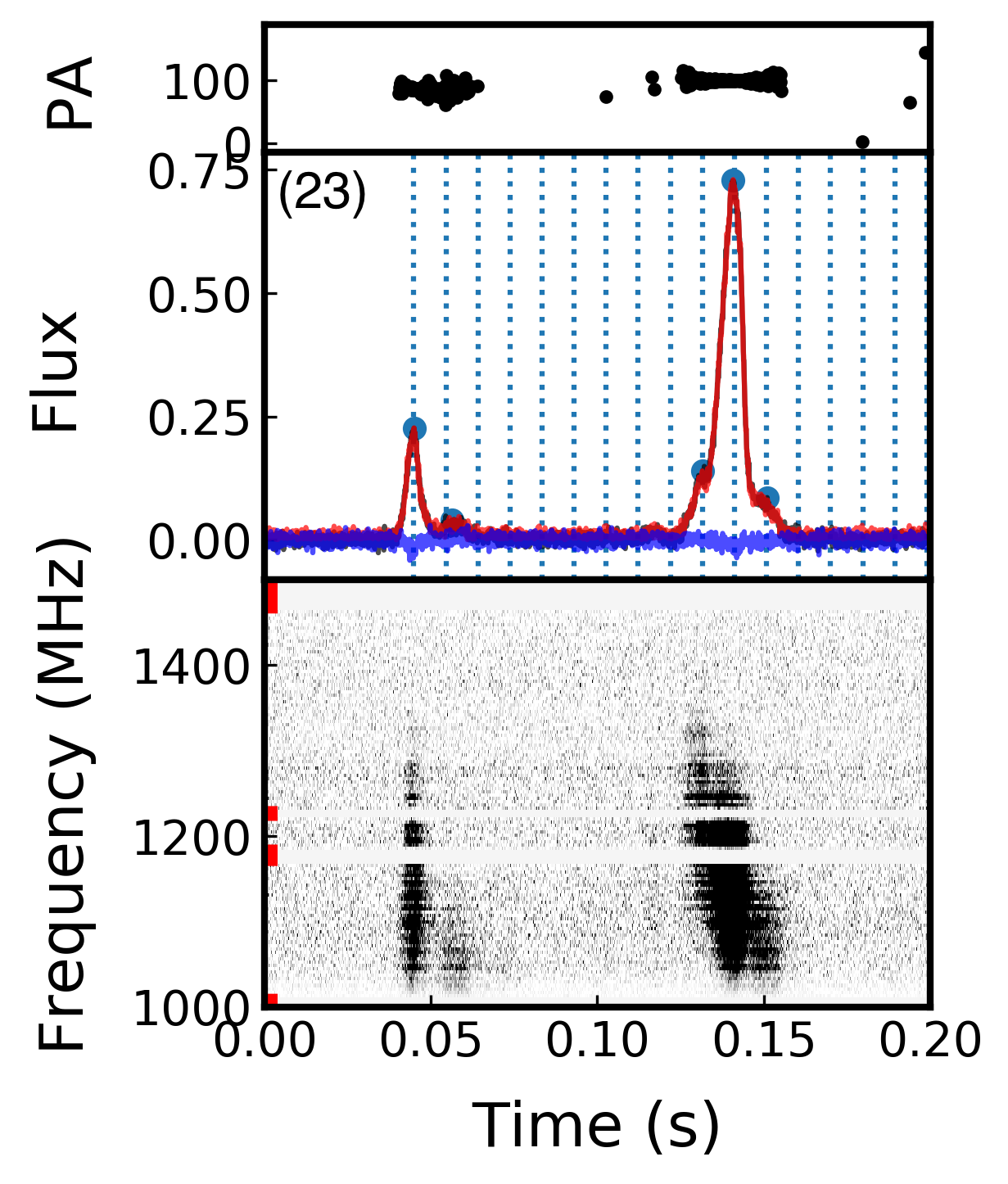}
     \end{subfigure}
     \hfill
     \begin{subfigure}[b]{0.3\textwidth}
         \centering
         \includegraphics[height=2.3in]{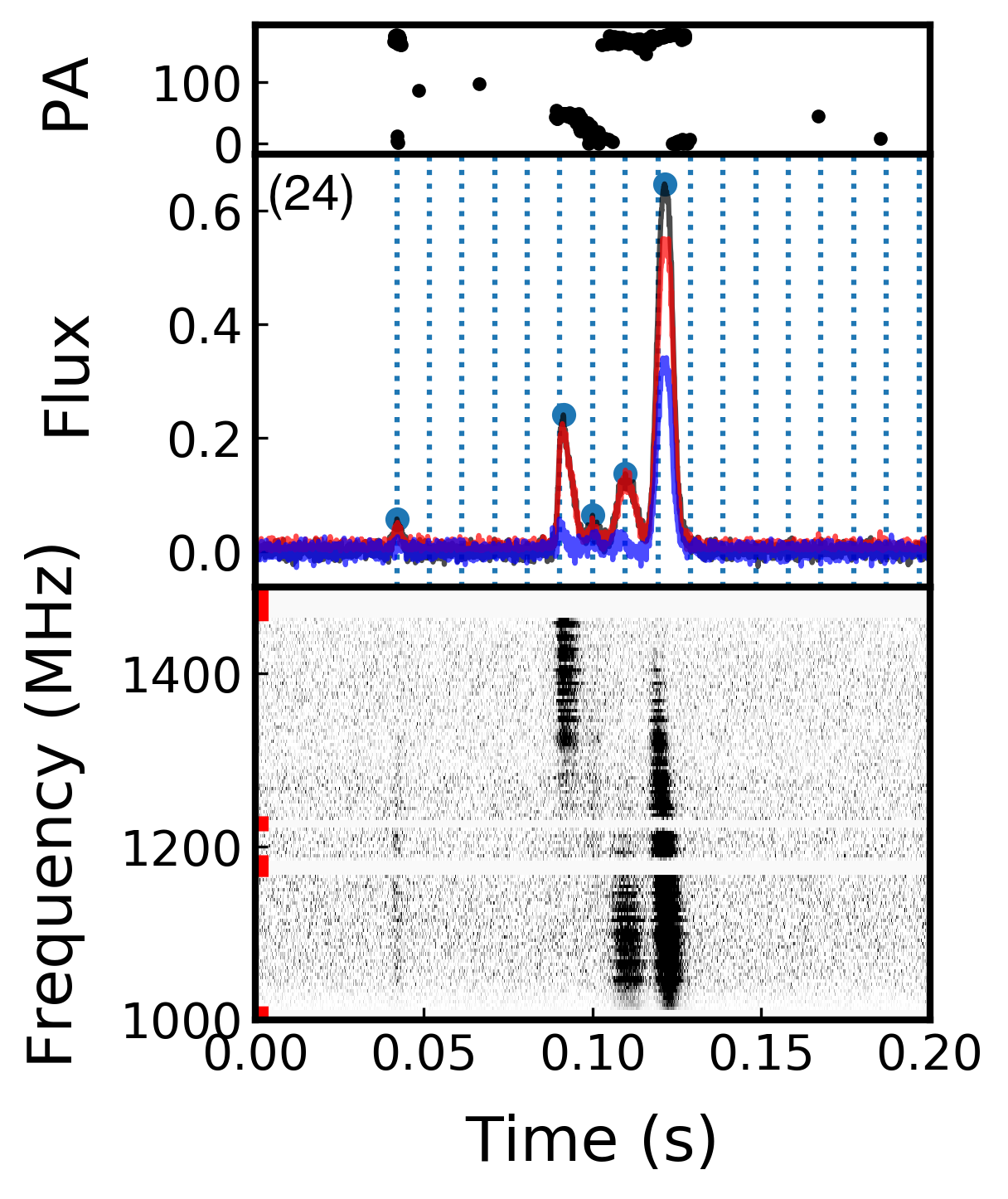}
     \end{subfigure}
 \end{figure*}

 \newpage
\begin{figure*}[t!]
 \ContinuedFloat
     \begin{subfigure}[b]{0.3\textwidth}
         \centering
         \includegraphics[height=2.3in]{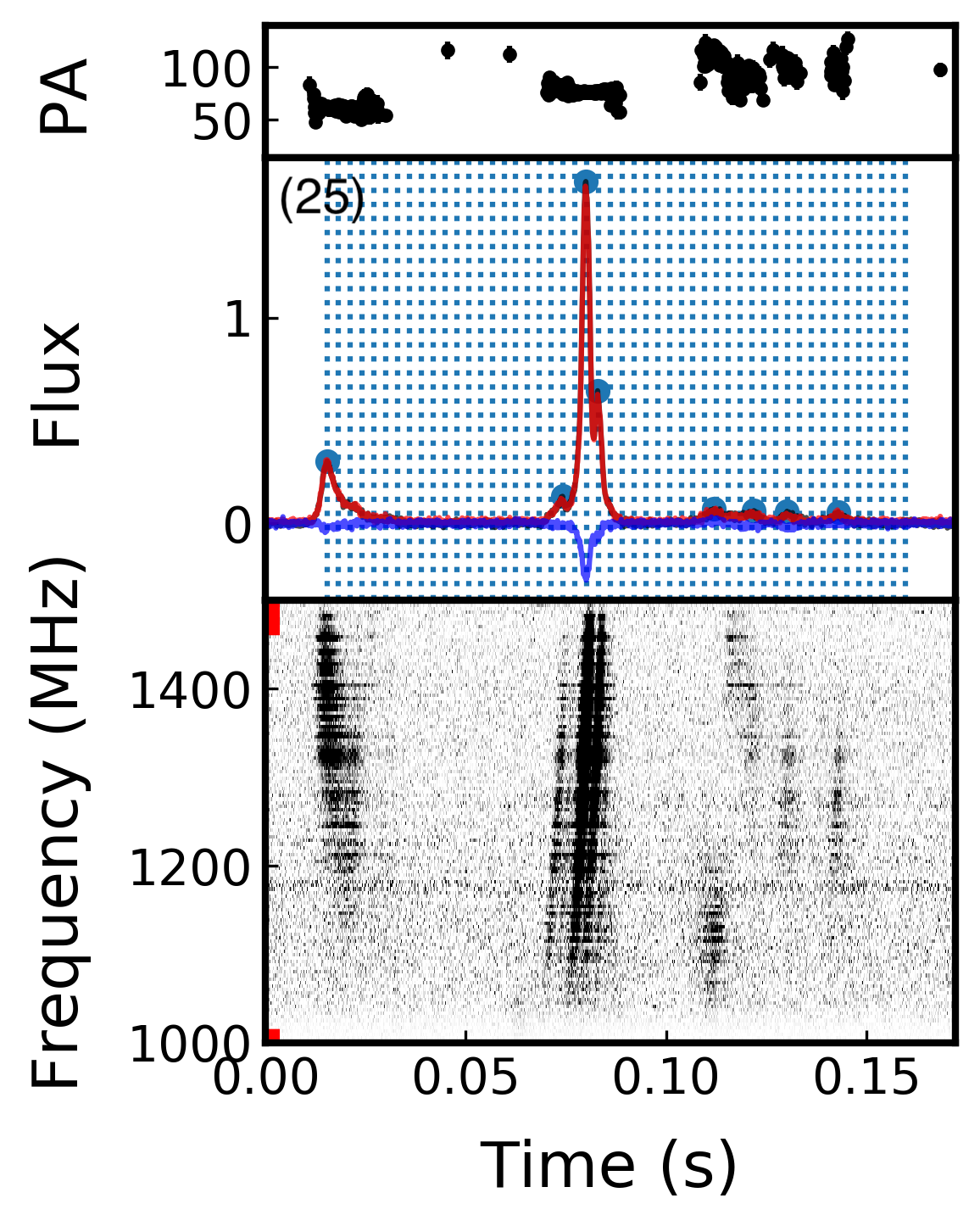}
     \end{subfigure}
     \hfill
     \begin{subfigure}[b]{0.3\textwidth}
         \centering
         \includegraphics[height=2.3in]{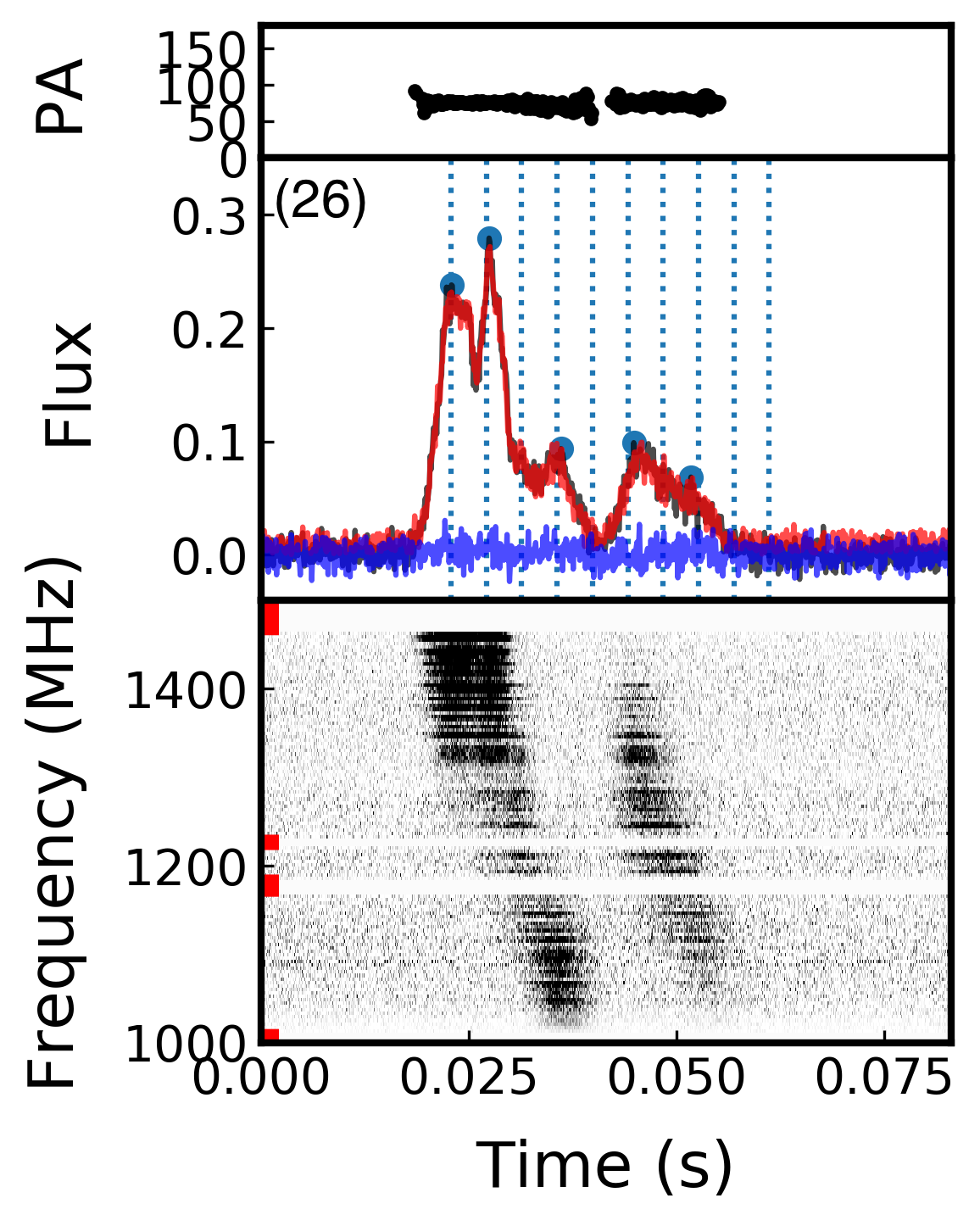}
     \end{subfigure}
     \hfill
     \begin{subfigure}[b]{0.3\textwidth}
         \centering
         \includegraphics[height=2.3in]{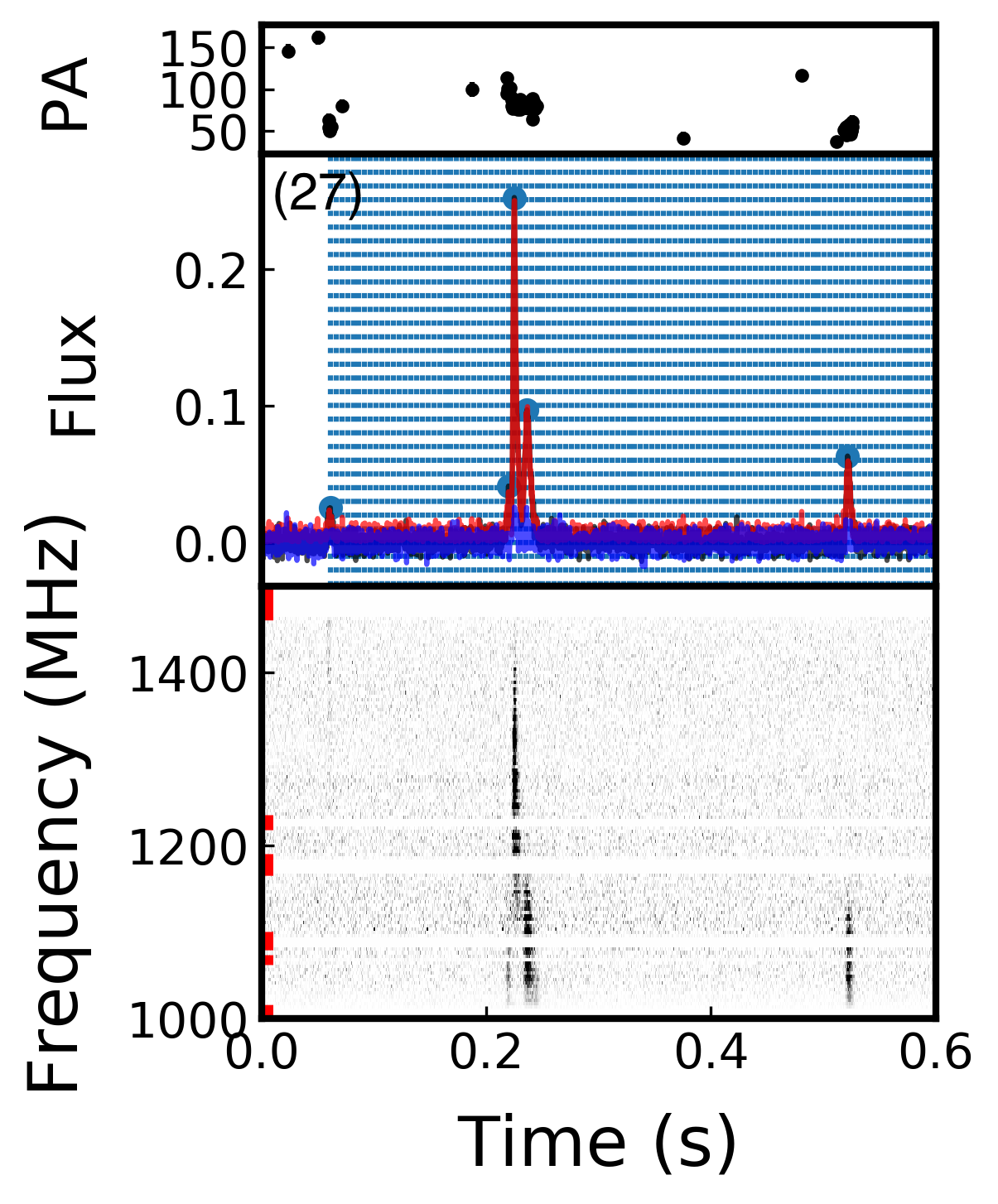}
     \end{subfigure}
     \hfill
     \begin{subfigure}[b]{0.3\textwidth}
         \centering
         \includegraphics[height=2.3in]{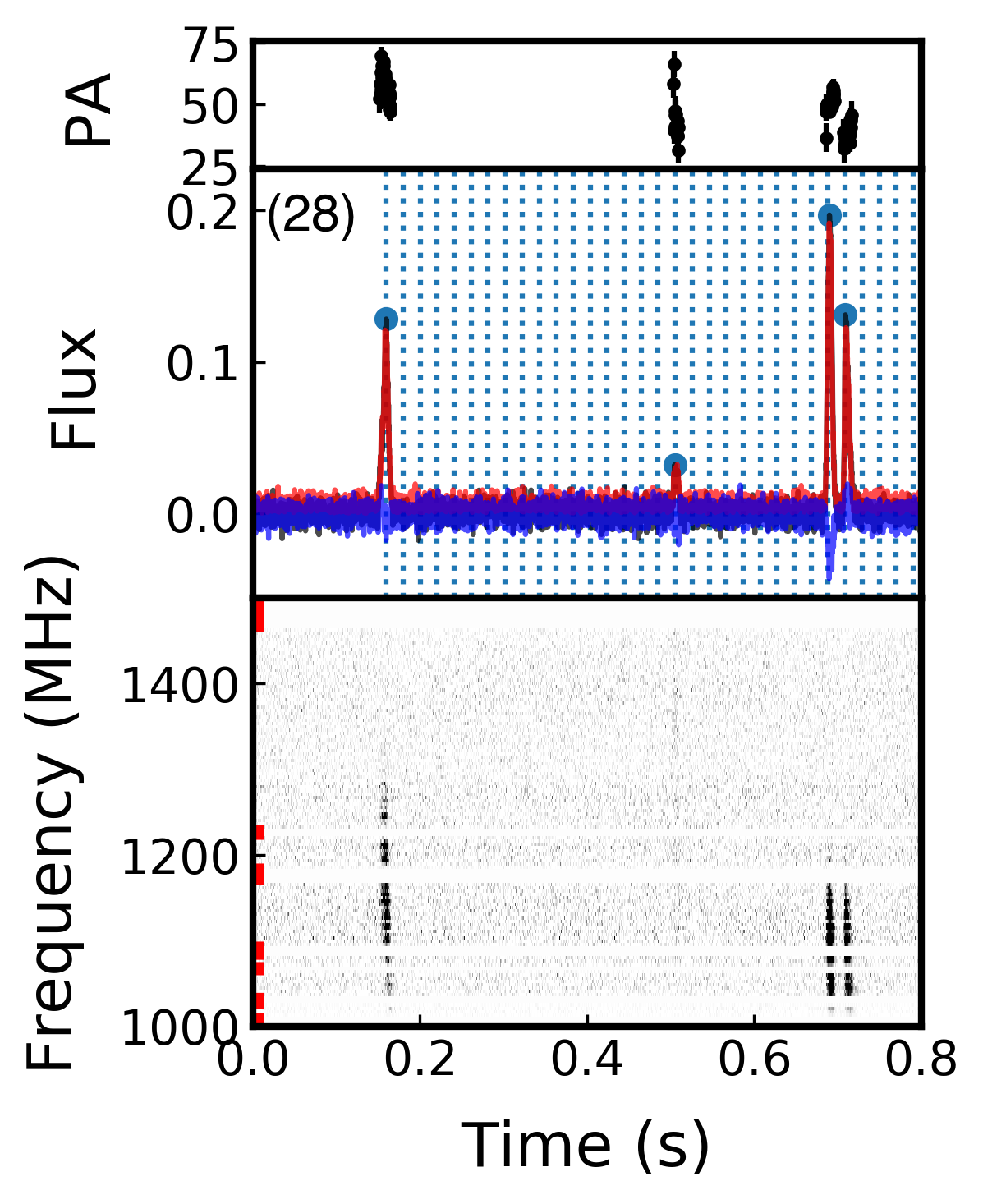}
     \end{subfigure}
     \hfill
     \begin{subfigure}[b]{0.3\textwidth}
         \centering
         \includegraphics[height=2.3in]{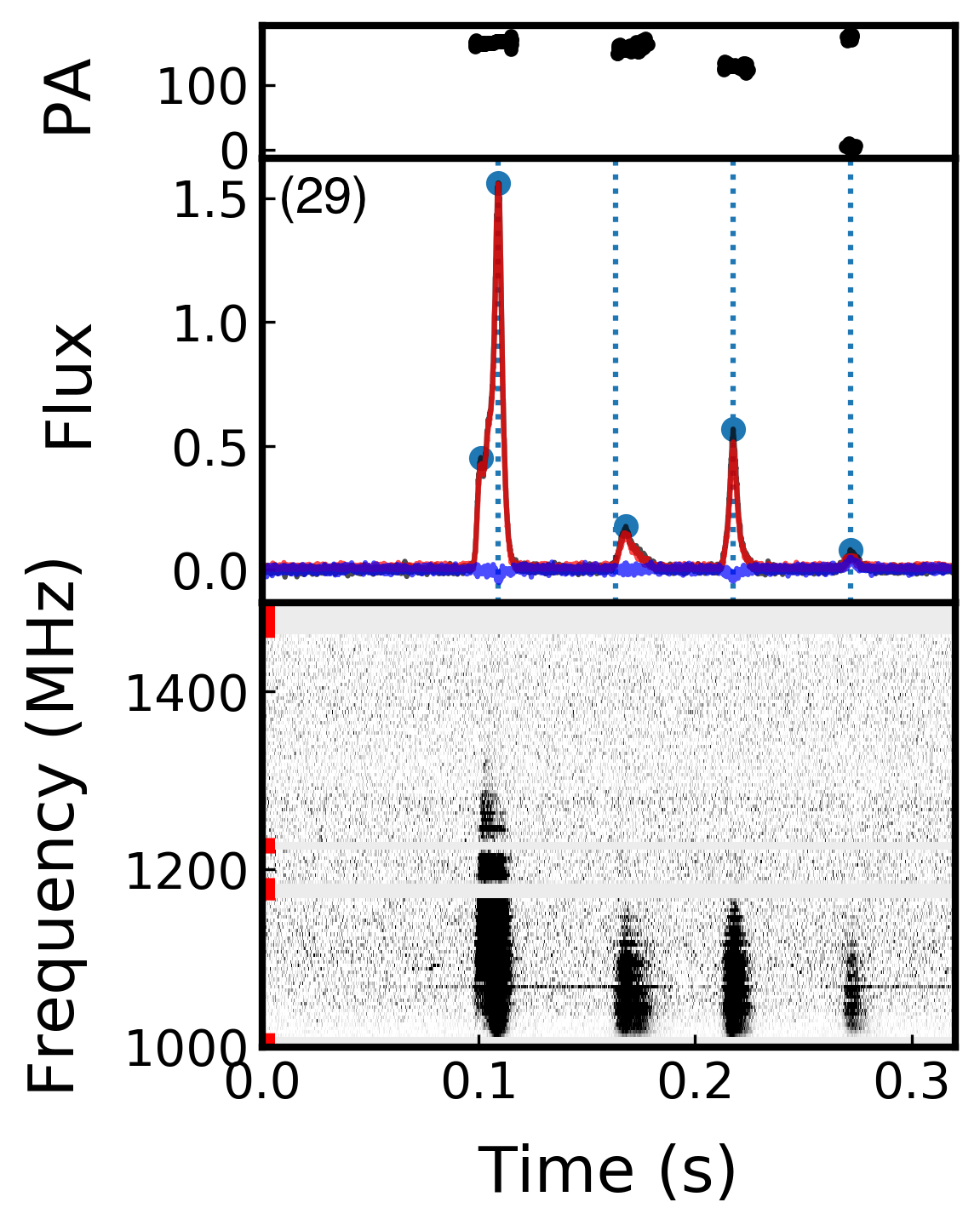}
     \end{subfigure}
     \hfill
     \begin{subfigure}[b]{0.3\textwidth}
         \centering
         \includegraphics[height=2.3in]{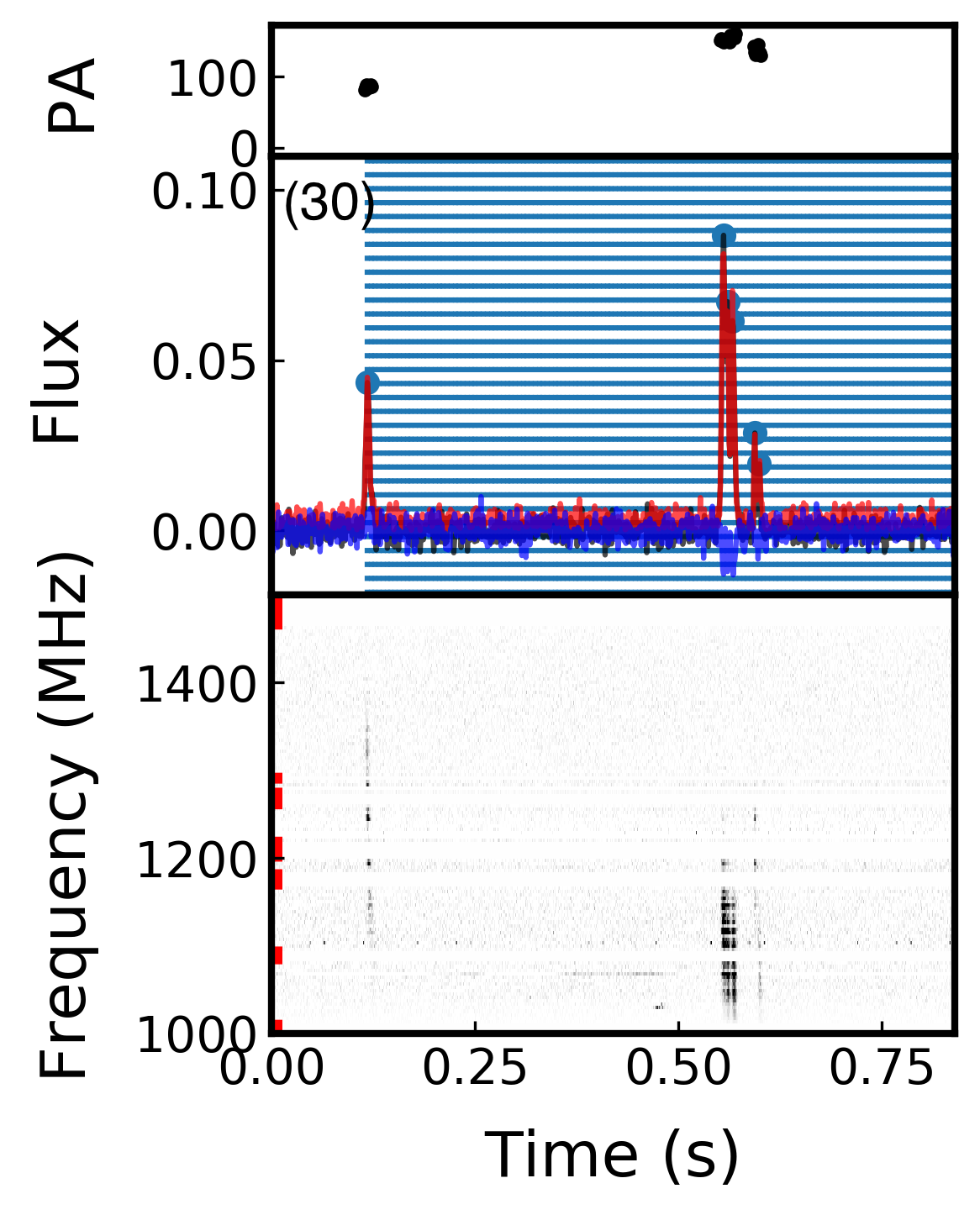}
     \end{subfigure}
     \hfill
        \begin{subfigure}[b]{0.3\textwidth}
         \centering
         \includegraphics[height=2.3in]{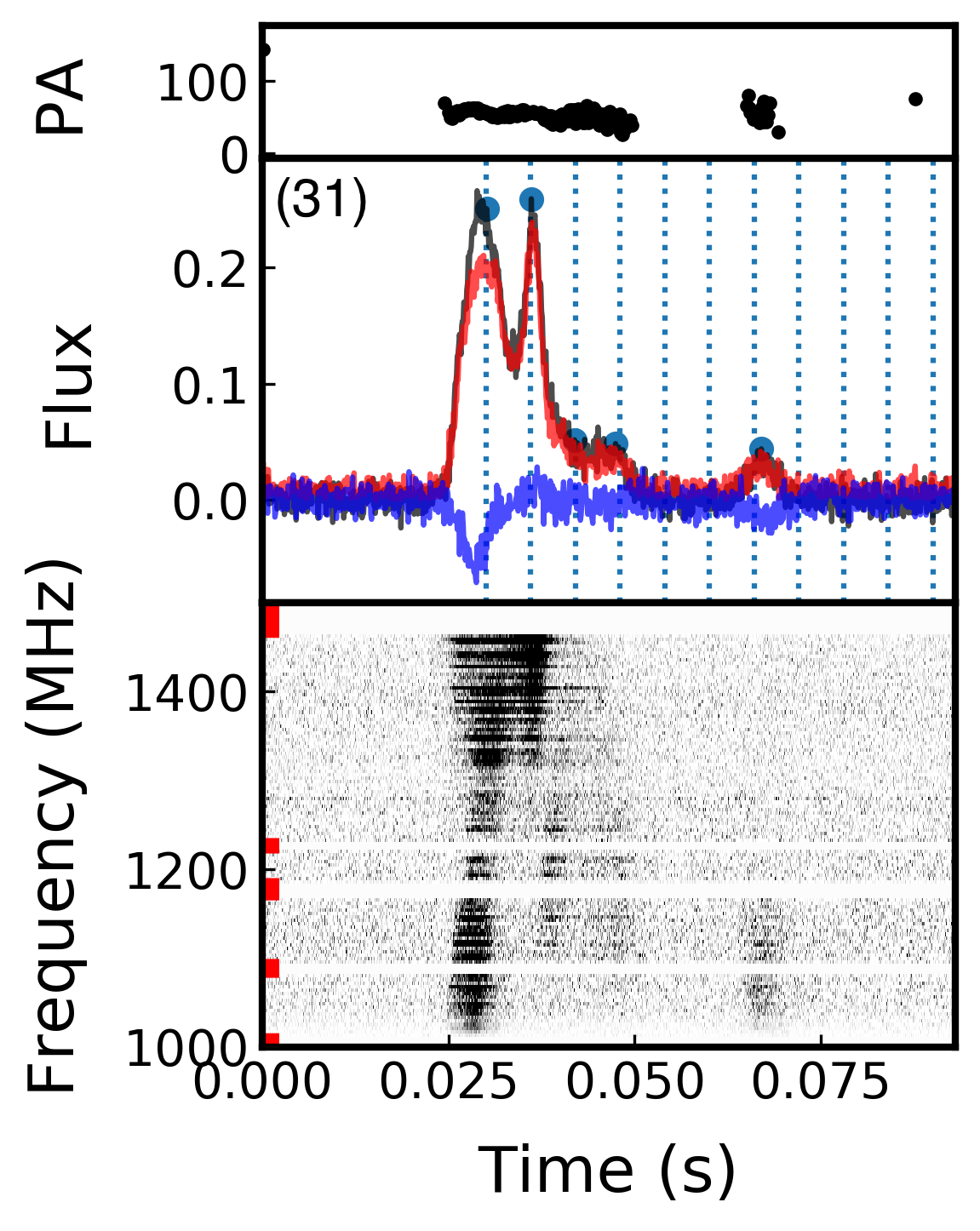}
     \end{subfigure}
     \hfill
     \begin{subfigure}[b]{0.3\textwidth}
         \centering
         \includegraphics[height=2.3in]{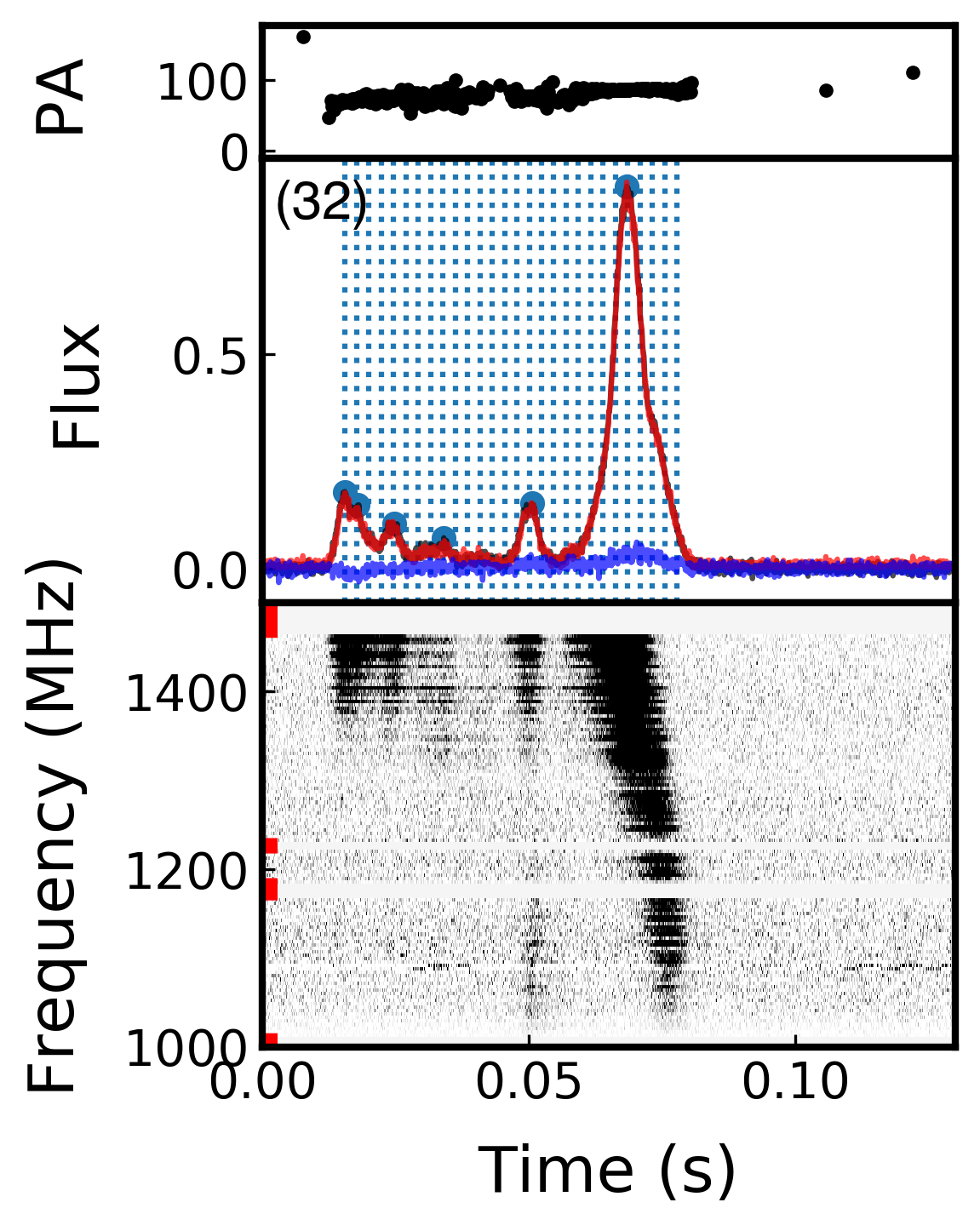}
     \end{subfigure}
     \hfill
     \begin{subfigure}[b]{0.3\textwidth}
         \centering
         \includegraphics[height=2.3in]{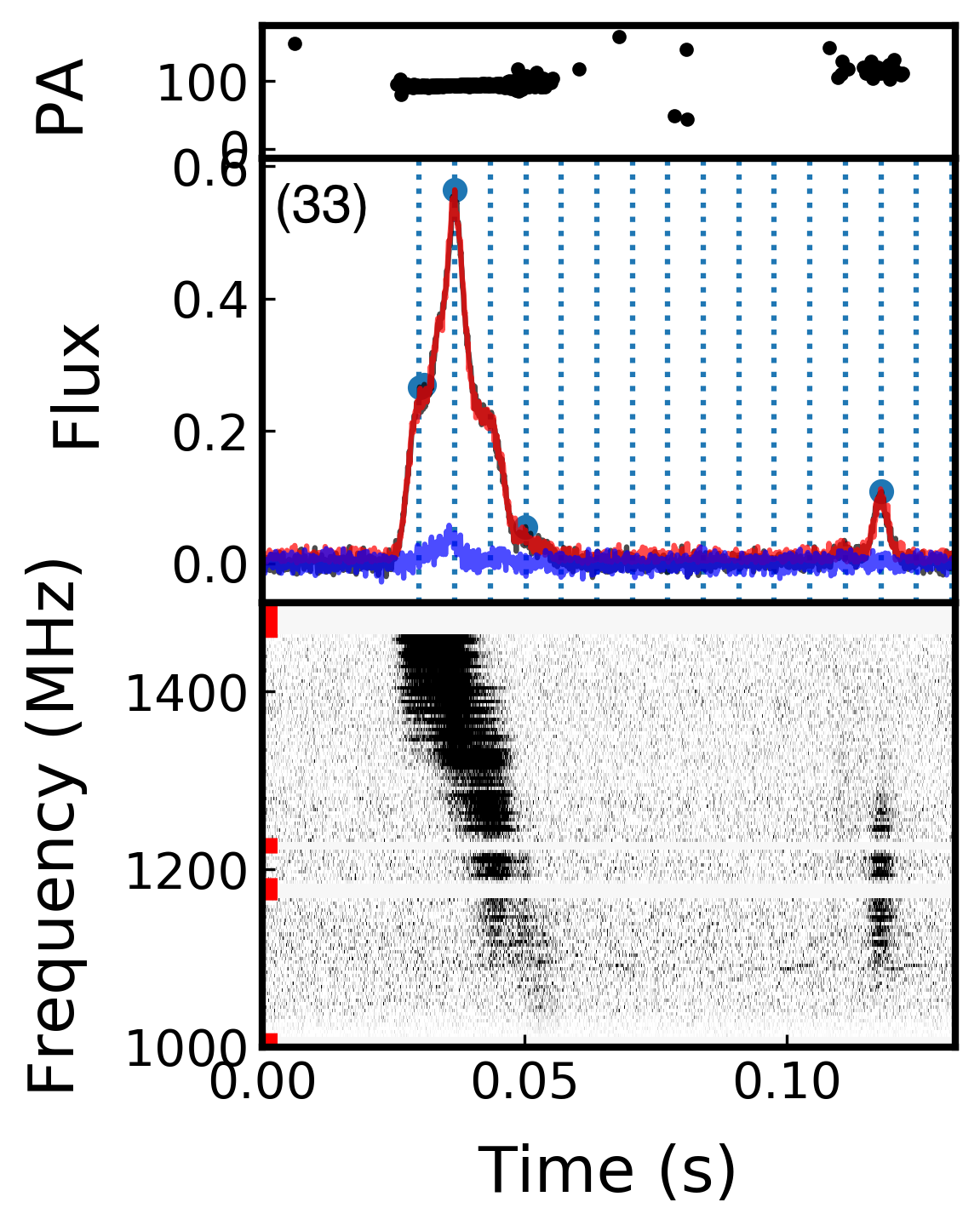}
     \end{subfigure}
     \hfill
     \begin{subfigure}[b]{0.3\textwidth}
         \centering
         \includegraphics[height=2.3in]{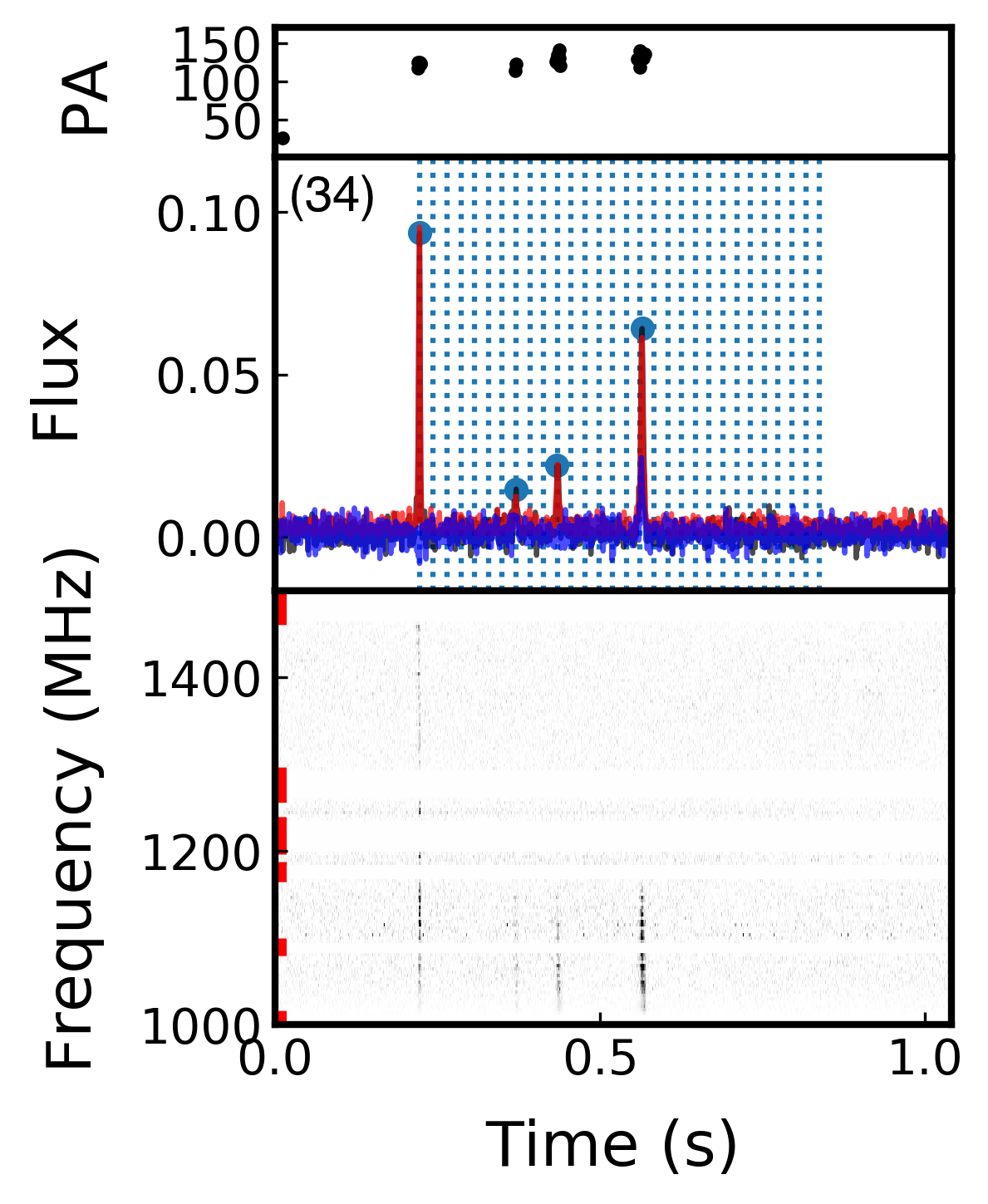}
     \end{subfigure}
     \hfill
     \begin{subfigure}[b]{0.3\textwidth}
         \centering
         \includegraphics[height=2.3in]{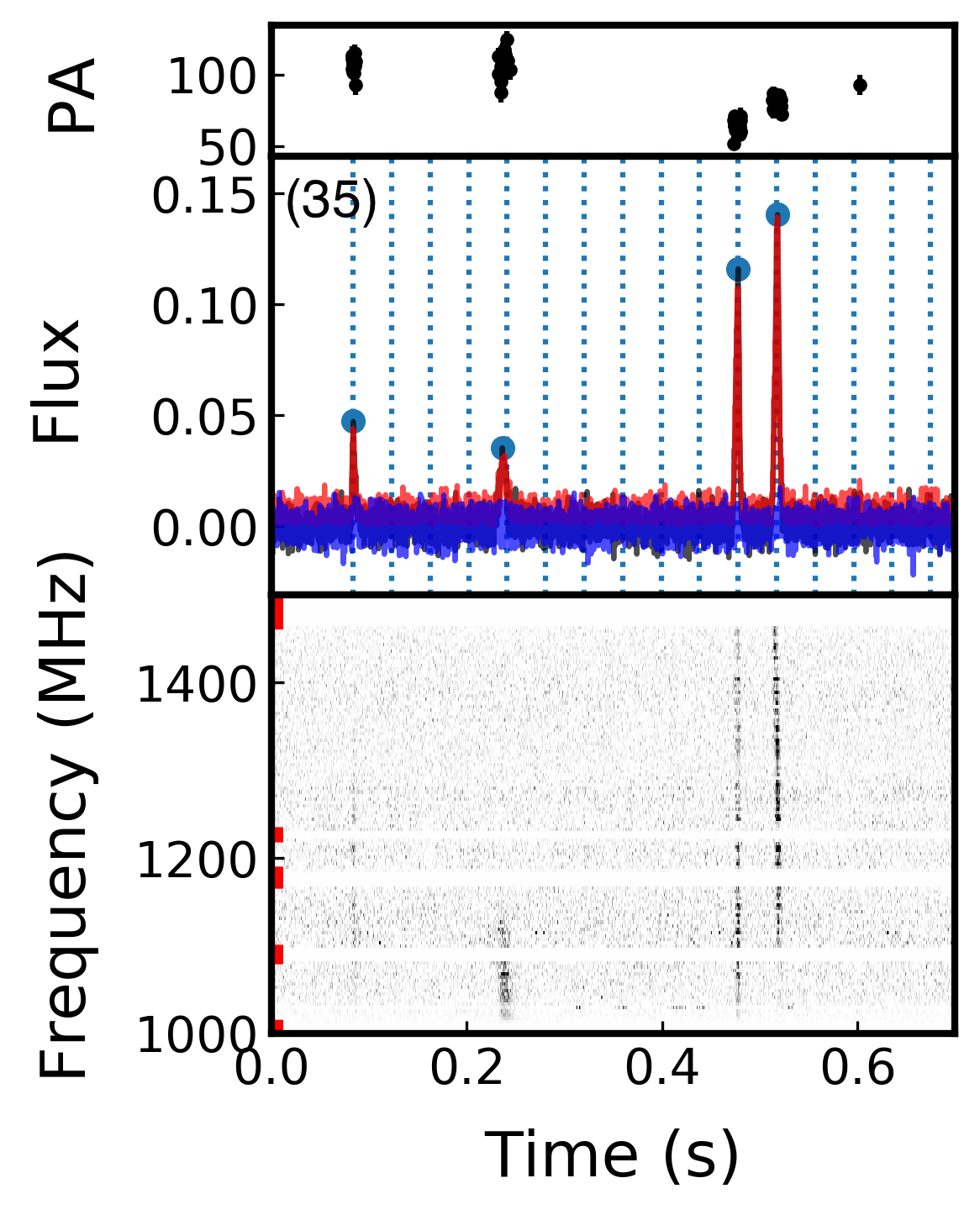}
     \end{subfigure}
     \hfill
     \begin{subfigure}[b]{0.3\textwidth}
         \centering
         \includegraphics[height=2.3in]{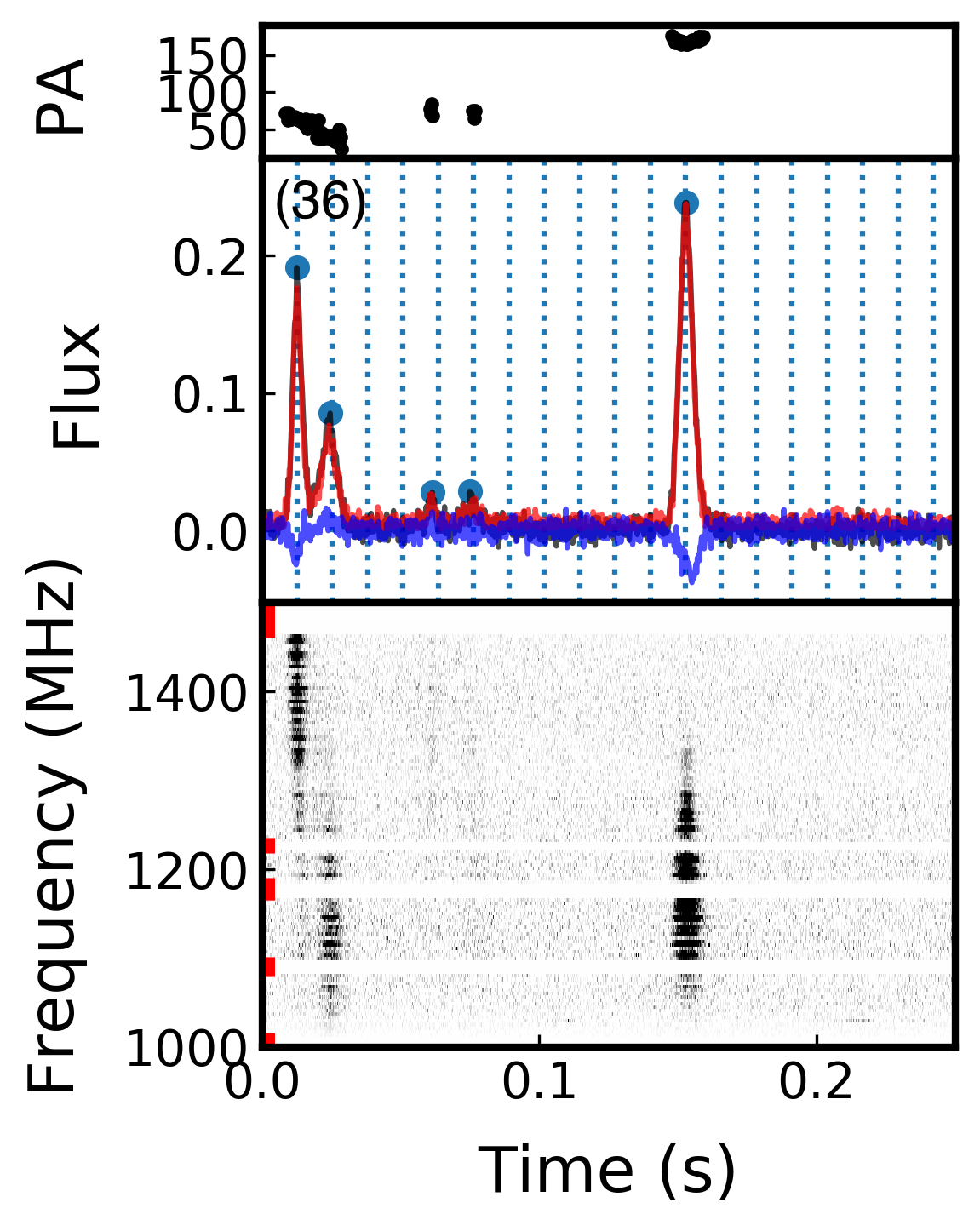}
     \end{subfigure}
 \end{figure*}

 \clearpage
 \begin{figure*}[t!]
  \ContinuedFloat
      \begin{subfigure}[b]{0.3\linewidth}
         \centering
         \includegraphics[height=2.3in]{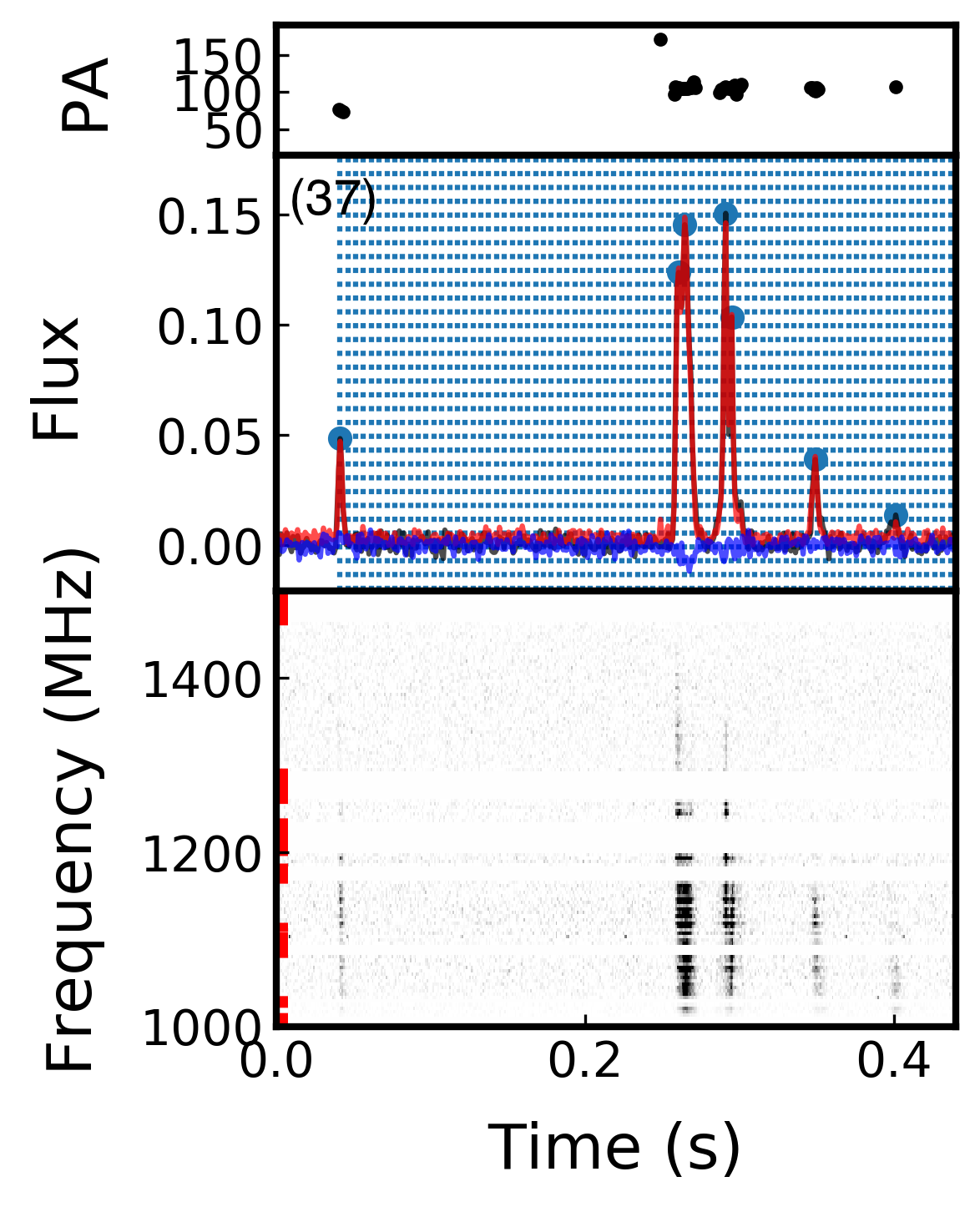}
     \end{subfigure}
     \hfill
     \begin{subfigure}[b]{0.3\linewidth}
         \centering
         \includegraphics[height=2.3in]{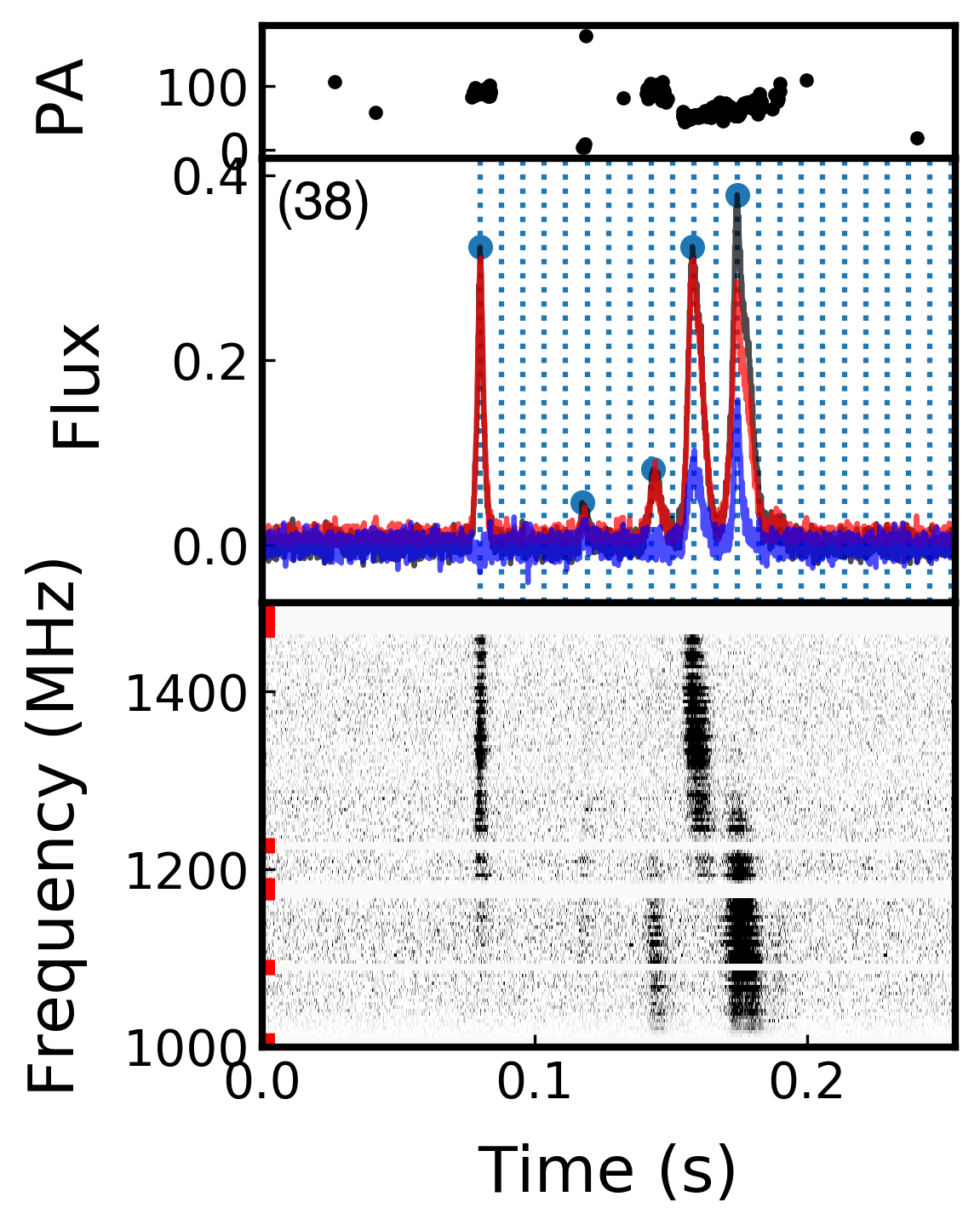}
     \end{subfigure}
     \hfill
     \begin{subfigure}[b]{0.3\linewidth}
         \centering
         \includegraphics[height=2.3in]{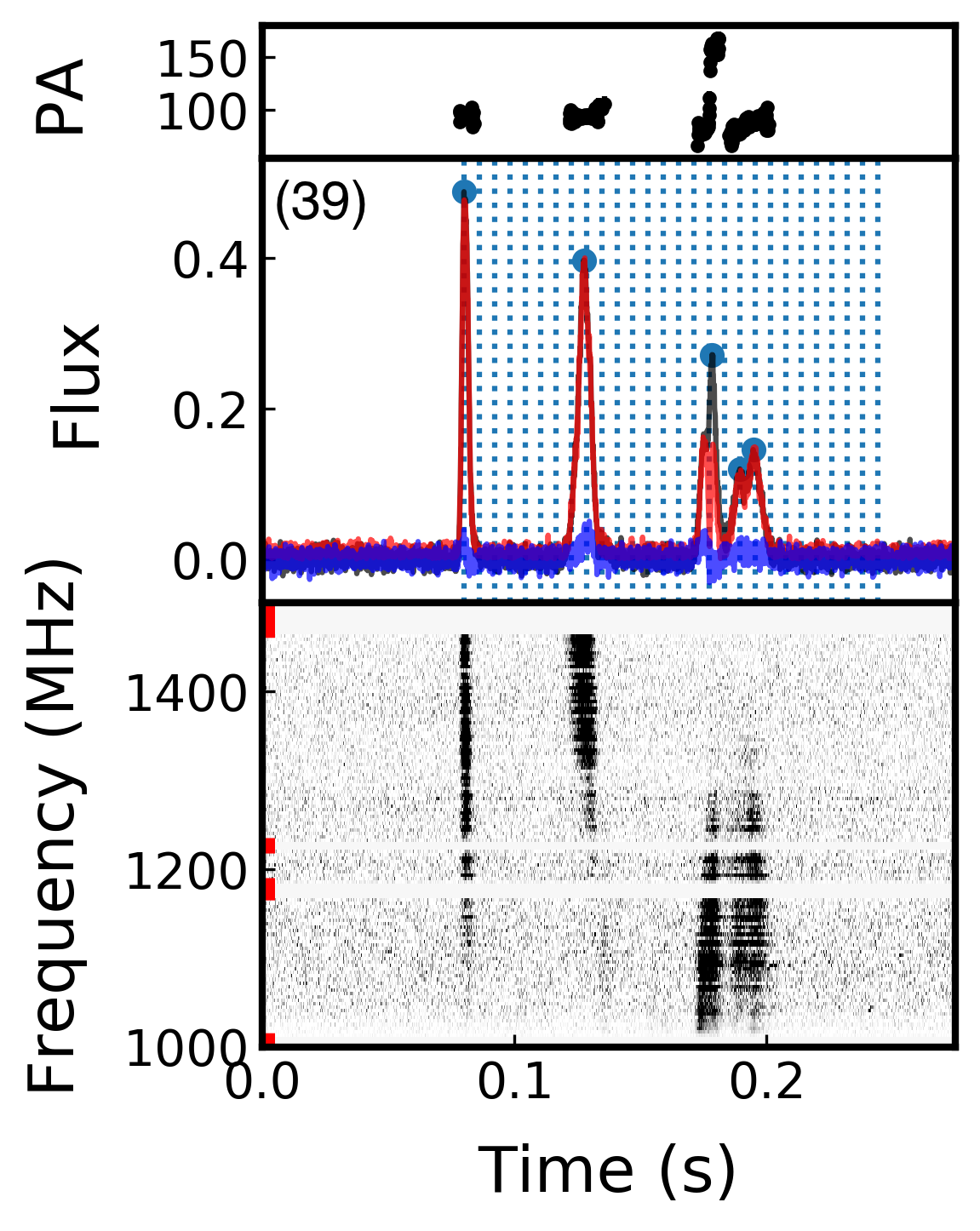}
     \end{subfigure}
    \hfill
     \begin{subfigure}[b]{0.3\textwidth}
         \centering
         \includegraphics[height=2.3in]{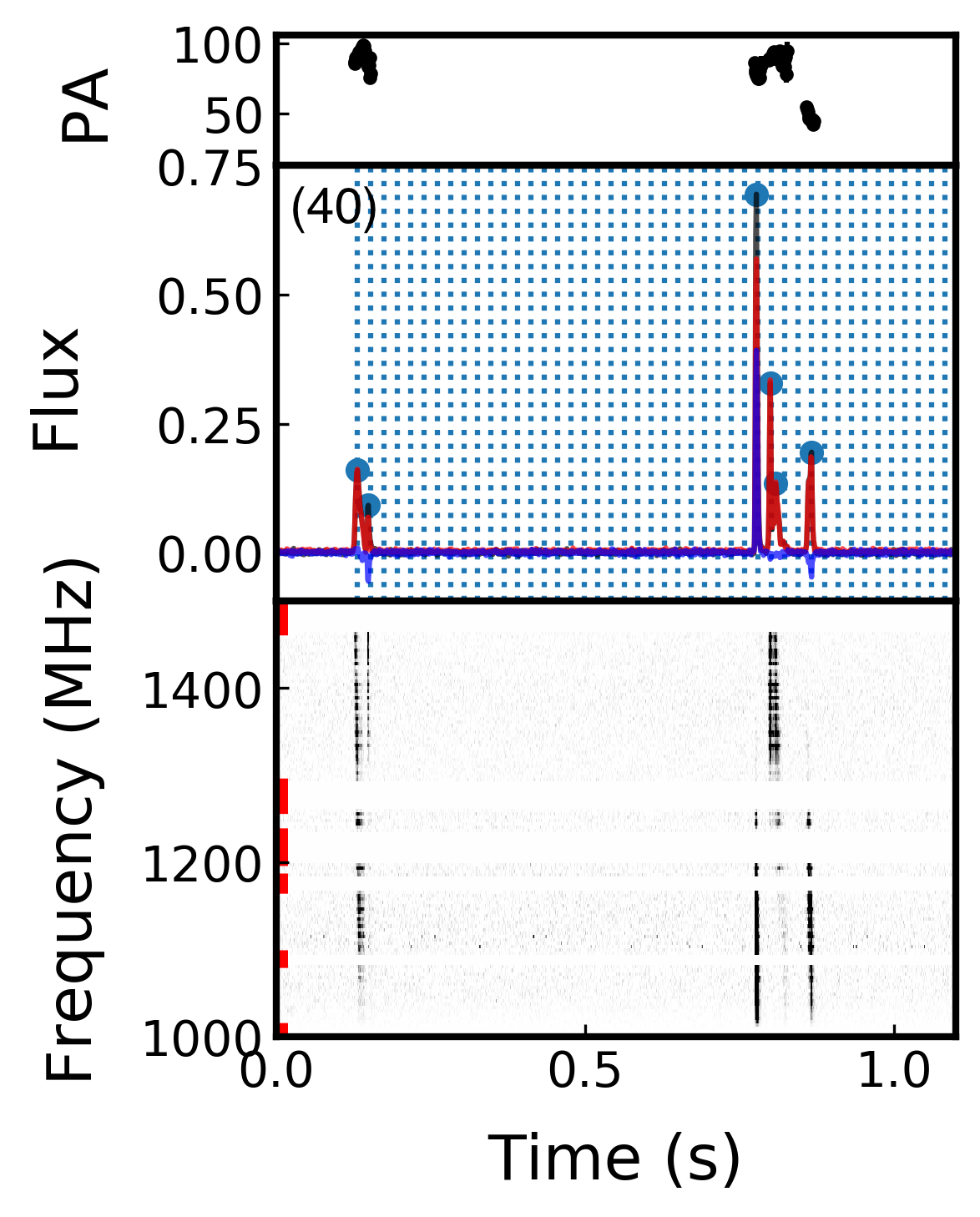}
     \end{subfigure}
     \hfill
     \begin{subfigure}[b]{0.3\textwidth}
         \centering
         \includegraphics[height=2.3in]{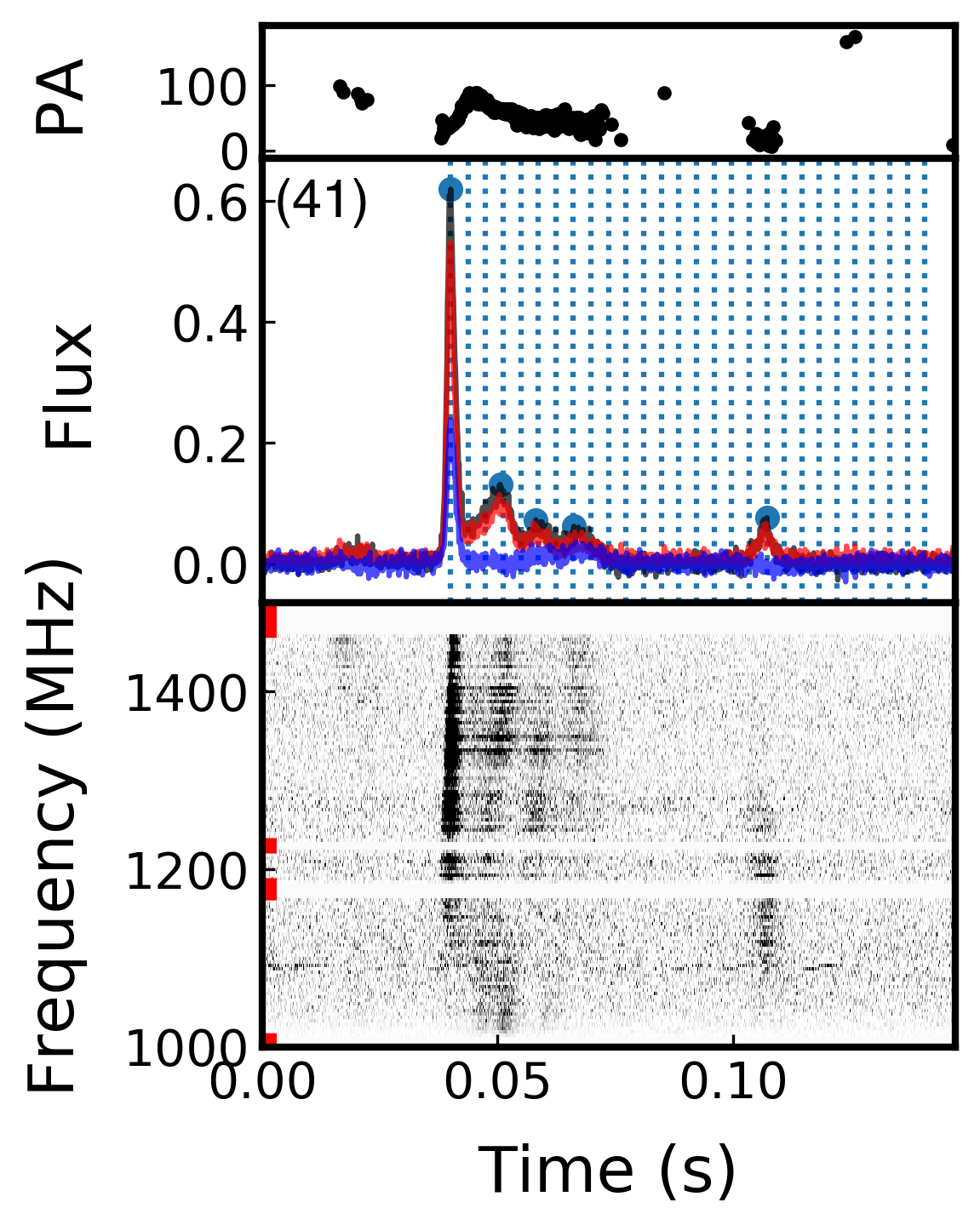}
     \end{subfigure}
     \hfill
     \begin{subfigure}[b]{0.3\textwidth}
         \centering
         \includegraphics[height=2.3in]{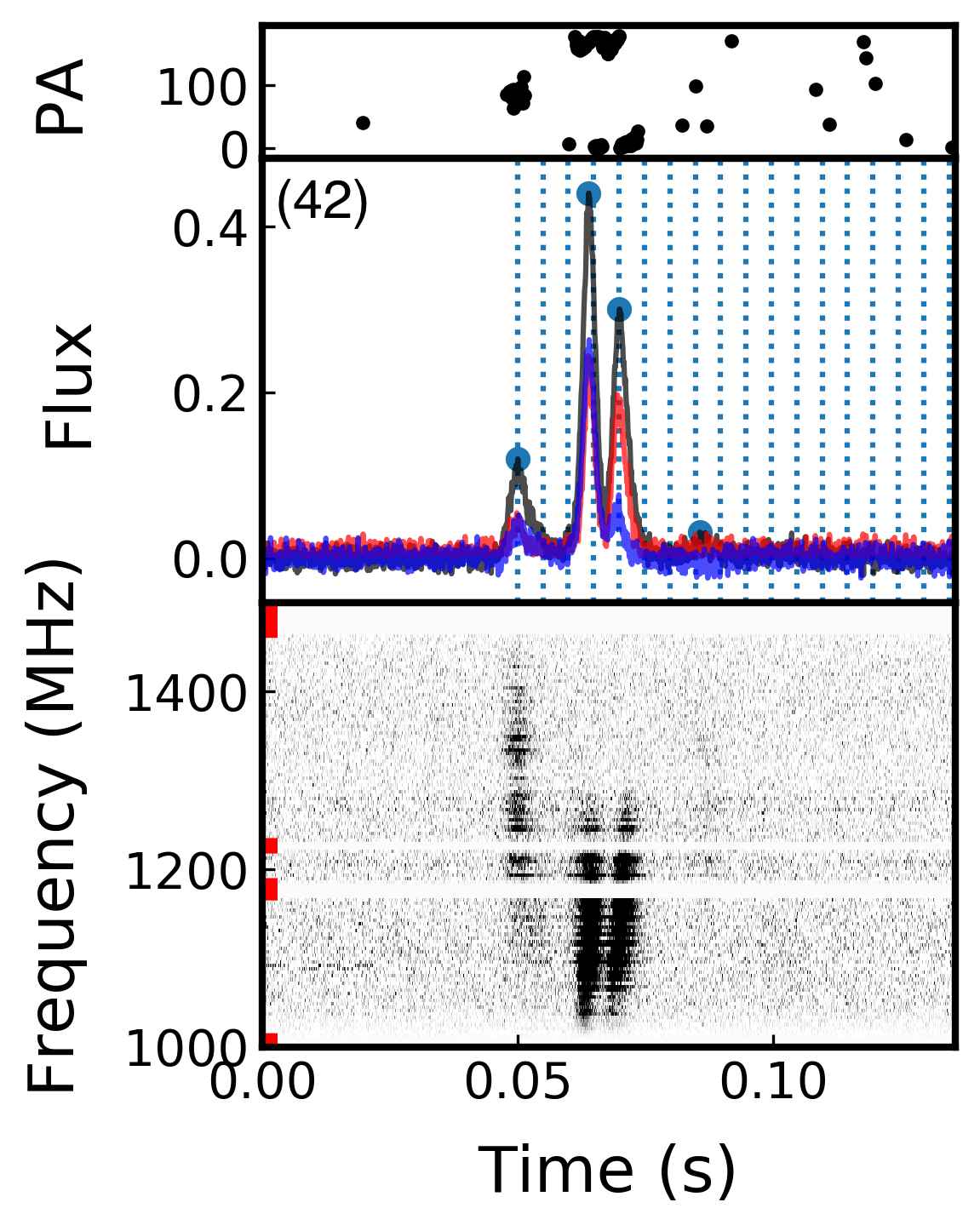}
     \end{subfigure}
     \hfill
     \begin{subfigure}[b]{0.3\textwidth}
         \centering
         \includegraphics[height=2.3in]{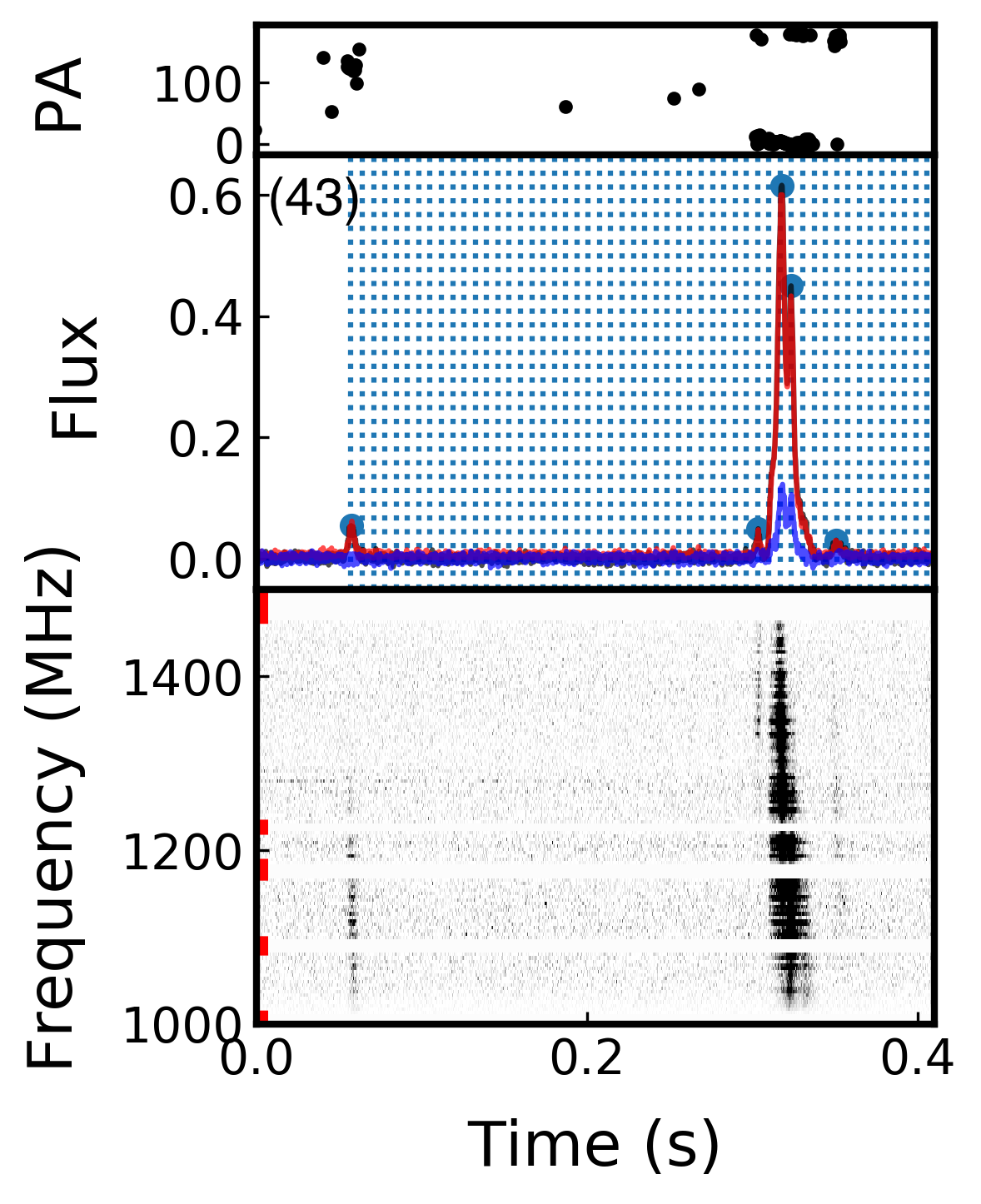}
     \end{subfigure}
     \hfill
     \begin{subfigure}[b]{0.3\textwidth}
         \centering
         \includegraphics[height=2.3in]{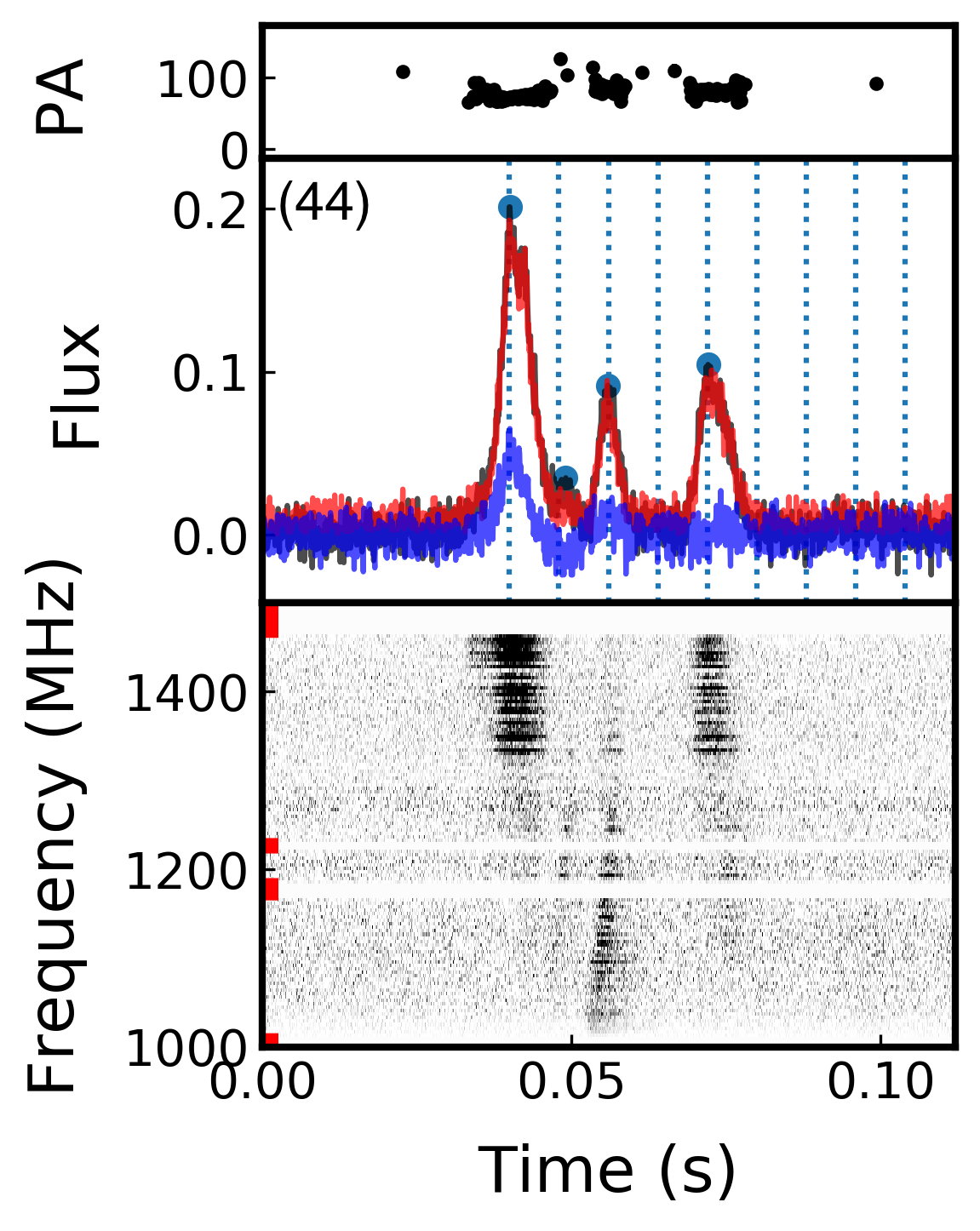}
     \end{subfigure}
     \hfill
     \begin{subfigure}[b]{0.3\textwidth}
         \centering
         \includegraphics[height=2.3in]{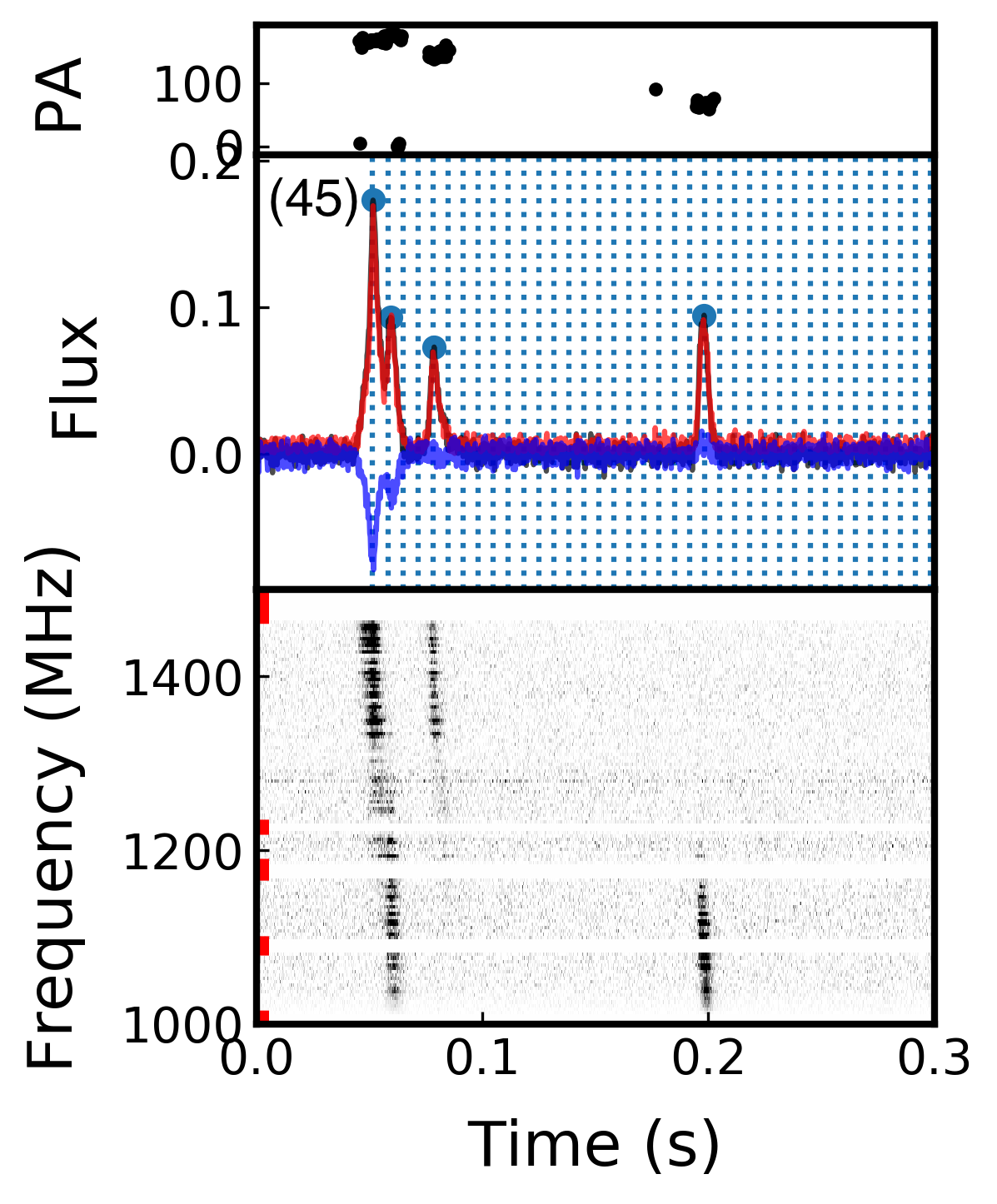}
     \end{subfigure}
     \hfill
     \begin{subfigure}[b]{0.3\textwidth}
         \centering
         \includegraphics[height=2.3in]{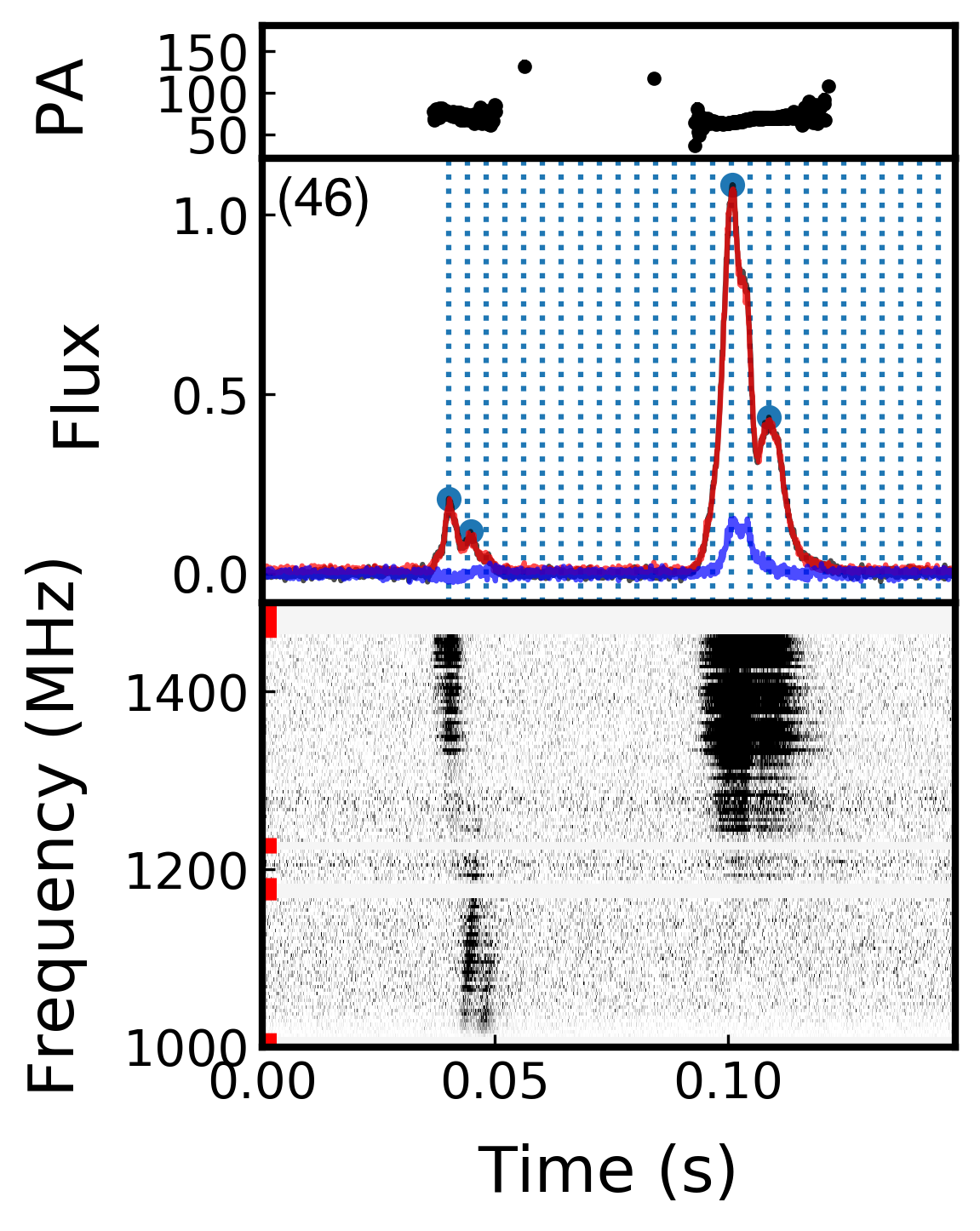}
     \end{subfigure}
     \hfill
     \begin{subfigure}[b]{0.3\textwidth}
         \centering
         \includegraphics[height=2.3in]{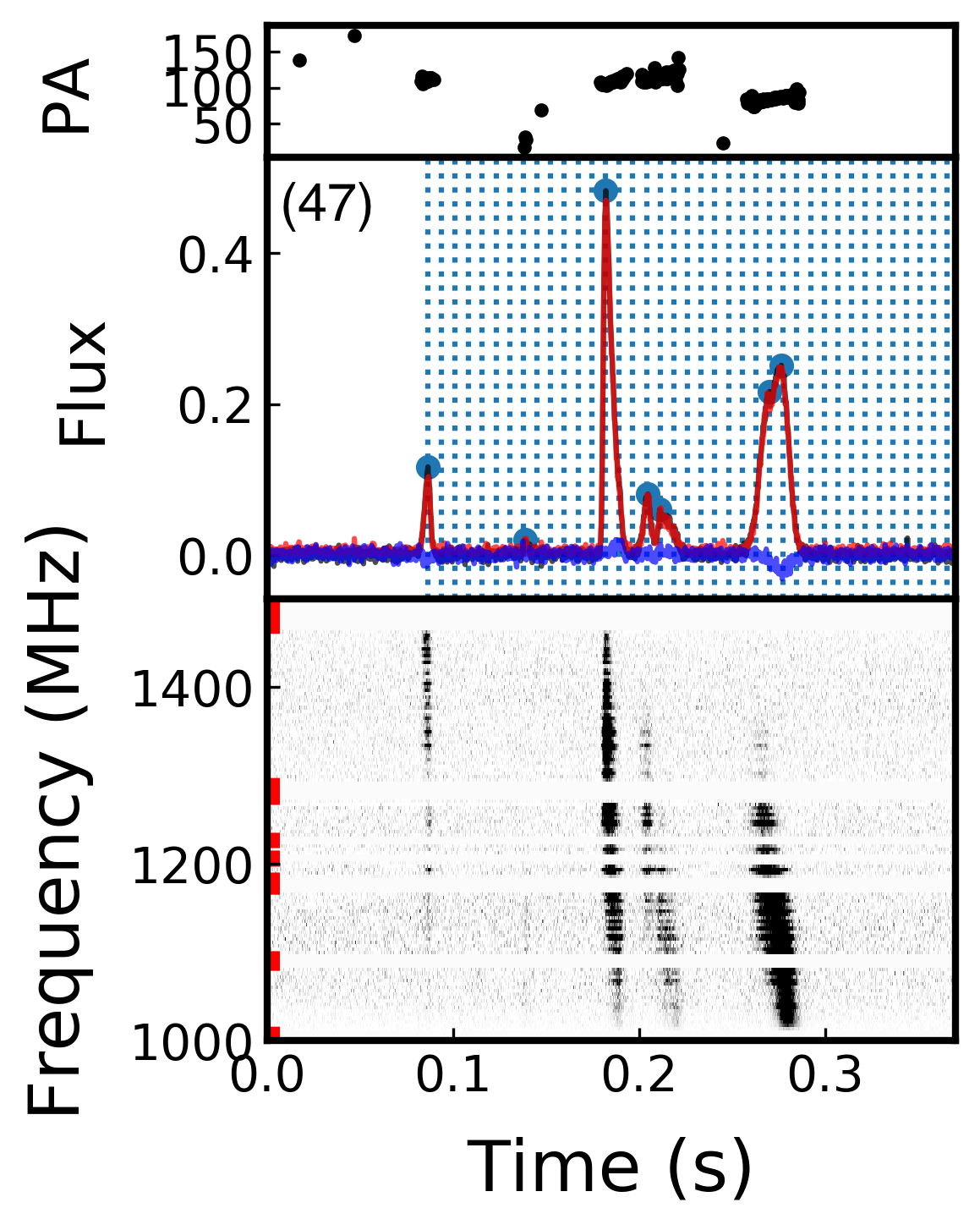}
     \end{subfigure}
     \hfill
     \begin{subfigure}[b]{0.3\textwidth}
         \centering
         \includegraphics[height=2.3in]{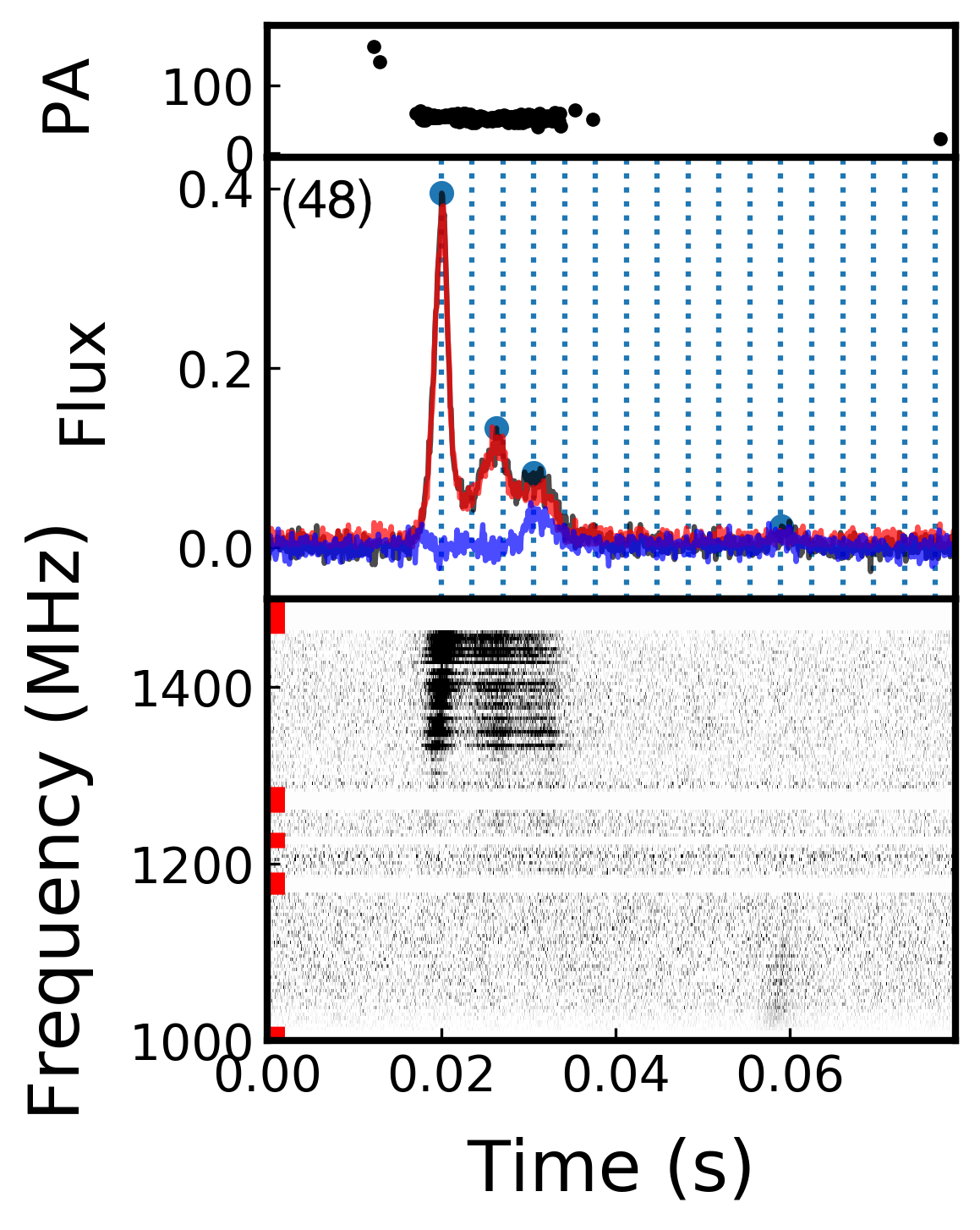}
     \end{subfigure}
 \end{figure*}   

 \clearpage
 \begin{figure*}[t!]
  \ContinuedFloat
     \begin{subfigure}[b]{0.3\textwidth}
         \centering
         \includegraphics[height=2.3in]{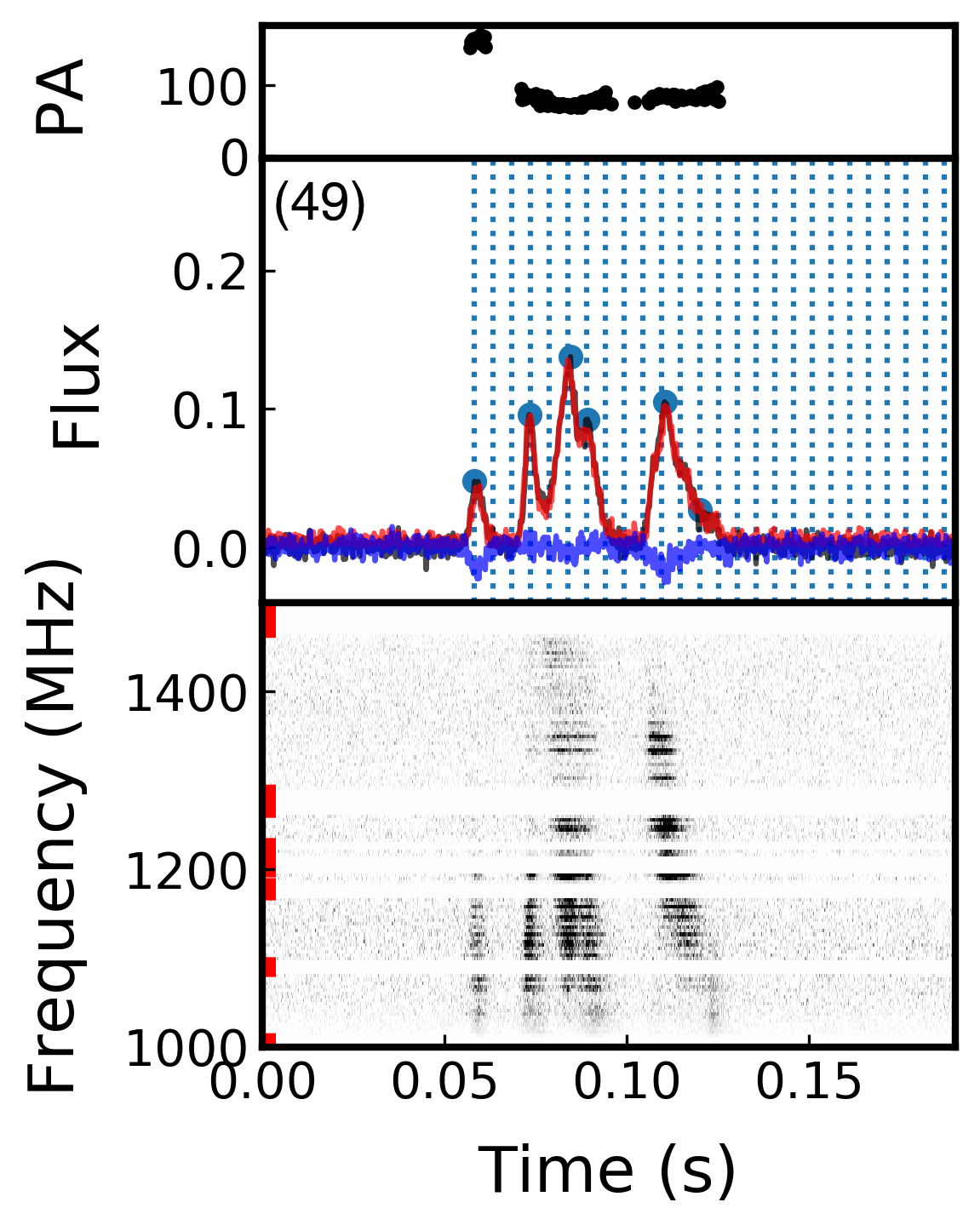}
     \end{subfigure}
     \hfill
     \begin{subfigure}[b]{0.3\textwidth}
         \centering
         \includegraphics[height=2.3in]{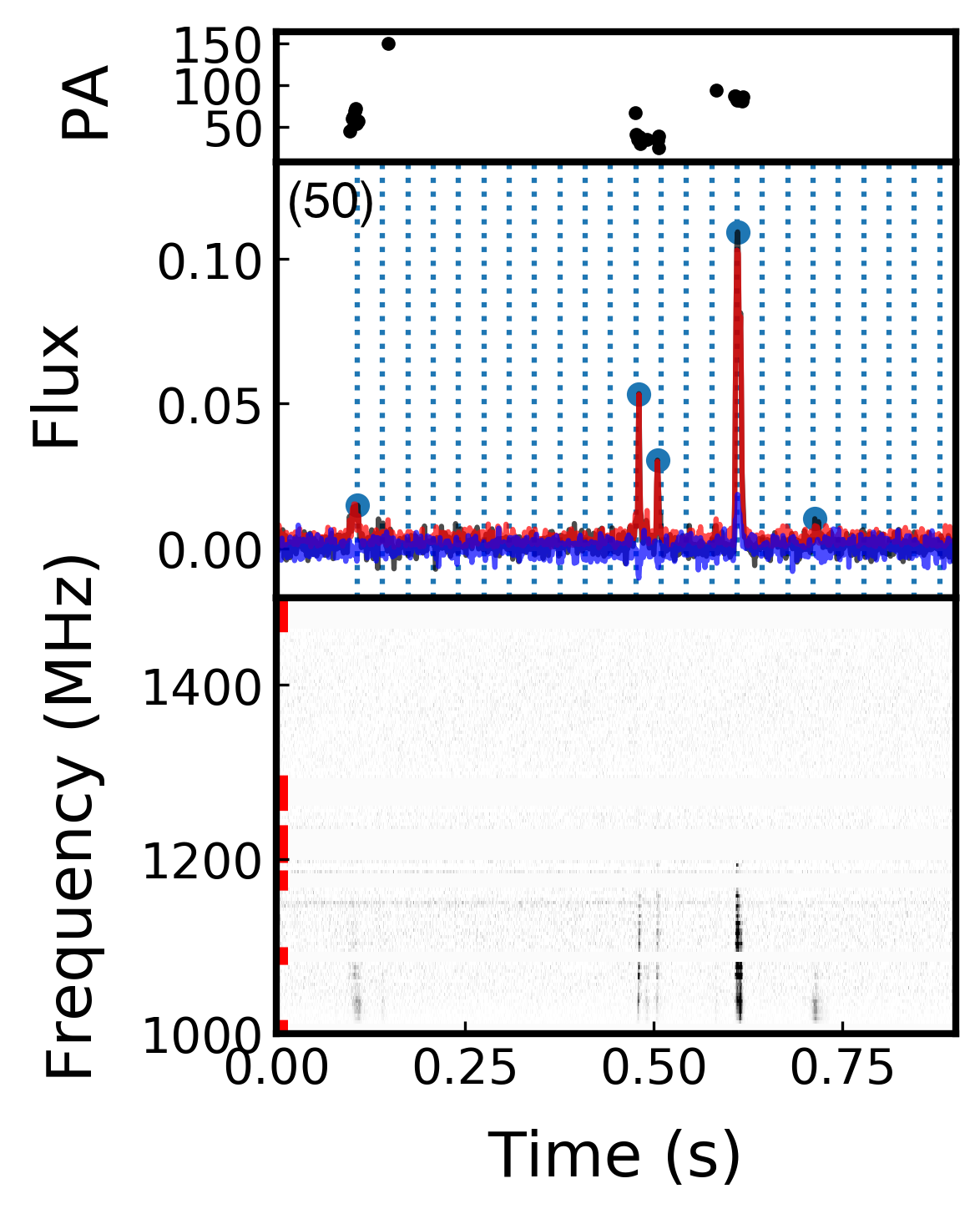}
     \end{subfigure}
     \hfill
     \begin{subfigure}[b]{0.3\textwidth}
         \centering
         \includegraphics[height=2.3in]{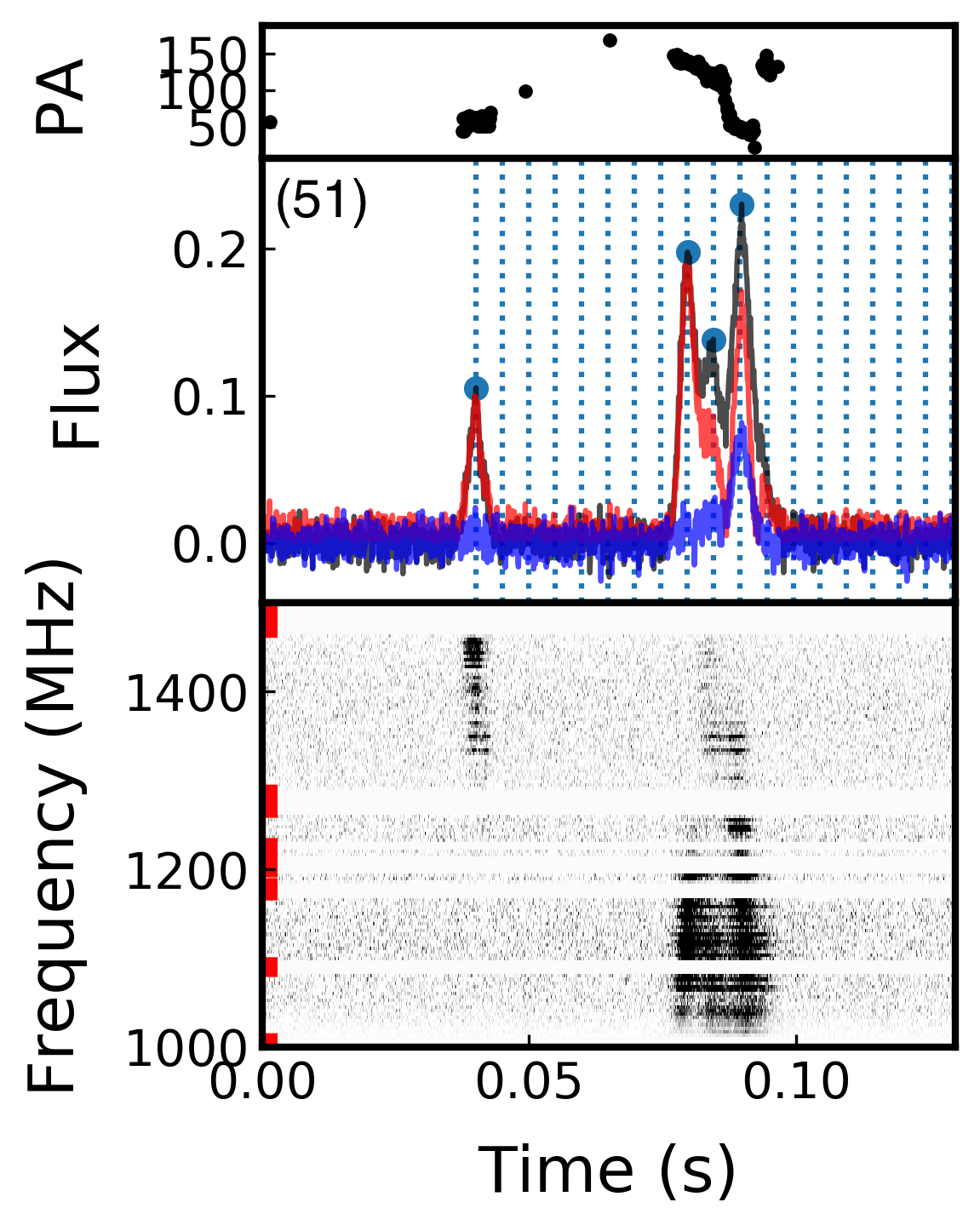}
     \end{subfigure}
     \hfill
     \begin{subfigure}[b]{0.3\textwidth}
         \centering
         \includegraphics[height=2.3in]{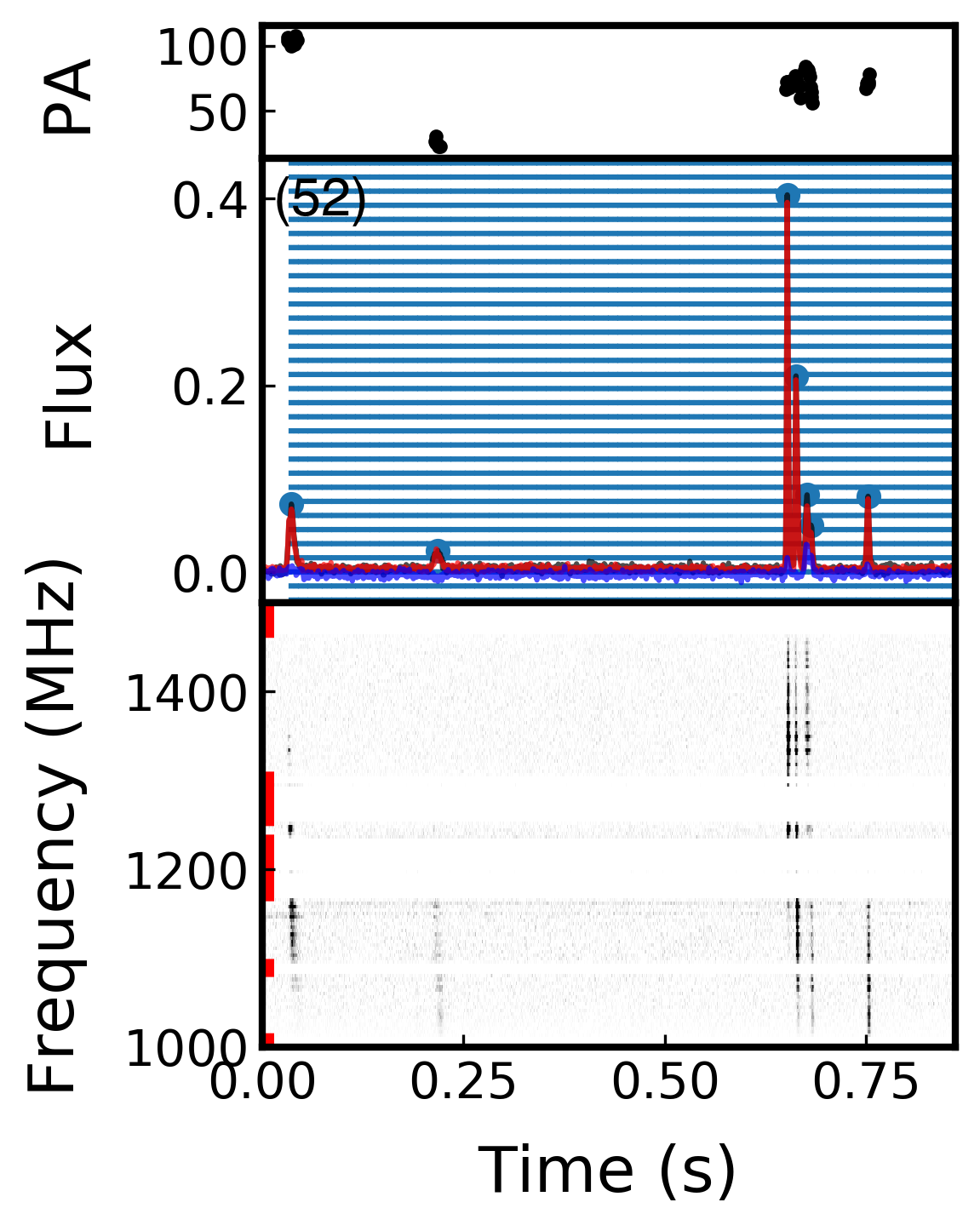}
     \end{subfigure}
     \hfill
     \begin{subfigure}[b]{0.3\textwidth}
         \centering
         \includegraphics[height=2.3in]{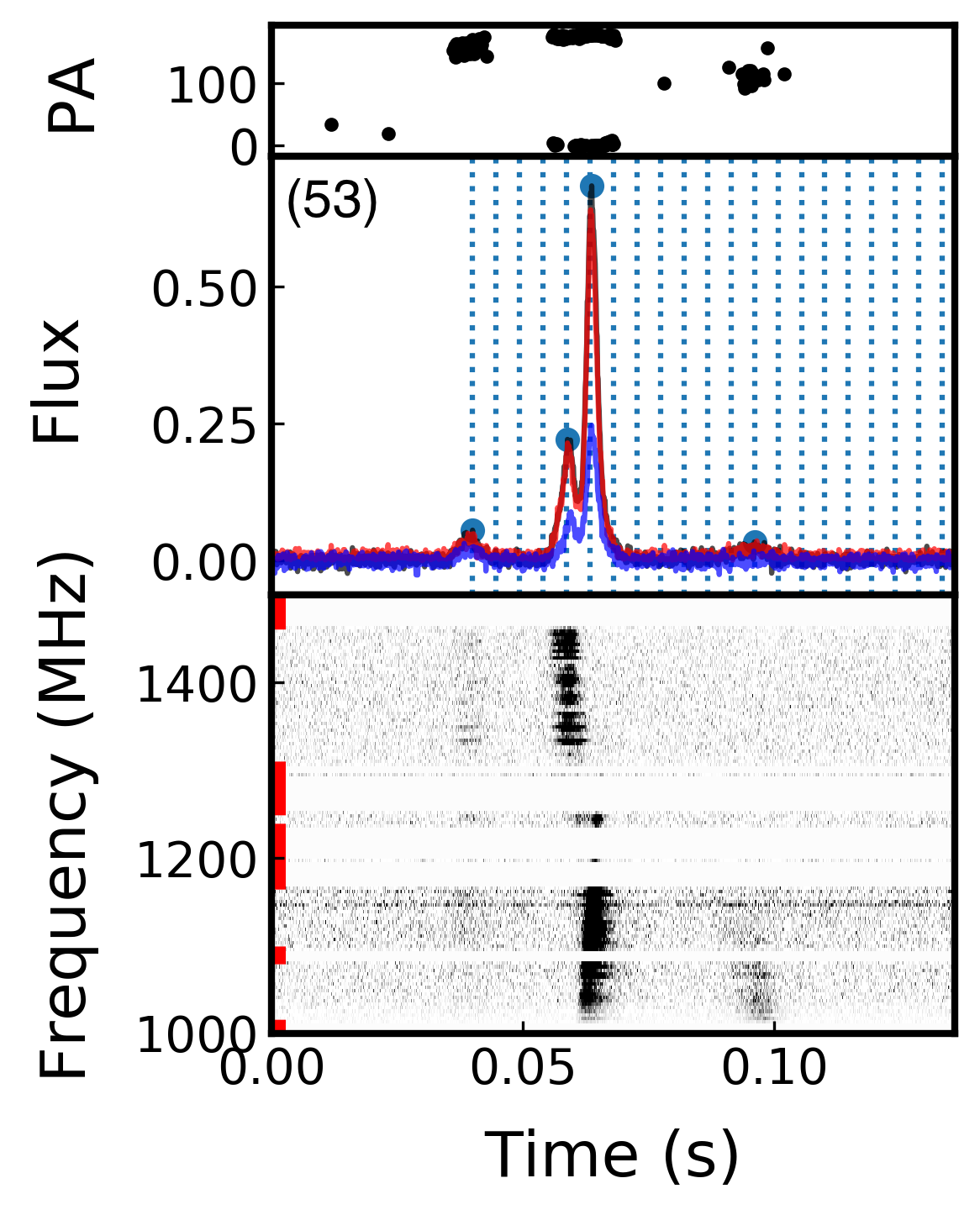}
     \end{subfigure}
      \hfill
     \begin{subfigure}[b]{0.3\textwidth}
         \centering
         \includegraphics[height=2.3in]{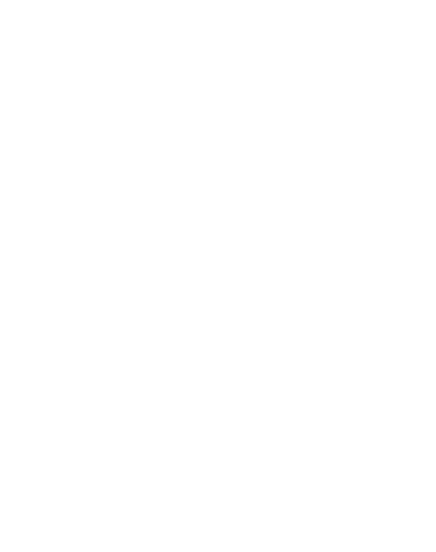}
     \end{subfigure}
          \caption{Examples of multi-components.}       
          \label{tab:mulall}
 \end{figure*}    

\clearpage

\begin{acknowledgements}
We thank the referee, Scott Ransom, for helpful comments. 
This work made use of data from the FAST, a Chinese national mega-science facility, built and operated by the National Astronomical Observatories, Chinese Academy of Sciences. 
This is work is supported by the National SKA Program of China No. 2020SKA0120200, 2020SKA0120100 
the National Nature Science Foundation grant No. 12041303, 11873067, 12041304 
the National Key R$\&$D Program of China No. 2017YFA0402600, 2021YFA0718500, 2017YFA0402602, 
the CAS-MPG LEGACY project, the Max-Planck Partner Group, the Key Research
Project of Zhejiang Lab no. 2021PE0AC0. 
J.~L. Han is supported by the National Natural Science Foundation of
China (NSFC, Nos. 11988101 and 11833009) and the Key Research
Program of the Chinese Academy of Sciences (Grant No. QYZDJ-SSW-SLH021);
D.~J. Zhou is supported by the Cultivation Project for the FAST scientific Payoff and Research Achievement of CAMS-CAS.

\end{acknowledgements}

\section*{Authors contributions}
Jia-Rui Niu did the main data processing and analysis and the article writing.
Wei-Wei Zhu led the writing of this paper and conducted the $P$-$\dot{P}$ search section of the data analysis.
Bing Zhang and Wei-Wei Zhu proposed and chaired the FAST FRB key science project. 
Bing Zhang, Wei-Wei Zhu, J.~L. Han, Di Li and Ke-jia Lee coordinated the teamwork, the observational campaign, co-supervised data analyses and interpretations.
Mao Yuan helped in interference mitigation.
De-Jiang Zhou analyzed burst morphology and the results are presented in paper I of this series.
Yong-Kun Zhang analyzed the energy distribution and the details are presented in paper II of this series.
Jin-Chen Jiang analyzed the polarization and the results are presented in paper III.
Pei Wang, Yi Feng, Dong-Zi Li, Chen-Chen Miao, Chen-Hui Niu, Heng Xu, Rui Luo, Chun-Feng Zhang, Wei-Yang Wang, Bo-Jun Wang, Jiang-Wei Xu participated in the discussion and revision of the paper.

\section*{Data availability} 
All data for the plots in this paper, including these in appendix, can be obtained from the authors with a kind request.

\appendix                 

\clearpage
\bibliographystyle{raa}
\bibliography{bibtex}

\label{lastpage}

\end{document}